\documentclass[aps, pra, superscriptaddress, longbibliography,  twocolumn, nofootinbib, bibnotes]{revtex4-2}

\usepackage[normalem]{ulem}
\usepackage{latexsym,exscale,enumerate,comment, bbold}
\usepackage{amsmath,amsthm,amscd,amsfonts,amssymb}
\usepackage{mathrsfs,float}
\usepackage{tikz}
\usepackage{tikz-cd}
\usetikzlibrary{shapes,snakes}
\usetikzlibrary{decorations.markings}
\usetikzlibrary{decorations.pathreplacing}
\tikzstyle directed=[postaction={decorate,decoration={markings,
    mark=at position #1 with {\arrow{>}}}}]
\usepackage{blkarray}
\input xy
\usepackage[all]{xy}
\newcommand{\hackcenter}[1]{
 \xy (0,0)*{#1}; \endxy}
\usepackage{braket}
\usepackage{caption}
\usepackage{subcaption}
\usepackage{ragged2e}

\newcommand{\scs}{\scriptstyle}

\usepackage{url}
\usepackage[bookmarks=true,%
    colorlinks=true,%
    linkcolor=blue,%
    citecolor=blue,%
    filecolor=blue,%
    menucolor=blue,%
    urlcolor=blue,%
    breaklinks=true]{hyperref}

\newcommand{\maps}{\colon}
\newcommand{\cat}{\mathscr{C}}

\usepackage{graphicx}
\usepackage{color}
 
\theoremstyle{plain}
\newtheorem{theorem}{Theorem}

\newtheorem{proposition}[theorem]{Proposition}
\newtheorem{notation}[theorem]{Notation}
\newtheorem{lemma}[theorem]{Lemma}

\theoremstyle{definition}
\newtheorem{example}[theorem]{Example}
\newtheorem{definition}[theorem]{Definition}

\theoremstyle{definition}
\newtheorem{remark}[theorem]{Remark}

\newcommand{\Hom}{{\rm Hom}}

\newcommand{\Tr}{{\rm Tr}}
\renewcommand{\to}{\rightarrow}

\newcommand{\Br}{{\rm Br}}

\def\Jz{K}
\def\Jp{E}
\def\Jm{F}

\def\Y{{Y_c^{ab}}}
\def\Yd{{Y_c^{ab}}^\dag}

\def\b{\mathsf{b}}
\def\tb{\tilde{\mathsf{b}}}

\def\Id{\mathrm{Id}}

\def\mf{\mathfrak}

\def\Br{{\mathrm{Br}}}

\newcommand{\ro}{r}
\newcommand{\qr}{{q}}
\newcommand{\qdim}{\operatorname{qdim}}

\numberwithin{equation}{section}
\newcommand{\wb}{\overline}

\newcommand{\slt}{{\mathfrak{sl}_2}}
\newcommand{\Uq}{{U_q\slt}}
\newcommand{\UqMed}{{\wb U_q\slt}}

\newcommand{\Ubar}{{\wb U_q^{H}(\slt)}}

\let\tilde=\widetilde

\let\epsilon=\varepsilon

\newcommand{\coev}{i}
\newcommand{\ev}{e}
\newcommand{\tev}{e'}
\newcommand{\tcoev}{i'}

\newcommand{\brk}[1]{{\left\langle{#1}\right\rangle}}

\newcommand{\Proj}{{\mathsf {Proj}}}

\newcommand{\ptr}{\operatorname{ptr}}
\newcommand{\End}{\operatorname{End}}

\newcommand{\md}{\operatorname{\mathsf{d}}}
\newcommand{\mt}{\operatorname{\mathsf{t}}}
\newcommand{\ob}{\operatorname{Ob}(\cat)}

\newcommand{\qn}[1]{{\left\{#1\right\}}}
\newcommand{\qN}[1]{{\left[#1\right]}}

\newcommand{\unit}{\ensuremath{\mathbb{I}}}

\usepackage{bbm}
\def\C{{\mathbb{C}}}
\def\N{{\mathbb N}}

\def\Z{{\mathbb Z}}

\def\H{{\mathcal{H}}}
\def\1{\mathbbm{1}}%
\usepackage{bm}

\allowdisplaybreaks

\begin{document}
\title{Robust Universal Braiding with Non-semisimple Ising Anyons}

\author{Filippo Iulianelli}
\thanks{Contact author: iulianel@usc.edu}
 \affiliation{Department of Physics, University of Southern California, Los Angeles, California 90089,USA}

\author{Sung Kim}
\affiliation{Department of Mathematics,
 University of Southern California,
  Los Angeles, California 90089, USA}
\author{Joshua Sussan}
\affiliation{Department of Mathematics,
  CUNY Medgar Evers,
  Brooklyn, NY 11225, USA}
\affiliation{Mathematics Program,
 The Graduate Center, CUNY,
  New York, NY 10016, USA}

\author{Aaron D. Lauda}
\thanks{Contact author: lauda@usc.edu}
\affiliation{Department of Mathematics,
 University of Southern California,
  Los Angeles, California 90089, USA}
\affiliation{Department of Physics, University of Southern California, Los Angeles, California 90089,USA}

\begin{abstract} 
Non-semisimple extensions of the Ising anyon model developed in our previous work~\cite{iulianelli2025} enable universal topological quantum computation via braiding alone, overcoming the Clifford-only limitation of semisimple theories.  The non-semisimple theory provides new anyon types indexed by a real parameter~$\alpha$, the \emph{neglecton}.  Braiding acts unitarily with respect to an indefinite Hermitian form, while the computational subspace sits in a positive-definite sector. We demonstrate that this universality is robust, persisting over an open interval of the neglecton parameter $\alpha$ where the computational subspace remains positive-definite. We identify special values of $\alpha$ where the physical subspace decouples exactly from negative-norm components, ensuring fully unitary evolution and suppressed leakage. We further present an alternative encoding supporting exact single-qubit Clifford gates alongside a non-Clifford phase gate. 
We show that high-precision tuning of $\alpha$ is not required for efficient gate compilation, significantly enhancing the physical plausibility of non-semisimple anyonic architectures.
\end{abstract}

\maketitle
\setcounter{tocdepth}{3}

%

\section{Introduction}

Topological quantum computation aims to harness the non-local properties of anyons in $(2+1)$-dimensional topological phases to build fault-tolerant quantum computers~\cite{Kit, FKLW}. One of the most experimentally promising candidates is the Ising anyon model, conjectured to arise in systems such as the $\nu = 5/2$ fractional quantum Hall state~\cite{MOORE1991362,NAYAK1996529,PhysRevB.83.075303,PhysRevLett.90.016802}. However, it is well known that Ising anyons fail to be computationally universal when restricted to braiding alone: they generate only the Clifford group and thus require supplementation by non-protected operations to achieve universal quantum computation~\cite{FKLW,BRAVYI2002210,MR1910833}.

The situation changes dramatically when Ising anyons are extended into the \emph{non-semisimple} regime. 
Within the framework of non-semisimple topological quantum field theory (TQFT),  new types of anyons appear that have no analog in the semisimple case.  These arise from quantum group representations  with quantum trace zero, which are traditionally termed \emph{negligible} in the semisimple literature due to their vanishing contribution to physical amplitudes.
However, in non-semisimple TQFT, these so-called negligible objects acquire physical significance through a renormalization of the quantum trace.   These new particle types lead to non-semisimple TQFTs having vastly greater ability to distinguish topology in $(2+1)$-dimensions, and, critically for our purposes, enable the construction of braid representations with rich computational properties.
Motivated by this newfound relevance, we propose a new term for these anyon types: \textit{neglectons}. While their quantum dimension remains zero, neglectons serve as essential ingredients in our extended Ising model for topological quantum computation, supporting universal quantum computation through braiding alone.

We demonstrate that Ising anyons, augmented by a single fixed neglecton parameterized by a real number $\alpha$, become computationally universal via braiding alone. 
To achieve universality, it is sufficient for the neglecton to remain fixed throughout the computation. 
In effect, the neglecton serves as a distinguished pole around which the Ising anyons braid. 
This naturally leads to representations of the affine braid group, rather than the ordinary braid group, since the fixed neglecton introduces a marked point that the other strands may wind around.

Unlike conventional TQFTs derived from unitary modular tensor categories, non-semisimple TQFTs assign state spaces equipped with a canonical, but indefinite, Hermitian form~\cite{GLPMS,GLPMS3} so that norms of vectors can be both positive or negative. This indefinite structure arises naturally from representation-theoretic considerations: the key feature is a renormalized quantum trace that enables anyons of vanishing quantum dimension to contribute nontrivially to the theory. The resulting dynamics are governed by operators that preserve this indefinite inner product, a property called \emph{indefinite unitarity}. While such structures are only recently appearing in condensed matter physics \cite{delcamp_donsemisimple_symm}, they have deep connections to pseudo-Hermitian quantum mechanics and appear in recent Hamiltonian lattice formulations of non-semisimple Levin-Wen models~\cite{GLPMS2,GLPMS3}.

A central advance of this article is the demonstration that
universality is \emph{robust} with respect to the neglecton parameter.
Our previous work~\cite{iulianelli2025} established universality only at a
single, finely tuned value of~$\alpha$, leaving open whether the scheme was
stable under perturbations of the parameter that would be expected in any 
experimental realization.  Here we show that universality persists throughout
an open interval of parameter values for which the computational Hilbert space
remains positive definite.  Rather than collapsing under small imperfections,
the model supports universal quantum computation for every non-integer
\[
  \alpha \in \bigl(2 - \tfrac{1}{3},\, 2 + \tfrac{1}{3}\bigr) \mod{4}.
\]
We furthermore provide a quantitative analysis of the robustness of the primitive gates under perturbation of the value $\alpha$ in this range,
establishing that the computational power of the neglecton-augmented Ising
theory is not a finely tuned phenomenon.

To establish universality, we adapt an iterative algorithm originally developed by Reichardt~\cite{Reichardt_2005} and extended by Cui et al.\ \cite{Cui-leak} to the affine braid representations relevant in our non-semisimple setting. While Reichardt's method provides efficient gate compilation in certain semisimple scenarios, our extension yields universality without general efficiency guarantees.  Nevertheless, for some values of $\alpha$, we recover efficient constructions, and recent work using Monte Carlo simulations provides numerical evidence of high fidelity gate compilation across a range of neglecton parameters~\cite{Long_2025_topological}.

Another contribution of this work is the identification of special values of $\alpha$ for which the computational subspace decouples completely from the indefinite sector. In these cases, all braid operations act within a genuinely positive-definite Hilbert space, with no leakage into states of indefinite norm, making these parameter values particularly appealing for physical realization.

The standard theory of Ising anyons allows for the implementation of exact Clifford gates, at the cost of universality. The non-semisimple encoding introduced in \cite{iulianelli2025} and used predominantly in this article supports universal computation but no longer produces exact Clifford gates. 
Section~\ref{subsec:encodings} introduces a new alternative encoding scheme for single qubit operations that retain the ability of Ising anyons to produce exact Clifford gates, while augmenting with a neglecton affording universal single qubit operations. 

We also include a clear account of the representation-theoretic background for the semisimple Ising model, since this provides the foundation for the non-semisimple theory and the Hermitian and modified-trace structures that emerge once projective and other indecomposable modules enter the picture. Even in the familiar semisimple case, the positive-definite inner products on fusion spaces are almost never written down explicitly. Their existence is typically inferred from the unitarity of the corresponding TQFTs, which in turn rests on deep results in the representation theory of quantum groups -- most notably work of Kirillov~\cite{Kir96}, Wenzl~\cite{Wen98}, and others establishing the compatibility of these Hermitian forms with tensor products.

Finally, we address the potential for physical realization of neglectons.  Semisimple anyon theories such as Ising and Fibonacci have long-standing theoretical motivation and partial experimental support.
Their modular tensor categories arise from rational conformal field theories (RCFTs)~\cite{MR990772,Bakalov2001}, which in turn produce unitary Chern–Simons theories with fixed, discrete anyon types. In the fractional quantum Hall setting, RCFT correlators generate trial wavefunctions whose quasihole counting, entanglement spectra, and energy gaps closely match numerical studies of realistic Coulomb Hamiltonians~\cite{MOORE1991362,PhysRevLett.80.1505,PhysRevLett.84.4685}. Although the chiral Ising and Fibonacci phases themselves are not known to arise from strictly local microscopic Hamiltonians~\cite{Kitaev_2012}, their doubled, non-chiral counterparts do admit exactly solvable realizations. Kitaev’s honeycomb model~\cite{Kit} and the Levin–Wen string-net Hamiltonians~\cite{LW} provide commuting-projector lattice models whose excitations realize doubled Ising and doubled Fibonacci anyons, establishing a concrete setting in which these semisimple phases emerge as bona fide topological orders.

 Non-semisimple phases have a more recent but increasingly plausible physical basis. On the field-theoretic side, non-semisimple TQFTs built from the unrolled quantum group $U_q^H(\mathfrak{sl}_2)$ give rise to consistent $(2+1)$-dimensional topological theories, and they are expected to appear as infrared limits of certain $3$D $\mathcal{N}=4$ Chern–Simons–matter theories~\cite{creutzig2021b}. In contrast to the semisimple case, the associated boundary theories are not rational CFTs but logarithmic CFTs, reflecting the underlying indecomposable representation theory. For the non-semisimple Ising model considered here, the corresponding edge modes are described by the $\mathcal{W}(2)$ triplet logarithmic CFT~\cite{MR2283660,MR3795642,Negron}, which is built from the same unrolled $U_q^H(\mathfrak{sl}_2)$ category that governs the bulk. This places neglectons within the well-developed correspondence between non-semisimple TQFTs and logarithmic CFTs, paralleling the familiar relationship between semisimple TQFTs and rational CFTs.

Recent work constructing non-semisimple Levin–Wen models~\cite{GLPMS2,GLPMS3} provides further support for non-semisimple topological phases. These are local, gapped commuting-projector lattice Hamiltonians whose input data is precisely the non-semisimple category generated by the unrolled quantum group. The resulting Hamiltonians are pseudo-Hermitian rather than Hermitian, but they nevertheless support doubled (non-chiral) neglecton quasiparticle excitations. Despite the indefinite inner product, pseudo-Hermitian systems possess real energy spectra, normalizable wavefunctions, and time evolution governed by a Hamiltonian that is self-adjoint with respect to the appropriate indefinite metric~\cite{Most-rep}. These models extend the semisimple paradigm and provide microscopic realizations of non-semisimple topological phases supporting neglecton excitations, similar to those studied in this work.

Taken together, our results show that non-semisimple extensions of the Ising model support robust universal quantum computation by braiding alone. This reveals that models traditionally viewed as non-universal in their semisimple incarnations may become powerful computational platforms once the full non-semisimple theory is taken into account. Our findings highlight non-semisimple TQFTs as both a novel mathematical framework and a promising resource in the search for physically realizable topological quantum computing architectures.

\section{Unrolled quantum groups and modified traces}

Most work in topological quantum computation begins by specifying the data of a semisimple modular tensor category: a finite collection of anyon types (i.e., simple objects), fusion rules $N^c_{ab}$, and the associated $F$- and $R$-symbols needed for the braiding of anyons. This categorical presentation, while convenient for computation, often obscures the underlying representation-theoretic origin of the data and the nontrivial steps required to obtain a semisimple theory.

Semisimplicity plays a central role in the construction of (2+1)-dimensional topological quantum field theories. It ensures that the fusion of simple objects $a \otimes b$ decomposes as a direct sum $ \bigoplus_c N^c_{ab} \, c $, where all $\Hom$-spaces remain finite-dimensional. This structure enables the definition of well-behaved TQFTs with finite-dimensional state spaces and modularity properties essential for topological applications.

To set the stage for the non-semisimple theory, we briefly recall how the modular fusion category data of the $SU(2)_k$ Chern--Simons theory emerges from the representation theory of the quantum group $U_q(\mathfrak{sl}_2) $, with $q = e^{\pi i / (k + 2)}$ a root of unity. This construction highlights where the standard semisimple framework truncates the full representation category, discarding representations with quantum dimension zero. For more details, see~\cite{MR2640343,MR4684298}.
The well-known theories of Ising anyons ($k = 2 $) and Fibonacci anyons ($ k = 3 $) can both be understood in this framework. It should be noted that there are several slightly different definitions of quantum $\mathfrak{sl}_2$. The one we consider in this work is known as the \emph{unrolled quantum group}, denoted by $\Ubar$. In Appendices \ref{sec:representations_of_Uqsl2} and \ref{sec:unrolled_Uqsl2} we review some of these constructions and the roles they play in the definition of simple and non-semisimple theories of anyons.

Set $q=e^{\frac{2\pi i}{2r}}$ to be a $2r^{th}$ root of unity with $r\geq 2$ and
let  $\Ubar$ be the $\C$-algebra given by generators $E, F, K, K^{-1},H$
and relations:
\begin{equation}\label{E:RelDCUqsl}
\begin{aligned}
  KK^{-1}&=1, & K^{-1}K&=1, & HK&=KH,  \\
  KEK^{-1}&=q^{-2} E, &KFK^{-1}&=q^{-2}F, & [E,F]&=\frac{K-K^{-1}}{q-q^{-1}},\\
  [H,E]&=2E, & [H,F]&=-2F, & E^r &= F^r=0.
\end{aligned}
\end{equation}

The algebra $\Ubar$  is a Hopf algebra where the coproduct, counit, and antipode are defined by
\begin{equation}\label{E:HopfAlgDCUqsl}
\begin{aligned}
  \Delta(E)&= 1\otimes E + E\otimes K,
  &\varepsilon(E)&= 0,
  &S(E)&=-EK^{-1},
  \\
  \Delta(F)&=K^{-1} \otimes F + F\otimes 1,
  &\varepsilon(F)&=0,& S(F)&=-KF,
    \\
  \Delta(K)&=K\otimes K,
  &\varepsilon(K)&=1,
  & S(K)&=K^{-1},\\
  \Delta(H) &= H\otimes 1 + 1 \otimes H,
  &\varepsilon(H)&=0,
  & S(H)&=-H.
\end{aligned}
\end{equation}

This structure equips the category of representations of $\Ubar$ with a tensor product enabling the fusion of anyons, and gives rise to dual representations controlling dual anyon types.   

An $R$-matrix can be defined in terms of the generators \ref{E:RelDCUqsl} thus endowing the category with a braiding operation (see Appendix \ref{sec:representations_of_Uqsl2} for a detailed guide on how to carry out such calculations). 
This category has infinitely many simple objects: for each $n$ $\in \{0,1 ...,r-1\}$ there is an $(n+1)$-dimensional representation $S_n$, and for each $\alpha \in (\C \backslash \Z) \cup r\Z$  there is an $r$-dimensional representation $V_\alpha$. Furthermore, many of these objects have zero quantum dimension (see Subsection \ref{subsec:quantum_dimensions} in the appendix). The standard procedure to obtain a semisimple category is to discard the representations with zero quantum dimension through a process often referred to as \emph{semisimplification}~\cite{semisimplification}. One is then left with $r-1$ number of simple objects $S_n$, $n\in \{0,1 ...,r-2\}$. At an eighth root of unity ($r=4$) the three simple objects $S_0, S_1, S_2$ correspond to the vacuum $\mathbb{1}$, the Ising anyon $\sigma$, and the fermion $\psi$ of the Ising theory.  
Here, we propose to work in the non-semisimple Ising category, where the neglecton $\alpha$ anyon types are now assigned non-zero quantum dimensions, called the \emph{modified quantum dimensions}, through a renormalization technique called the \emph{modified trace}~\cite{CGP2}. Using the modified quantum dimensions together with the Hermitian forms on modules defined in Appendix~\ref{sec:hermitian_structures}, the Hom-spaces are endowed with Hermitian forms for which the braiding is unitary.

\subsection{Renormalizing the quantum trace } \label{subsec:modtrace}
The key to leveraging non-semisimple representations for topological applications lies in renormalizing the quantum dimensions, and more generally, quantum traces so that the objects $V_{\alpha}$ no longer contribute negligibly to the theory.  For this reason, we have dubbed the corresponding anyon types as \textit{neglectons}.   The objects are renormalized using the modified trace from~\cite{GKP1, GPT}. The modified trace gives rise to the inner products on $\Hom$-spaces that govern the Hilbert spaces of the non-semisimple anyon theories.  

Unlike the usual semisimple theory, where quantum dimensions of simple objects are strictly positive, the modified dimensions of many objects in the non-semisimple theory are real, but not positive.  This means that there is no hope that the TQFT constructed in \cite{BCGP1} is unitary.  However, in the next section, we will recall the results from \cite{GLPMS} showing that the TQFT arising from a non-semisimple category of representations of the unrolled quantum group for $\mathfrak{sl}_2$  is Hermitian.  This means that the TQFT will produce non-degenerate bilinear forms with an indefinite signature. 
 
Given an endomorphism $f \maps V \to V$, the \textit{quantum trace} is defined by  
\begin{equation} \label{eq:diag-trace}
\hackcenter{\begin{tikzpicture}[scale = 0.8]
    \draw[very thick, ->] (0,0) -- (0,1.5);
    \draw [very thick] (0,1.5) .. controls ++(0,.35) and ++(0,.35) .. (1,1.5) to (1,0)
    .. controls ++(0,-.35) and ++(0,-.35) .. (0,0);
     \node () at (-.25,.05) {$V$};
    \node[draw, thick, fill=black!20,rounded corners=4pt,inner sep=3pt] () at (0,.75) {$f$};
\end{tikzpicture}}
\quad = {\rm tr}(f)
\end{equation}
so that the quantum dimension is the quantum trace of the identity map on $V$.  However, if $V$ is an object of vanishing quantum dimension, the quantum trace of any endomorphism of $V$ will always vanish as well.   
When $V$ is simple, $f \in \End(V)=\C\Id_V$ and we write  $f = \brk{f}_V \Id_V$.  

Intuitively, the modified trace is obtained from \eqref{eq:diag-trace} by cutting along an object $V$.  If $V$ is simple, then this procedure produces an endomorphism of $V$, and we set 
\begin{equation} \label{eq:mod-trace}
\hackcenter{\begin{tikzpicture}[scale = 0.8]
    \draw[very thick, ->] (0,0) -- (0,1.5);
    \draw [very thick] (0,1.5) .. controls ++(0,.35) and ++(0,.35) .. (1,1.5) to (1,0)
    .. controls ++(0,-.35) and ++(0,-.35) .. (0,0);
     \node () at (-.25,.05) {$V$};
    \node[draw, thick, fill=black!20,rounded corners=4pt,inner sep=3pt] () at (0,.75) {$f$};
\end{tikzpicture}}
\quad \longrightarrow 
\hackcenter{\begin{tikzpicture}[scale = 0.8]
    \draw[very thick, ->] (0,-.25) -- (0,1.75);
     \node () at (-.2,-.2) {$ \scriptstyle V$};
     \node () at (-.35,1.55) {$\scriptstyle V$};
    \node[draw, thick, fill=black!20,rounded corners=4pt,inner sep=3pt] () at (0,.75) {$f$};
\end{tikzpicture}}
\;\; = \;\; \brk{f}_V \Id_V
\end{equation}
so that the modified trace of $f$ is $\mt_V(f) := \md(V)\brk{f}_V$ and is nonzero as desired. 
The scalar $\md(V)$ is the modified dimension of $V$ and equals the modified trace of the identity morphism of $V$. The modified trace on the category
of finite-dimensional weight $\Ubar$-representations in this paper is normalized so that 
\begin{equation} \label{eq:moddimValpha}
    \md(V_\alpha)=(-1)^{r-1}\frac{r\qn\alpha}{\qn{r\alpha}} 
\end{equation}
where $\qn x = q^x-q^{-x}$.
Making this intuitive notion rigorous requires that the cutting process is independent of where the cut takes place, and it requires compatibility with the usual notion of trace when part of a larger diagram involving an endomorphism of a neglecton, see Appendix~\ref{app:trace}.

The modified trace is non-degenerate (cf Theorem 5.5 of~\cite{GKP3}), in the following way. Let
$V_\alpha,V_\beta\in\cat$. Then the pairing
$\brk{\cdot,\cdot}_{V_\alpha,V_\beta}: \Hom(V_\alpha,V_\beta)\otimes\Hom(V_\alpha,V_\beta)\to\C$
given by $$\brk{f,g}_{V_\alpha,V_\beta}=\mt_{V_\alpha}({f^\dag}g)$$ is non-degenerate. The Hermitian conjugation $\dag$ is an involution that is compatible with the structure $\Ubar$, as discussed in Appendix \ref{sec:hermitian_structures}.
The pairing is
symmetric in the sense that 
\begin{equation}
  \label{eq:pairing}
  \brk{g,f}_{V_\alpha,V_\beta}=\brk{f,g}_{V_\alpha,V_\beta} \ .
\end{equation}
%

\section{Data for non-semisimple Ising anyons}\label{sec:non-semisimple_ising}

With the representation theory in place, we can now extract the data needed to specify our computational model.  The theory of Ising anyons is closely connected with the modular tensor category for $SU(2)_2$ described by the quantum group $\Uq$ at a primitive $8^{th}$ root of unity. For the remainder of this article, we fix $q=e^{2\pi i/8}$.  

\subsection{Non-semisimple fusion rules}\label{sec:fusion_rules}

At $r=4$, there are only three simple objects that survive the semisimplification process:   
$S_0$, $S_1$, and $S_2$. We will employ the convention of abbreviating these modules by their highest weights, 0, 1, and 2, respectively.  

The Ising anyons in the non-semisimple theory are the simple modules cataloged in Proposition~\ref{prop:classification} when $r=4$. Table~\ref{tab:NSS-fusion-rules} provides a translation between the notation used in the representation-theoretic literature and the non-semisimple Ising model that are relevant to our work. Note that we introduce $\zeta$ to be identified with the simple $S_3 = V_0$.

For the $\alpha$ anyon type appearing in the non-semisimple Ising framework, the modified quantum dimension is given by
\begin{equation} \label{eq:Ising-mdalpha}
    \md_\alpha:=\md(V_\alpha)=-\frac{4\qn\alpha}{\qn{4\alpha}}=-4\frac{\sin(\pi \alpha / 4)}{\sin (\pi \alpha)}.
\end{equation}

While a complete set of fusion rules for $\Ubar$ is available in~\cite{CGP2}, we summarize here only those relevant to our analysis. In Table~\ref{tab:NSS-fusion-rules}, we write “zero” to denote the trivial $0$-dimensional module, avoiding notational conflict with $V_0$ and the vacuum $S_0$. This zero module appears because certain fusion rules involve representations of quantum dimension zero, which becomes trivial upon semisimplification. Fusion rules involving $V_\alpha$ differ depending on whether $\alpha \in \mathbb{C} \setminus \mathbb{Z}$ or $\alpha \in 4\mathbb{Z}$. Table~\ref{tab:NSS-fusion-rules} assumes $\alpha \in \mathbb{C} \setminus \mathbb{Z}$; when $\alpha = 0 \in 4\mathbb{Z}$, the corresponding object is $\zeta = V_0 = S_3$.   $P_2$ is the non-semisimple projective cover of $S_2$. This module (discussed in Appendix \ref{subsec:reps}) does not show up explicitly in this work, but is necessary for the internal consistency of the theory. Finally, the symbol $V$ in $\1 \times V$ denotes an arbitrary object.
\begin{table}[htp]
\begin{center}
\begin{tabular}{ |c|c|c|c|c|c|c| } 
\hline
Representation-theoretic notation & $S_0$ & $S_1$ & $S_2$ & $S_3$ & $V_\alpha$ & $P_2$ \\
\hline
Simplified notation & 0 & 1 & 2   & $\zeta$ & $\alpha$ & $\xi$  \\ \hline 
Traditional Ising notation & $\1$ & $\sigma$ & $\psi$ & \slash  &  \slash &  \slash  \\
\hline
\end{tabular}
\smallskip

\begin{tabular}{ |c|c|c| } 
\hline
fusion & semisimple  & non-semisimple  \\
\hline
$0\times V$ & $V$ & $V$ \\ 
$1 \times 1$ & $0 + 2$ & $0 + 2$ \\
$1 \times 2$ & $1$ & $1 + \zeta$\\
$1 \times \zeta$ & zero & $\xi$ \\
$2 \times 2$ & $0$ & $0 + \xi$\\
$\alpha \times 1$ & zero & $(\alpha{-}1) + (\alpha{+}1)$\\
$\alpha \times 2$ & zero & $(\alpha{-}2)  + (\alpha) + (\alpha{+}2)$\\
$\alpha \times \zeta$ & zero & $(\alpha{-}3) + (\alpha{-}1) + (\alpha{+}1) + (\alpha{+}3)$\\
\hline
\end{tabular}
\end{center}
\caption{\justifying 
Notational translation between the representation theory literature and its shorthand, as well as the more traditional Ising anyon notation from the literature.  Note that our shorthand notation indexes traditional semisimple anyons by their topological charge, or highest weight, while we reserve Greek letters for new non-semisimple anyon types.  The second table gives the fusion rules in the semisimple and non-semisimple framework. In our previous work on the non-semisimple Ising model~\cite{iulianelli2025}, we used $\sigma$ to denote the anyon $1$ used in this article. 
} \label{tab:NSS-fusion-rules} 
\end{table}

Notice that the fusion rules in the non-semisimple framework are an augmentation of the fusion rules that appear in the semisimple setting. Moreover, one should observe that the augmented objects are exactly those with vanishing quantum dimensions and therefore vanish during semisimplification.

\subsection{Normalization, $F$ and $R$ symbols}\label{sec:nonIsing-graph}

 In this section, we continue our conventions from  Notation~\ref{not:upper_and_lowercase} denoting simple objects by lowercase letters.
The fusion rules give rise to the maps $ Y_c^{ab}: c \rightarrow a \otimes b$ introduced in \eqref{eq:embedding}.
Recall that these maps are determined by mapping the highest weight vector of $c$ into a corresponding highest weight vector of the subrepresentation of $a \otimes b$ that is isomorphic to $c$. By Schur's lemma, there is a one-dimensional space of such maps for each summand of $a\otimes b$ isomorphic to $c$. 

In what follows we sort the basis vectors of tensor product spaces in lexicographic order: if $a$ and $b$ are two simples with bases $\{v_0^a, ...v_k^a\}$ and $\{v^b_0, ..., v_l^b\}$ then we order the basis of $a\otimes b$ as $\{ v_0^a \otimes v_0^b, \dots, v^a_0 \otimes v^b_l, v^a_1 \otimes v^b_0, \dots, v^a_k \otimes v^b_l \}$.
We normalize trivalent vertices by imposing that the first non-zero coefficient of $v$ (using the lexicographic order described earlier) be $1$. The image of all the other basis vectors of $V$ is determined from the action of  the lowering operator  $\Jm$.    This choice of normalization can be thought of as a \textit{gauge symmetry}, and is only constrained by the pentagon equation 
\eqref{eq:pentagon}.

\begin{example}\label{ex:two_into_11}
The embedding $Y_2^{11}: 2 \rightarrow 1 \otimes 1$ is constructed by finding a highest-weight vector $v \in 1 \otimes 1$ with weight 2 and setting
\begin{equation*}
\begin{split}
    Y_2^{11}(s_0^2) &= v = s_0^1\otimes s_0^1,\\
    Y_2^{11}(s_1^2) &= \Delta\Jm v = q^{-1} s_0\otimes s_1^1 + s_1^1 \otimes s_0^1,\\
    Y_2^{11}(s_2^2) &= (\Delta\Jm)^2 v   = (q+q^{-1})s_1^1 \otimes s_1^1.\\
\hackcenter{\begin{tikzpicture}[ scale=1.1]
  \draw[ultra thick, black] (0,0) to (.6,-.6) to (.6,-1.2);
  \draw[ultra thick, black] (1.2,0) to (.6,-.6);
   \node at (0,0.2) {$\scs 1$};
   \node at (1.2,0.2) {$\scs 1$};
   \node at (.8,-1.1) {$\scs 2$};
\end{tikzpicture} } & = \left(
\begin{array}{ccc}
 1 & 0 & 0 \\
 0 & \frac{1}{q} & 0 \\
 0 & 1 & 0 \\
 0 & 0 & q+\frac{1}{q} \\
\end{array}
\right)
\end{split}
\end{equation*}
\end{example}

\begin{example} 
The fusion rules from Table~\ref{tab:NSS-fusion-rules} state that $\alpha \otimes 1 \cong (\alpha {+}1) \oplus (\alpha{-}1)$. The map $Y^{\alpha 1}_{\alpha -1}: \alpha {-}1 \rightarrow \alpha \otimes 1$ is determined by mapping the highest weight vector of $(\alpha{-}1)$ to a highest weight vector in $v \in \alpha \otimes 1$ of the same highest weight $\alpha +2$. (Recall that $V_{\alpha}$ has highest weight $\alpha +r -1$).
Solving the equations $\Jz v = (\alpha+2)v$ and $\Jp v = 0$ we find $v = v^\alpha_0 \otimes s_1^1 - \frac{[1]}{q[1-a]} v^\alpha_1 \otimes s_0^1$. Notice that the basis vectors are ordered lexicographically, and that the first coefficient is $1$. The image of all other vectors in $(\alpha{-}1)$ is then determined by applying sequences of $\Delta F$:
\begin{widetext}
\begin{equation}
  \vcenter{\hbox{%
    \hackcenter{\begin{tikzpicture}[ scale=1.1]
  \draw[ultra thick, black] (0,0) to (.6,-.6) to (.6,-1.2);
  \draw[ultra thick, black] (1.2,0) to (.6,-.6);
   \node at (0,0.2) {$\alpha$};
   \node at (1.2,0.2) {$1$};
   \node at (.6,-1.4) {$\alpha{-}1$};
\end{tikzpicture} }}}
\;:\;
\left\{
  \begin{aligned}
    & v^{\alpha{-}1}_0 \mapsto  v\\[2pt]
    & v^{\alpha{-}1}_1 \mapsto \Delta\Jm v\\[2pt]
    & v^{\alpha{-}1}_2 \mapsto (\Delta\Jm)^2v  \\[2pt]
    & v^{\alpha{-}1}_3 \mapsto (\Delta\Jm)^3 v
  \end{aligned}
  \right.
~=~
\begin{blockarray}{ccccc}
\scs{v_0^{\alpha{-}1}} & \scs{v_1^{\alpha{-}1}} & \scs{v_2^{\alpha{-}1}} & \scs{v_3^{\alpha{-}1}} \\
\begin{block}{[cccc]c}
 0 & 0 & 0 & 0 & \scs{v_0^\alpha \otimes s_0^1}\\
 1 & 0 & 0 & 0 & \scs{v_0^\alpha \otimes s_1^1}\\
 \frac{q^2-1}{q^{\alpha +1}-q^{3-\alpha }} & 0 & 0 & 0 & \scs{v_1^\alpha \otimes s_0^1}\\
 0 & \frac{-q^{2 \alpha +2}+q^4-q^2+1}{q^4-q^{2 \alpha +2}} & 0 & 0 & \scs{v_1^\alpha \otimes s_1^1}\\
 0 & \frac{q^2-1}{q^{\alpha +1}-q^{3-\alpha }} & 0 & 0 & \scs{v_2^\alpha \otimes s_0^1}\\
 0 & 0 & \frac{1-q^{2 \alpha +2}}{q^4-q^{2 \alpha +2}} & 0 & \scs{v_2^\alpha \otimes s_1^1}\\
 0 & 0 & \frac{q^2-1}{q^{\alpha +1}-q^{3-\alpha }} & 0 & \scs{v_3^\alpha \otimes s_0^1}\\
 0 & 0 & 0 & \frac{-q^{2 \alpha +2}-q^6+q^4+1}{q^4-q^{2 \alpha +2}} & \scs{v_3^\alpha \otimes s_1^1}\\
\end{block}
\end{blockarray}
\end{equation}
\end{widetext}
\end{example}

The $F$-symbols describe the change of basis
\begin{equation}\label{eq:F_move}
\begin{split}
\hackcenter{\begin{tikzpicture}[ scale=1.1]
  \draw[ultra thick, black] (0,0) to (.6,-.6) to (.6,-1.2);
  \draw[ultra thick, black] (0.6,0) to (.3,-.3);
  \draw[ultra thick, black] (1.2,0) to (.6,-.6);
   \node at (0,0.2) {$a$};
   \node at (.6,0.2) {$b$}; 
   \node at (1.2,0.2) {$c$};
   \node at (.8,-1) {$ d$};
   \node at (.2,-.6) {$\scriptstyle m$};
\end{tikzpicture} }
~=\sum\limits_n ~[F_{d}^{abc}]_{nm}
\hackcenter{\begin{tikzpicture}[ scale=1.1]
  \draw[ultra thick, black] (0,0) to (.6,-.6) to (.6,-1.2);
  \draw[ultra thick, black] (0.6,0) to (.9,-.3);
  \draw[ultra thick, black] (1.2,0) to (.6,-.6);
   \node at (0,0.2) {$a$};
   \node at (.6,0.2) {$b$}; 
   \node at (1.2,0.2) {$c$};
   \node at (.8,-1) {$ d$};
   \node at (.9,-.6) {$\scriptstyle n$};
\end{tikzpicture} } .
\end{split}
\end{equation}
These symbols must satisfy the pentagon equation
\begin{equation} \label{eq:pentagon}
    (F^{ncd}_e)_{\ell m}(F^{abm}_e)_{np}
    = \sum_t (F^{abc}_\ell)_{nt} (F^{atd}_e)_{\ell r} (F^{bcd}_r)_{tm}
\end{equation}
Sample computations and specific $F$-moves relevant to our construction appear in Appendix~\ref{app:F}. 

The braiding map $c_{ab}: a\otimes b \rightarrow b\otimes a$ from \eqref{eq:braiding} acts on the anyonic Hilbert spaces.  To compute the action in our chosen bases, note that both $Y^{ba}_c$ and $c_{ab}Y^{ab}_c$ map the simple $c$ as a submodule in $b \otimes a$; Schur's lemma implies that $Y^{ba}_c$ and $c_{ab}Y^{ab}_c$ must only differ by a scalar, which we call $R^{ba}_c$.
\begin{equation}  \label{eq:R-move}
\begin{split}
\hackcenter{\begin{tikzpicture}[ scale=-0.8]
    \draw[ultra thick] (0,0) to (0,1);
    \draw[ultra thick] (-.5,-1)to [out=90, in=210] (0,0);
    \draw[ultra thick] (.5,-1) to [out=90, in=-30] (0,0);
    \node at (.75,-.8) {$ a$};
    \node at (-.75,-.8) {$ b$};
    \node at (-.35,.6) {$ c$};
    \draw [black, ultra thick]    (-.5,-2) .. controls +(0,.25) and +(0,-.25) ..  (.5,-1) ;
    \path [fill=white] (-.25,-1.4) rectangle (.75,-1.6);
    \draw [black, ultra thick]   (.5,-2)  .. controls +(0,.25) and +(0,-.25) ..  (-.5,-1);
\end{tikzpicture} } = \left(\, \hackcenter{\begin{tikzpicture}[ scale=.9]
   \draw [black, ultra thick]    (-.5,-2) .. controls +(0,.25) and +(0,-.25) ..  (.5,-1) ;
    \path [fill=white] (-.25,-1.4) rectangle (.25,-1.6);
    \draw [black, ultra thick]   (.5,-2)  .. controls +(0,.25) and +(0,-.25) ..  (-.5,-1);
    \end{tikzpicture} }\, \right) \circ
\hackcenter{\begin{tikzpicture}[ scale=.9]
    \draw[ultra thick, black] (0,0) to (.6,-.6) to (.6,-1.2);
    \draw[ultra thick, black] (1.2,0) to (.6,-.6);
    \node at (0,0.2) {$a$};
    \node at (1.2,0.2) {$b$};
    \node at (.8,-1) {$ c$};
\end{tikzpicture} }  ~=~R^{ba}_c
\hackcenter{\begin{tikzpicture}[ scale=.9]
  \draw[ultra thick, black] (0,0) to (.6,-.6) to (.6,-1.2);
  \draw[ultra thick, black] (1.2,0) to (.6,-.6);
   \node at (0,0.2) {$b$};
   \node at (1.2,0.2) {$a$};
   \node at (.8,-1) {$ c$};
\end{tikzpicture} }  
\end{split}
\end{equation}

\subsection{Hermitian conjugates and bubble pops} \label{subsec:daggerY}

We have so far described the action of the braiding and the associator on the Hom-spaces of the category of representations of $\Ubar$. In order to discuss the unitarity of such operations, we must first define a Hermitian form on the Hom-spaces. Wenzl \cite{Wen98} defined a Hermitian form on $\Ubar$-modules and their products that is compatible with the action of $\Ubar$ (see Appendix~\ref{sec:hermitian_structures}). If $c$ and $a\otimes b$ are modules with their respective Hermitian forms $\langle\cdot,\cdot\rangle_c$ and $\langle\cdot,\cdot \rangle_{a\otimes b}$, we define $\Yd: a\otimes b\rightarrow c$ as the map that satisfies
\begin{equation*}
    \langle \Y v^c, v^{a\otimes b} \rangle_{a\otimes b} = \langle v^c, \Yd v^{a \otimes b} \rangle_{c}
\end{equation*}
for all $v^c \in c$ and $v^{a\otimes b} \in a\otimes b$. In practice, we compute $\Yd$ by picking bases $\{v^a_j\}$, $\{v^b_k\}$, and $\{v^c_i\}$ for $a, b,$ and $c$ respectively and computing the components of $\Yd$
\begin{equation*}
    \langle \Y v^c_i, v^a_j\otimes v^b_k \rangle_{a\otimes b} = \langle v^c_i, \Yd v^a_j \otimes v^b_k \rangle_{c}.
\end{equation*}
Let $\Y v^c_i=\sum\limits_{mn} \left[\Y\right]_{i, mn}v^a_m\otimes v^b_n$ and $\Yd v^a_j \otimes v^b_k = \sum\limits_l \left[\Yd\right]_{jk, l}v^c_l$.  Note that the index $mn$ and $jk$ in the matrices $\left[\Y\right]_{i, mn}$ and $\left[\Yd\right]_{jk, l}$ correspond to the position of $v_m^a \otimes v_n^b$ and $v_j^a \otimes v_k^b$, respectively, in the lexicographic ordering of the tensor product basis.  Then the matrix coefficients of $\Yd$ are given by
\begin{equation}\label{ex:trivalent_conjugate}
    \left[\Yd\right]_{jk, i} = \frac{1}{\langle v^c_i,  v^c_i \rangle_c} \sum_{mn}\left[\Y\right]_{i,mn}^* \langle  v^a_m\otimes v^b_n, v^a_j\otimes v^b_k \rangle_{a\otimes b}
\end{equation}
where it is important to note that our bases are assumed orthogonal, but not orthonormal in general.  

We shall denote $\Yd$ by an  upside-down  trivalent vertex: 
\begin{equation} \label{eq:trivalent_conjugate}
\Yd = 
\hackcenter{
\begin{tikzpicture}[ scale=1.0, yscale=-1.0]
  \draw[ultra thick, black] (0,0) to (.6,-.6) to (.6,-1.2);
  \draw[ultra thick, black] (1.2,0) to (.6,-.6);
   \node at (0,0.2) {$a$};
   \node at (1.2,0.2) {$b$};
   \node at (.6,-1.4) {$c$};
\end{tikzpicture} } =
\left( \hackcenter{\begin{tikzpicture}[ scale=1.0]
  \draw[ultra thick, black] (0,0) to (.6,-.6) to (.6,-1.2);
  \draw[ultra thick, black] (1.2,0) to (.6,-.6);
   \node at (0,0.2) {$a$};
   \node at (1.2,0.2) {$b$};
   \node at (.8,-1) {$c$};
\end{tikzpicture} }  \right)^{\dagger}.  
\end{equation}
Note that $c$ is simple, so Schur's lemma implies that $\Yd \Y = B^{ab}_c \Id_c$ for some coefficient $B^{ab}_c$. We call this relation the \textit{bubble-pop} move
\begin{equation}  \label{eq:bubble_pop}
\hackcenter{\begin{tikzpicture}[ scale=1.1]
  \draw[ultra thick, black] (.6,-.6) to (.6,-1.2);
   \node at (.1,-.2) {$a$};
   \node at (1.1,-.2) {$b$};
   \node at (.8,-1) {$c$};
   \node at (.8,.6) {$c'$};
   \draw[ultra thick, black] (.9,-.2) to [out=270, in=45] (.6,-.6);
   \draw[ultra thick, black] (.9,-.2) to [out=90, in=315] (.6,.2);
   \draw[ultra thick, black] (.3,-.2) to [out=270, in=135] (.6,-.6);
   \draw[ultra thick, black] (.3,-.2) to [out=90, in=225] (.6,.2);
   \draw[ultra thick, black] (.6,.2) to (.6,.8);
\end{tikzpicture} }~=~\delta_{cc'}B_{c}^{ab}~
\hackcenter{\begin{tikzpicture}[scale=1.1]
  \draw[ultra thick, black] (.6,-1.2) to (.6,.8);
   \node at (.8,-.2) {$c$};
\end{tikzpicture} } . 
\end{equation}
It is easy to see that $B^{ab}_c$ is real since $\left( \Yd \Y\right)^\dag  =  \Yd  \Y$,  so $(B^{ab}_{c})^* = B^{ab}_{c}$.  These coefficients are computed in Appendix~\ref{app:bubble}.

\section{Non-semisimple Topological Quantum Computation}

We are now in a position to specify the Hilbert space and encoding for the non-semisimple Ising model.  
We will use a shorthand $(a,b)_c$  where fusion channels of $a\otimes b$ are represented by parentheses with a subscript $c$ indicating the fusion channel.   
\subsection{Traditional Ising encoding}
In the traditional semisimple theory describing Ising anyons, qubits are encoded into the state of a collection of anyons with a fixed topological charge.  The single qubit space corresponds to three Ising anyons in a state with topological charge $1$. This two-dimensional space spanned by $((1,1)_0, 1)_1$ and $((1,1)_2,1)_1$ is where we encode logical qubits, see Figure~\ref{fig:fullwidth-figs}(a).  
Notice that the only possibility for the total topological spin is $1$. This is closely related to the fact that Ising anyons can achieve leakage-free entangling of multiple qubits~\cite{Cui-leak}.

\begin{figure*}[t]
  \centering
  \begin{subfigure}[b]{0.45\textwidth}
\begin{equation}\label{eq:basis_single_qubit}
\begin{split}
 \ket{0} &= \hackcenter{\begin{tikzpicture}[scale=0.35]
          \draw[thick, fill=gray!30] (1.75,0,0) ellipse (3.4 and 1.45);
        \draw[thick,   fill=gray!30] (1,0,0) ellipse (2 and .7);
    \shade[ball color = red!40, opacity = 0.5] (0,0,0) circle (.35);
    \shade[ball color = red!40, opacity = 0.5] (2,0,0) circle (.35);
    \shade[ball color = red!40, opacity = 0.5] (4,0,0) circle (.35);
        \node at (0,0) {$\scriptstyle 1$};
    \node at (2,0) {$\scriptstyle 1$};
    \node at (4,0) {$\scriptstyle 1$};
            \node at (2.7,-.8) {$\scriptscriptstyle 0$};
            \node at (5,-1.15) {$\scriptstyle 1$};
\end{tikzpicture}} =
\hackcenter{\begin{tikzpicture}[ scale=1.1]
  \draw[ultra thick, black] (0,0) to (.6,-.6) to (.6,-1.2);
  \draw[ultra thick, black] (0.6,0) to (.3,-.3);
  \draw[ultra thick, black] (1.2,0) to (.6,-.6); 
   \node at (0,0.2) {$1$};
   \node at (.6,0.2) {$1$}; 
   \node at (1.2,0.2) {$1$};  
    \node at (.8,-1) {$\scriptstyle 1$};
    \node at (.2,-.6) {$\scriptstyle 0$}; 
\end{tikzpicture} }
\\
\ket{1} &= \hackcenter{\begin{tikzpicture}[scale=0.35]
          \draw[thick, fill=gray!30] (1.75,0,0) ellipse (3.4 and 1.45);
        \draw[thick,   fill=gray!30] (1,0,0) ellipse (2 and .7);
    \shade[ball color = red!40, opacity = 0.5] (0,0,0) circle (.35);
    \shade[ball color = red!40, opacity = 0.5] (2,0,0) circle (.35);
    \shade[ball color = red!40, opacity = 0.5] (4,0,0) circle (.35);
    \node at (0,0) {$\scriptstyle 1$};
    \node at (2,0) {$\scriptstyle 1$};
    \node at (4,0) {$\scriptstyle  1$};
            \node at (2.7,-.8) {$\scriptscriptstyle 2$};
            \node at (5,-1.15) {$\scriptstyle 1$};
\end{tikzpicture}} =
\hackcenter{\begin{tikzpicture}[ scale=1.1]
  \draw[ultra thick, black] (0,0) to (.6,-.6) to (.6,-1.2);
  \draw[ultra thick, black] (0.6,0) to (.3,-.3);
  \draw[ultra thick, black] (1.2,0) to (.6,-.6); 
   \node at (0,0.2) {$1$};
   \node at (.6,0.2) {$1$}; 
   \node at (1.2,0.2) {$1$};  
    \node at (.8,-1) {$\scriptstyle 1$};
    \node at (.2,-.6) {$\scriptstyle 2$}; 
\end{tikzpicture} } 
\end{split}
\end{equation}
    \caption{\justifying Encoding a qubit into a configuration of three Ising anyons with total topological charge $1$. }
  \end{subfigure}
  \hfill
  \begin{subfigure}[b]{0.45\textwidth}
\begin{equation}\label{eq:basis_single_qubit}
\begin{split}
 \ket{0} &= \hackcenter{\begin{tikzpicture}[scale=0.35]
          \draw[thick, fill=gray!30] (1.75,0,0) ellipse (3.4 and 1.45);
        \draw[thick,   fill=gray!30] (1,0,0) ellipse (2 and .7);
    \shade[ball color = red!40, opacity = 0.5] (0,0,0) circle (.35);
    \shade[ball color = red!40, opacity = 0.5] (2,0,0) circle (.35);
    \shade[ball color = red!40, opacity = 0.5] (4,0,0) circle (.35);
    \node at (0,0) {$\scriptstyle \alpha$};
    \node at (2,0) {$\scriptstyle 1$};
    \node at (4,0) {$\scriptstyle 1$};
            \node at (2.7,-.8) {$\scriptscriptstyle \alpha{+}1$};
            \node at (5,-1.15) {$\scriptstyle \alpha$};
\end{tikzpicture}} =
\hackcenter{\begin{tikzpicture}[ scale=1.1]
  \draw[ultra thick, black] (0,0) to (.6,-.6) to (.6,-1.2);
  \draw[ultra thick, black] (0.6,0) to (.3,-.3);
  \draw[ultra thick, black] (1.2,0) to (.6,-.6); 
   \node at (0,0.2) {$\alpha$};
   \node at (.6,0.2) {$1$}; 
   \node at (1.2,0.2) {$1$};  
    \node at (.8,-1) {$\scriptstyle \alpha$};
    \node at (.2,-.6) {$\scriptstyle \alpha{+}1$}; 
\end{tikzpicture} },
\\
\ket{1} &= \hackcenter{\begin{tikzpicture}[scale=0.35]
          \draw[thick, fill=gray!30] (1.75,0,0) ellipse (3.4 and 1.45);
        \draw[thick,   fill=gray!30] (1,0,0) ellipse (2 and .7);
    \shade[ball color = red!40, opacity = 0.5] (0,0,0) circle (.35);
    \shade[ball color = red!40, opacity = 0.5] (2,0,0) circle (.35);
    \shade[ball color = red!40, opacity = 0.5] (4,0,0) circle (.35);
    \node at (0,0) {$\scriptstyle \alpha$};
    \node at (2,0) {$\scriptstyle 1$};
    \node at (4,0) {$\scriptstyle 1$};
            \node at (2.7,-.8) {$\scriptscriptstyle \alpha{-}1$};
            \node at (5,-1.15) {$\scriptstyle \alpha$};
\end{tikzpicture}} =
\hackcenter{\begin{tikzpicture}[ scale=1.1] 
  \draw[ultra thick, black] (0,0) to (.6,-.6) to (.6,-1.2);
  \draw[ultra thick, black] (0.6,0) to (.3,-.3);
  \draw[ultra thick, black] (1.2,0) to (.6,-.6); 
   \node at (0,0.2) {$\alpha$};
   \node at (.6,0.2) {$1$}; 
   \node at (1.2,0.2) {$1$};  
    \node at (.8,-1) {$\scriptstyle \alpha$};
    \node at (.2,-.6) {$\scriptstyle \alpha - 1$}; 
\end{tikzpicture} }
\end{split}
\end{equation}
    \caption{\justifying Encoding a qubit into a configuration consisting of a single neglecton and two Ising anyons. }
  \end{subfigure}

  \vspace{0.3cm}

  \begin{subfigure}[b]{0.45\textwidth}
\begin{equation} \label{eq:nIsing-braids}
\begin{split}
\b_1  &= \; 
\hackcenter{\begin{tikzpicture}[scale=0.35]
          \draw[thick, fill=gray!30] (1.75,0,0) ellipse (3.4 and 1.45);
        \draw[thick,   fill=gray!30] (1,0,0) ellipse (2 and .7);
    \shade[ball color = red!40, opacity = 0.5] (0,0,0) circle (.35);
    \shade[ball color = red!40, opacity = 0.5] (2,0,0) circle (.35);
    \shade[ball color = red!40, opacity = 0.5] (4,0,0) circle (.35);
    \node at (0,0) {$\scriptstyle 1$};
    \node at (2,0) {$\scriptstyle 1$};
    \node at (4,0) {$\scriptstyle 1$};
            \node at (2.7,-.8) {$\scriptscriptstyle 0$};
            \node at (5,-1.15) {$\scriptstyle 1$};
    \draw[ultra thick, black,-] (4,0.5) to (4,5);  
     \draw[ultra thick, black,-] (2,.5) to (2,2) .. controls ++(0,.8) and ++(0,-0.8) ..   (0,4) to (0,5) ;
    \path [fill=white] (0.8,1.6) rectangle (1.2,3.5);
\draw[ultra thick, black,-] (0,.5) to (0,2) .. controls ++(0,.8) and ++(0,-0.8) ..   (2,4) to (2,5);
    \end{tikzpicture}}
    =
  e^{i\pi/8}  \left(
  \begin{array}{cc}
    -1 & 0 \\
    0& i \\
  \end{array}
\right) 
\\
  \b_2 &= \;
\hackcenter{\begin{tikzpicture}[scale=0.35]
          \draw[thick, fill=gray!30] (1.75,0,0) ellipse (3.4 and 1.45);
        \draw[thick,   fill=gray!30] (1,0,0) ellipse (2 and .7);
    \shade[ball color = red!40, opacity = 0.5] (0,0,0) circle (.35);
    \shade[ball color = red!40, opacity = 0.5] (2,0,0) circle (.35);
    \shade[ball color = red!40, opacity = 0.5] (4,0,0) circle (.35);
    \node at (0,0) {$\scriptstyle 1$};
    \node at (2,0) {$\scriptstyle 1$};
    \node at (4,0) {$\scriptstyle 1$};
            \node at (2.7,-.8) {$\scriptscriptstyle 2 $};
            \node at (5,-1.15) {$\scriptstyle 1$};
    \draw[ultra thick, black,-] (0,0.5) to (0,5);  
     \draw[ultra thick, black,-] (4,.5) to (4,2) .. controls ++(0,.8) and ++(0,-0.8) ..   (2,4) to (2,5) ;
    \path [fill=white] (2.8,1.6) rectangle (3.2,3.5);
\draw[ultra thick, black,-] (2,.5) to (2,2) .. controls ++(0,.8) and ++(0,-0.8) ..   (4,4) to (4,5);
    \end{tikzpicture}}
        =
  \frac{e^{-i\pi/8}}{\sqrt{2}}  \left(
  \begin{array}{cc}
    1& i\\
    i& 1 \\
  \end{array}
\right) 
\end{split}
\end{equation}
    \caption{\justifying Braiding operations in the standard Ising anyon model and their corresponding matrices on the encoded qubit space. }
  \end{subfigure}
  \hfill
  \begin{subfigure}[b]{0.45\textwidth}
\begin{equation} \label{eq:nIsing-braids2}
\begin{split}
\b_2 =& 
\hackcenter{\begin{tikzpicture}[scale=0.3]
          \draw[thick, fill=gray!30] (1.75,0,0) ellipse (3.4 and 1.45);
        \draw[thick,   fill=gray!30] (1,0,0) ellipse (2 and .7);
    \shade[ball color = red!40, opacity = 0.5] (0,0,0) circle (.35);
    \shade[ball color = red!40, opacity = 0.5] (2,0,0) circle (.35);
    \shade[ball color = red!40, opacity = 0.5] (4,0,0) circle (.35);
    \node at (0,0) {$\scriptstyle 1$};
    \node at (2,0) {$\scriptstyle 1$};
    \node at (4,0) {$\scriptstyle 1$};
            \node at (2.7,-.8) {$\scriptscriptstyle a$};
            \node at (5,-1.15) {$\scriptstyle 1$};
    \draw[ultra thick, black,-] (0,0.5) to (0,5);  
     \draw[ultra thick, black,-] (4,.5) to (4,2) .. controls ++(0,.8) and ++(0,-0.8) ..   (2,4) to (2,5) ;
    \path [fill=white] (2.8,1.6) rectangle (3.2,3.5);
\draw[ultra thick, black,-] (2,.5) to (2,2) .. controls ++(0,.8) and ++(0,-0.8) ..   (4,4) to (4,5);
    \end{tikzpicture}}
\stackrel{F}{\longrightarrow}
\hackcenter{\begin{tikzpicture}[scale=0.3]
          \draw[thick, fill=gray!30] (1.75,0,0) ellipse (3.4 and 1.45);
        \draw[thick,   fill=gray!30] (2.85,0,0) ellipse (2 and .7);
    \shade[ball color = red!40, opacity = 0.5] (0,0,0) circle (.35);
    \shade[ball color = red!40, opacity = 0.5] (2,0,0) circle (.35);
    \shade[ball color = red!40, opacity = 0.5] (4,0,0) circle (.35);
    \node at (0,0) {$\scriptstyle 1$};
    \node at (2,0) {$\scriptstyle 1$};
    \node at (4,0) {$\scriptstyle 1$};
            \node at (1.6,-.8) {$\scriptscriptstyle b $};
            \node at (5,-1.15) {$\scriptstyle 1$};
    \draw[ultra thick, black,-] (0,0.5) to (0,5);  
     \draw[ultra thick, black,-] (4,.5) to (4,2) .. controls ++(0,.8) and ++(0,-0.8) ..   (2,4) to (2,5) ;
    \path [fill=white] (2.8,1.6) rectangle (3.2,3.5);
\draw[ultra thick, black,-] (2,.5) to (2,2) .. controls ++(0,.8) and ++(0,-0.8) ..   (4,4) to (4,5);
    \end{tikzpicture}}
  \\
   \quad     \stackrel{R}{\longrightarrow}& 
\hackcenter{\begin{tikzpicture}[scale=0.3]
          \draw[thick, fill=gray!30] (1.75,0,0) ellipse (3.4 and 1.45);
        \draw[thick,   fill=gray!30] (2.85,0,0) ellipse (2 and .7);
    \shade[ball color = red!40, opacity = 0.5] (0,0,0) circle (.35);
    \shade[ball color = red!40, opacity = 0.5] (2,0,0) circle (.35);
    \shade[ball color = red!40, opacity = 0.5] (4,0,0) circle (.35);
    \node at (0,0) {$\scriptstyle 1$};
    \node at (2,0) {$\scriptstyle 1$};
    \node at (4,0) {$\scriptstyle 1$};
            \node at (1.6,-.8) {$\scriptscriptstyle b $};
            \node at (5,-1.15) {$\scriptstyle 1$};
    \draw[ultra thick, black,-] (0,0.5) to (0,5);  
     \draw[ultra thick, black,-] (4,.5) to (4,2) .. controls ++(0,.8) and ++(0,-0.8) ..   (2,4) to (2,5) ;
    \path [fill=white] (2.8,1.6) rectangle (3.2,3.5);
\draw[ultra thick, black,-] (2,.5) to (2,2) .. controls ++(0,.8) and ++(0,-0.8) ..   (4,4) to (4,5);
    \end{tikzpicture}}
    \stackrel{F^{-1}}{\longrightarrow}
    \hackcenter{\begin{tikzpicture}[scale=0.3]
          \draw[thick, fill=gray!30] (1.75,0,0) ellipse (3.4 and 1.45);
        \draw[thick,   fill=gray!30] (1,0,0) ellipse (2 and .7);
    \shade[ball color = red!40, opacity = 0.5] (0,0,0) circle (.35);
    \shade[ball color = red!40, opacity = 0.5] (2,0,0) circle (.35);
    \shade[ball color = red!40, opacity = 0.5] (4,0,0) circle (.35);
    \node at (0,0) {$\scriptstyle 1$};
    \node at (2,0) {$\scriptstyle 1$};
    \node at (4,0) {$\scriptstyle 1$};
            \node at (2.7,-.8) {$\scriptscriptstyle c$};
            \node at (5,-1.15) {$\scriptstyle 1$};
    \draw[ultra thick, black,-] (0,0.5) to (0,5);  
     \draw[ultra thick, black,-] (4,.5) to (4,2) .. controls ++(0,.8) and ++(0,-0.8) ..   (2,4) to (2,5) ;
    \path [fill=white] (2.8,1.6) rectangle (3.2,3.5);
\draw[ultra thick, black,-] (2,.5) to (2,2) .. controls ++(0,.8) and ++(0,-0.8) ..   (4,4) to (4,5);
    \end{tikzpicture}}
    \end{split}
\end{equation}
    \caption{\justifying Computing the braid generator $\b_2$ by changing basis, using the $R$-symbols, and changing back to the computational basis.}
  \end{subfigure}
  \caption{\justifying Encoding computational qubits into the collective state of a group of anyons.  Unitaries are implemented on the computational basis by braiding the anyons around one another. }
  \label{fig:fullwidth-figs}
\end{figure*}

Braiding Ising anyons then achieves unitary transformations on our single qubit space, see Figure~\ref{fig:fullwidth-figs}(c).  
Computing $\b_2$ requires a composite of changing basis via the $F$-symbols, simplifying the braiding using the $R$-symbols, and changing back to the original basis as show in Figure~\ref{fig:fullwidth-figs}(d).
It is easy to see that the braiding of Ising anyons will not be sufficient to achieve single-qubit density.  The braid matrices from \eqref{eq:nIsing-braids} only generate Clifford gates.  However, one nice feature of this theory is that it is possible to achieve leakage-free entangling gates~\cite{Fan_2010,Cui}.    Our non-semisimple Ising model discussed in the next section is capable of achieving single-qubit density, though the non-semisimple theory loses the leakage-free entangling property.

\subsection{Non-semisimple encoding}
For the non-semisimple anyon model,  our encoding makes use of a single neglecton $\alpha$ together with pairs of Ising anyons with a total topological charge $\alpha$.  Our single qubit Hilbert space is given by the space $\Hom(\alpha, \alpha \otimes 1 \otimes 1)$.  
This space is two-dimensional with basis $\ket{0} =((\alpha,1)_{\alpha+1},1)_\alpha$ and $\ket{1}=((\alpha,1)_{\alpha-1},1)_\alpha$ given by the ``left-hanging tree" basis, in which all strands fuse on the left as in Figure~\ref{fig:fullwidth-figs}(b).  
Unlike the semisimple encoding, the the total topological charge of an $\alpha$ neglecton and two Ising anyons can take three values: $\alpha{-}2$, $\alpha$, and $\alpha{+}2$.  The $\alpha{\pm}2$ channels will cause leakage in multiple qubit encodings.

More generally, the Hilbert space of $n$ qubits $\mathcal{H}_n$ can be formed from a single $\alpha$ anyon and $2n$ Ising anyons with a total topological charge $\alpha$, so that $\mathcal{H}_n = \Hom(\alpha, \alpha \otimes 1^{\otimes 2n})$. 
We fix a basis of $\H_n$ where we denote a basis vector as a $2n$-tuple $\underline{v}=(v_1,v_1',v_2,v_2',\dots, v_{n}, v_{n}'=\alpha)$ representing the allowed left hanging fusion channels from $(\alpha 1^{2n})$ to $\alpha$. 

\begin{equation} \label{eq:leftbasis}
\underline{v} = \hackcenter{\begin{tikzpicture}[scale=0.9]
  \draw[ultra thick, black] (0,0) to (-1.0,1.0);
  \draw[ultra thick, black] (-.6,.6) to (-.2,1.0);
  \draw[ultra thick, black] (-.3,.3) to (.4,1.0);
    \draw[ultra thick, black] (.6,-.6) to (1.2,-1.2);
  \draw[ultra thick, black] (2.8,1.0) to (1.8,0) to (.9,-.9);
  \draw[ultra thick, black] (3.4,1.0) to (2.4,0) to (1.2,-1.2) to (1.2,-1.8);
   \node at (-1.0,1.2) {$\alpha$};
   \node at (.4,1.2) {$1$};
   \node at (-.2,1.2) {$1$};
   \node at (2.8,1.2) {$1$};
   \node at (3.4,1.2) {$1$};
   \node at (1.2,-2) {$\alpha$};
   \node at (.3,-.2) {$\ddots$};
   \node at (-.7,.3) {$\scriptstyle v_1$};
   \node at (-.35,0) {$\scriptstyle v_1'$};
    \node at (.4,-.9) {$\scriptstyle v_{n-1}'$};
    \node at (.8,-1.2) {$\scriptstyle v_n$};
\end{tikzpicture}}
\end{equation}
The computational space $\mathcal{H}_n^{{\rm comp}}$ corresponds to the subspace of $\mathcal{H}_n$ where all $v_i'=\alpha$.  The $i^{th}$ qubit is encoded into the value of $v_i$ with $v_i = \alpha{+}1$ encoding $\ket{0}$ and $v_i = \alpha{-}1$ encoding $\ket{1}$.  To simplify notation, we denote computational basis vectors using the notation $\ket{v_1 \cdots v_n}$.   Nontrivial left-hanging fusion trees with some $v_i' \neq \alpha$ generate the noncomputational space $\mathcal{H}_n^{NC}$.  The basis for $\H_2$ is given in \eqref{eq:H2basis}. 
%

\subsection{Normalization} \label{subsec:normal}
While the representation theory of quantum $\slt$ guarantees that braids and $F$-moves are unitary operators with respect to the Hermitian form from Section~\ref{sec:hermitian_structures}, the matrices representing these maps are not manifestly unitary since our basis of fusion trees is not normalized.    The lexicographic normalization that we have been employing, where $Y$-symbols are normalized so that the first entry in lexicographic order has coefficient 1, while convenient for calculations, leads to unnormalized basis vectors. 

Bra's $\bra{v_1\cdots v_n}$ of basis vectors are given by the Hermitian structure as explained in Section~\ref{subsec:daggerY}. These basis vectors are orthogonal by \eqref{eq:bubble_pop}, but they are not orthonormal even for single qubits as  $\braket{0|0}=B_0\md_\alpha$ and $\braket{1|1}=B_1\md_\alpha$, which are computed in Example~\ref{ex:norm_single_qubit_states}.    Renormalizing this basis makes the $F$-symbols more complex; we will compute the change of basis matrix and transform operators into the normalized basis. 

Given a fusion tree basis vector \(\underline{v} \in \Hom(a_0, a_1 \otimes \dots \otimes a_n)\), the inner product \(\braket{\underline{v} | \underline{v}}\) (as defined in Section~\ref{sec:hermitian_structures}) is computed using the normalized trace. When \(a_0\) is simple, the endomorphism \(v^\dag v\) is proportional to the identity, i.e., \(v^\dag v = \lambda_v \Id_{a_0}\) for some scalar \(\lambda_v\). Hence,
\[
\braket{\underline{v} | \underline{v}} = \mathsf{t}(v^\dag v) = \lambda_v \, \mathsf{t}(\Id_{a_0}) = \lambda_v \, \mathsf{d}_{a_0},
\]
where \(\mathsf{d}_{a_0}\) denotes the modified quantum dimension of \(a_0\) from \eqref{eq:Ising-mdalpha}.

\begin{example}\label{ex:norm_single_qubit_states}
As an example, we compute the norm of the basis vectors in $\Hom(\alpha, \alpha\otimes 1\otimes 1)$ 
\begin{equation}\label{eq:norm_of_ket0}
\begin{split}
&\braket{0|0}= 
\\ & \quad \left\langle 
\hackcenter{\begin{tikzpicture}[ scale=1.1]
    \draw[ultra thick, black] (0,0) to (.6,-.6) to (.6,-1.2);
    \draw[ultra thick, black] (0.6,0) to (.3,-.3);
    \draw[ultra thick, black] (1.2,0) to (.6,-.6); 
    \node at (0,0.2) {$\alpha$};
    \node at (.6,0.2) {$1$}; 
    \node at (1.2,0.2) {$1$};  
    \node at (.8,-1) {$\scriptstyle \alpha$};
    \node at (.2,-.6) {$\scriptstyle \alpha{+}1$}; 
\end{tikzpicture}} ,
\hackcenter{\begin{tikzpicture}[ scale=1.1]
    \draw[ultra thick, black] (0,0) to (.6,-.6) to (.6,-1.2);
    \draw[ultra thick, black] (0.6,0) to (.3,-.3);
    \draw[ultra thick, black] (1.2,0) to (.6,-.6); 
    \node at (0,0.2) {$\alpha$};
    \node at (.6,0.2) {$1$}; 
    \node at (1.2,0.2) {$1$};  
    \node at (.8,-1) {$\scriptstyle \alpha$};
    \node at (.2,-.6) {$\scriptstyle \alpha{+}1$}; 
\end{tikzpicture} } \right\rangle =
\mt\left(
\hackcenter{\begin{tikzpicture}[ scale=1.1]
    \begin{scope}[shift={(0,0 )},scale={-1}]
       \draw[ultra thick, black] (0,-.4) to (0,-.6);
        \draw[ultra thick, black] (.3,0) to [out=270, in=45] (0,-.4);
        \draw[ultra thick, black] (-.3,0) to [out=270, in=135] (0,-.4);
   \end{scope}
   \begin{scope}[shift={(0.3, .6 )},scale={-1}]
       \draw[ultra thick, black] (0,-.4) to (0,-.8);
        \draw[ultra thick, black] (.3,0) to [out=270, in=45] (0,-.4);
        \draw[ultra thick, black] (-.3,0) to [out=270, in=135] (0,-.4);
   \end{scope}
   \begin{scope}[shift={(0,0)}]
       \draw[ultra thick, black] (0,-.4) to (0,-.6);
        \draw[ultra thick, black] (.3,0) to [out=270, in=45] (0,-.4);
        \draw[ultra thick, black] (-.3,0) to [out=270, in=135] (0,-.4);
   \end{scope}
   \begin{scope}[shift={(0.3,-.6)}]
       \draw[ultra thick, black] (0,-.4) to (0,-.8);
        \draw[ultra thick, black] (.3,0) to [out=270, in=45] (0,-.4);
        \draw[ultra thick, black] (-.3,0) to [out=270, in=135] (0,-.4);
   \end{scope}
   \draw[ultra thick, black] (.6,-.6) -- (.6,.6);  
   \node at (-.5,0) {$\scriptstyle \alpha$ };
   \node at (-.25,-.85) {$\scriptstyle \alpha{+}1$};
   \node at (.45,0) {$\scriptstyle 1$ };
   \node at (.75,0) {$\scriptstyle 1$ };
   \node at (-.25,.85) {$\scriptstyle \alpha{+}1$};
   \node at (.5, -1.2) {$\scriptstyle \alpha$};
   \node at (.5, 1.2) {$\scriptstyle \alpha$};
\end{tikzpicture} } \right)
\\
&=~ B^{\alpha 1}_{\alpha{+}1}
\mt \left(
\hackcenter{\begin{tikzpicture}[ scale=1.1]
   \begin{scope}[shift={(0.3, .6 )},scale={-1}]
       \draw[ultra thick, black] (0,-.4) to (0,-.8);
        \draw[ultra thick, black] (.3,0) to [out=270, in=45] (0,-.4);
        \draw[ultra thick, black] (-.3,0) to [out=270, in=135] (0,-.4);
   \end{scope}
   \begin{scope}[shift={(0.3,-.6)}]
       \draw[ultra thick, black] (0,-.4) to (0,-.8);
        \draw[ultra thick, black] (.3,0) to [out=270, in=45] (0,-.4);
        \draw[ultra thick, black] (-.3,0) to [out=270, in=135] (0,-.4);
   \end{scope}
   \draw[ultra thick, black] (.6,-.6) -- (.6,.6);  
    \draw[ultra thick, black] (0, .6) -- (0,-.6);
   \node at (-.3,0) {$\scriptstyle \alpha{+}1$ };
   \node at (.75,0) {$\scriptstyle 1$ };
   \node at (.5, -1.2) {$\scriptstyle \alpha$};
   \node at (.5, 1.2) {$\scriptstyle \alpha$};
\end{tikzpicture} }\right) =~ B^{\alpha 1}_{\alpha{+}1}B^{(\alpha{+}1) 1}_{\alpha}
\mt\left(
\hackcenter{\begin{tikzpicture}[ scale=1.1]
   \draw[ultra thick, black] (0,-1.2) -- (0,1.2); 
   \node at (0.2,0) {$\scriptstyle \alpha$ };
   \node at (-0.2,0) {$\;$ };
\end{tikzpicture} } \right)
\\
&=~B^{\alpha 1}_{\alpha{+}1}B^{(\alpha{+}1) 1}_{\alpha}\md_\alpha.
\end{split}
\end{equation}

One can similarly find that 
\begin{equation}\label{eq:norm_of_ket1}
    \braket{1|1} = B^{\alpha 1}_{\alpha{-}1}B^{(\alpha{-}1) 1}_{\alpha}\md_\alpha.
\end{equation}
Inspection of the bubble pop coefficients~\eqref{eq:B-data} and the modified dimensions~\eqref{eq:moddimValpha} reveals that the inner products $\braket{0|0}$ and $\braket{1|1}$, while real for real values of $\alpha$, are not always positive.  Figure~\ref{fig:norm_basis_vectors} gives a graph of the values of the inner products as a function of $\alpha$.  
\begin{figure}[htp!]\label{fig:two_qubit_basis}
    \centering
    \includegraphics[width=8cm]{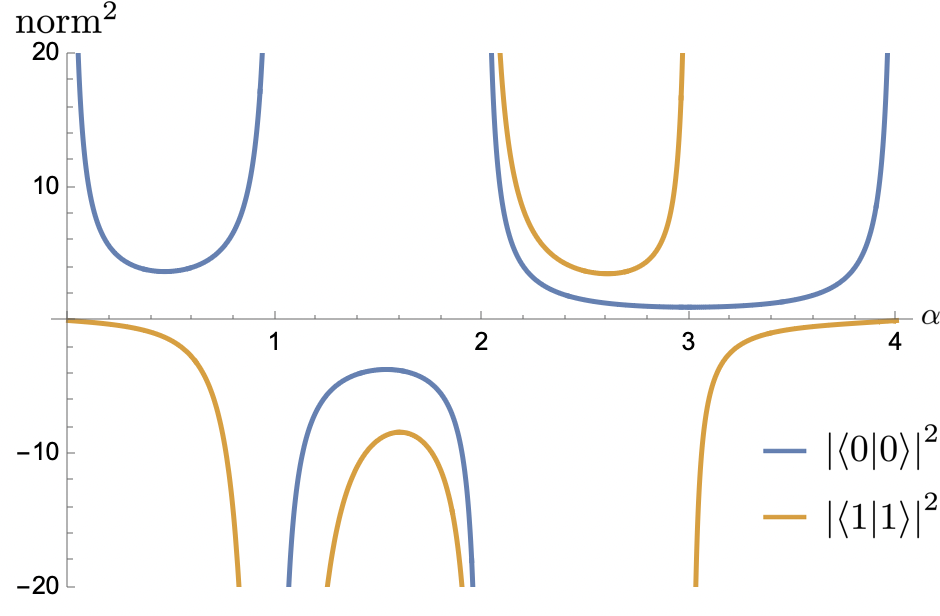} 
    \caption{\justifying Norms of the basis vectors $\ket{0}$ and $\ket{1}$. The graph is periodic with period 8, but only $\alpha \in (0,4)$ is shown for clarity. The graph for $\alpha \in (4,8)$ is the same as in the figure, but flipped upside-down.}  
    \label{fig:norm_basis_vectors}
\end{figure}
This plot shows that the signature of $\mathcal{H}_1$ will be positive definite for $\alpha \in (2,3) \mod 4$, negative definite for $\alpha \in (1,2) \mod 4$, and will have mixed signature otherwise.  
\end{example}

This example computation of the norm for our single-qubit encoding plays an important role in determining the signature of the inner product on the multi-qubit computational space.  It will be convenient to define
\begin{equation}
\label{eq:B0B1}
    B_0 :=B^{\alpha 1}_{\alpha{+}1}B^{(\alpha{+}1) 1}_{\alpha}, \qquad 
    B_1:=B^{\alpha 1}_{\alpha{-}1}B^{(\alpha{-}1) 1}_{\alpha}.
\end{equation}

To properly normalize $\H_n$, we will need the change of basis matrices.  In general, 
given  basis $\{v_1,...,v_k\}$ of $\mathcal{H}=\Hom(a_0, a_1\otimes \dots\otimes a_n)$, let 
\begin{equation}
M_\H 
= \text{diag}\left(\sqrt{\braket{v_1|v_1}},\dots,\sqrt{\braket{v_k|v_k}}\right) .
\end{equation}
In this paper, taking the square root of a negative number $-x$ (with $x > 0 $) refers to $\sqrt{-x} = i\sqrt{x}$, where $i$ is the imaginary unit satisfying $i^2 = -1$. This follows the standard branch of the complex square root, where $\sqrt{z}$ is defined with argument in $(-\pi, \pi]$.
Then any operator $\mathsf{b}: \mathcal{H} \rightarrow \H$ can be expressed in the normalized basis via a change of basis
\begin{equation}\label{eq:normalized_braid}
    \mathsf{b}_{\text{norm}} = M_\H \mathsf{b}M_\H^{-1}.
\end{equation}
The resulting operator $\mathsf{b}_{\text{norm}}$ will be unitary with respect to the possibly indefinite metric $\braket{\cdot|\cdot}$.  In particular, if $D\maps \H \to \H$ is the diagonal operator acting on a basis vector $\underline{v}$ by ${\rm sign}(\braket{\underline{v}|\underline{v}})$ then the operator will satisfy
$
\mathsf{b}_{\text{norm}}^{\dagger} D \mathsf{b}_{\text{norm}} = D.
$ The matrix $\mathsf{b}_{\text{norm}}$ is unitary in the standard sense as long as the norms of the basis vectors have the same sign: a negative-definite metric can be rescaled by a negative number to yield an equivalent positive metric.

\subsection{Braiding and Jucys-Murphy operators}

The affine braid group $\Br_n^{\rm aff}$ is the group of braids that wrap around a fixed `flag pole' in the first position~\cite{halverson2007}.  It can be regarded as a subgroup of the braid group on $n+1$ strands generated by the braid $\mathsf{b}_1^2$ and the elementary braids $\mathsf{b}_i$ for $2 \leq i \leq n$.  
\begin{equation}\label{eq:affine-gen}
\mathsf{b}_1^2 = \; 
\hackcenter{\begin{tikzpicture}[scale=0.3]
    \draw[ultra thick, red,-] (0,3) to (0,5);  
                \path [ultra thick, fill=white] (-.2,3.5) rectangle (.2,4.1);
    \draw[ultra thick, black,-] (2,.5) to (2,.75).. controls ++(0,1.5) and ++(0,-1.2) ..   (-.75,3);
            \path[ultra thick, fill=white] (-.25,2) rectangle (.25,2.4);
\draw[ultra thick, black,-] (-.75,3) .. controls ++(0,1.2) and ++(0,-1.2) ..   (2,5);
    \draw[ultra thick, red,-] (0,.5) to (0,3);  
     \draw[ultra thick, black,-] (4,.5) to (4,5);
     \draw[ultra thick, black,-] (6.5,.5) to (6.5,5);
     \node at (5.5, 3) {$  \dots$};
         \node at (0,0) {$\scriptstyle 1$};
    \node at (2,0) {$\scriptstyle 2$};
    \node at (4,0) {$\scriptstyle 3$};
    \node at (6.8,0) {$\scriptstyle n+1$};
    \end{tikzpicture}}
    \qquad   
    \mathsf{b}_i = \;
\hackcenter{\begin{tikzpicture}[scale=0.3]
    \draw[ultra thick, red,-] (0,0.5) to (0,5); 
    \draw[ultra thick, black,-] (2,0.5) to (2,5);
    \draw[ultra thick, black,-] (10,0.5) to (10,5);
     \draw[ultra thick, black,-] (7,.5) to (7,2) .. controls ++(0,.8) and ++(0,-0.8) ..   (5,4) to (5,5) ;
    \path [fill=white] (5.8,1.6) rectangle (6.2,3.5);
\draw[ultra thick, black,-] (5,.5) to (5,2) .. controls ++(0,.8) and ++(0,-0.8) ..   (7,4) to (7,5);
\node at (3.5, 3) {$  \dots$};
\node at (8.5, 3) {$  \dots$};
         \node at (0,0) {$\scriptstyle 1$};
    \node at (2,0) {$\scriptstyle 2$};
    \node at (5,0) {$\scriptstyle i$};
    \node at (7,0) {$\scriptstyle i+1$};
    \node at (10,0) {$\scriptstyle n+1$};
    \end{tikzpicture}}
\end{equation}
The categorical braiding from \eqref{eq:braiding} induces an action of $\Br_{2n}^{\rm aff}$ on $\mathcal{H}_n$ by unitary operators arising from braiding the anyon configuration $(\alpha 1^{2n})$ fixing the initial $\alpha$ anyon. (Note our indexing is slightly different  than the standard description of $\Br_{2n}^{\rm aff}$ where our $\mathbf{b}_1^2$ is sometimes written as $J_1$ and our $\mathsf{b}_i$ for $2 \leq i \leq n$ is shifted to $\mathsf{b}_{i-1}$.)   

Consider the \textit{Jucys-Murphy} operators $J_i$, for $i=1,\ldots,n$ in the affine braid group $\Br_n^{\rm aff}$:
\begin{equation} \label{eq:JMdef}
J_i =
 \mathsf{b}_{i}\mathsf{b}_{i-1}\dots \mathsf{b}_{2}\mathsf{b}_{1}^2 \mathsf{b}_{2}\dots \mathsf{b}_{i-1}\mathsf{b}_{i} =
 \hackcenter{
\begin{tikzpicture}[yscale=-1, scale=0.5,  decoration={markings, mark=at position 0.6 with {\arrow{>}};},]
       \draw [black, very thick]    (-.5,-2) .. controls +(0,.25) and +(0,-.25) ..  (.5,-1) ;
      \path [fill=white] (-.25,-1.4) rectangle (.75,-1.6);
       \draw [black, very thick]   (.5,-2)  .. controls +(0,.25) and +(0,-.25) ..  (-.5,-1);
       \draw[very thick, postaction={}] (-1.5,-1) to (-1.5,-2);
       \draw[very thick, red] (-2.5,-1) to (-2.5,-3);
          \draw [black, very thick]    (-1.5,-3) .. controls +(0,.25) and +(0,-.25) ..  (-.5,-2) ;
      \path [fill=white] (-1.25,-2.4) rectangle (-.25,-2.6);
       \draw [black, very thick]   (-.5,-3)  .. controls +(0,.25) and +(0,-.25) ..  (-1.5,-2);
          \draw[very thick, postaction={}] (.5,-2) to (.5,-6);
          \draw [black, very thick]    (-2.5,-4) .. controls +(0,.25) and +(0,-.25) ..  (-1.5,-3) ;
      \path [fill=white] (-2.25,-3.4) rectangle (-1.25,-3.6);
       \draw [red, very thick]   (-1.5,-4)  .. controls +(0,.25) and +(0,-.25) ..  (-2.5,-3);
          \draw [red, very thick]    (-2.5,-5) .. controls +(0,.25) and +(0,-.25) ..  (-1.5,-4) ;
      \path [fill=white] (-2.25,-4.4) rectangle (-1.25,-4.6);
       \draw [black, very thick]   (-1.5,-5)  .. controls +(0,.25) and +(0,-.25) ..  (-2.5,-4);
          \draw[very thick, postaction={}] (-.5,-5) to (-.5,-3);
                    \draw [black, very thick]    (-1.5,-6) .. controls +(0,.25) and +(0,-.25) ..  (-.5,-5) ;
      \path [fill=white] (-1.25,-5.4) rectangle (-.25,-5.6);
       \draw [black, very thick]   (-.5,-6)  .. controls +(0,.25) and +(0,-.25) ..  (-1.5,-5);
                    \draw [black, very thick]    (-.5,-7) .. controls +(0,.25) and +(0,-.25) ..  (.5,-6) ;
      \path [fill=white] (-.25,-6.4) rectangle (.75,-6.6);
       \draw [black, very thick]   (.5,-7)  .. controls +(0,.25) and +(0,-.25) ..  (-.5,-6);
          \draw[very thick, red] (-2.5,-5) to (-2.5,-7);
          \draw[very thick, postaction={}] (-1.5,-6) to (-1.5,-7);
          \draw[very thick, postaction={}] (1.5,-1) to (1.5,-7);
          \node at (1,-4) {$\cdots$};
          \node at (-2,-1.5) {$\cdots$};
                  \node at (-2.5,-.5) {$\scs {1}$};
         \node at (-.5,-.5) {$\scs {i}$};
            \node at (1.5,-.5) {$\scs {n}$};
\end{tikzpicture} }  
\end{equation}
Observe that $J_1 = \mathsf{b}_{1}^2$.   From the braid representations of these operators, it follows that they are mutually commuting operators on $\mathcal{H}_n$.   Hence, the Hilbert space $\mathcal{H}_n$ can be simultaneously diagonalized into eigenstates of the Jucys-Murphy operators.  
It is not hard to see that the basis of left-oriented fusion trees is an eigenbasis for these operators.  See for example, Figure~\ref{fig:jucys_murphy} and observe that $Y$-diagrams are an eigenbasis for the action of the braiding as in \eqref{eq:R-move}.


We now show that the affine braid group $\Br_2^{\rm aff}$ acting on our single qubit Hilbert space $\H_1$ generates a universal set of single qubit operations for all values where $\alpha$ in the form $\braket{\cdot | \cdot }$ is positive-definite, extending the results from~\cite{iulianelli2025}. 
The braid generators we consider are the following: 
\begin{equation}\label{eq:primitives}
J_1 = \mathsf{b}_1^2 = \; 
\hackcenter{\begin{tikzpicture}[scale=0.35]
          \draw[thick, fill=gray!30] (1.75,0,0) ellipse (3.4 and 1.45);
        \draw[thick,   fill=gray!30] (1,0,0) ellipse (2 and .7);
    \shade[ball color = red!40, opacity = 0.5] (0,0,0) circle (.35);
    \shade[ball color = red!40, opacity = 0.5] (2,0,0) circle (.35);
    \shade[ball color = red!40, opacity = 0.5] (4,0,0) circle (.35);
    \node at (0,0) {$\scriptstyle \alpha$};
    \node at (2,0) {$\scriptstyle 1$};
    \node at (4,0) {$\scriptstyle 1$};
            \node at (2.7,-.8) {$\scriptscriptstyle \alpha\pm 1$};
            \node at (5,-1.15) {$\scriptstyle \alpha$};
    \draw[ultra thick, red,-] (0,3) to (0,5);  
                \path [ultra thick, fill=white] (-.2,3.5) rectangle (.2,4.1);
    \draw[ultra thick, black,-] (2,.5) to (2,.75).. controls ++(0,1.5) and ++(0,-1.2) ..   (-.75,3);
            \path[ultra thick, fill=white] (-.25,2) rectangle (.25,2.4);
\draw[ultra thick, black,-] (-.75,3) .. controls ++(0,1.2) and ++(0,-1.2) ..   (2,5);
    \draw[ultra thick, red,-] (0,.5) to (0,3);  
     \draw[ultra thick, black,-] (4,.5) to (4,5);
    \end{tikzpicture}}
    \qquad   
    \mathsf{b}_2 = \;
\hackcenter{\begin{tikzpicture}[scale=0.35]
          \draw[thick, fill=gray!30] (1.75,0,0) ellipse (3.4 and 1.45);
        \draw[thick,   fill=gray!30] (1,0,0) ellipse (2 and .7);
    \shade[ball color = red!40, opacity = 0.5] (0,0,0) circle (.35);
    \shade[ball color = red!40, opacity = 0.5] (2,0,0) circle (.35);
    \shade[ball color = red!40, opacity = 0.5] (4,0,0) circle (.35);
    \node at (0,0) {$\scriptstyle \alpha$};
    \node at (2,0) {$\scriptstyle q$};
    \node at (4,0) {$\scriptstyle q$};
            \node at (2.7,-.8) {$\scriptscriptstyle \alpha\pm1$};
            \node at (5,-1.15) {$\scriptstyle \alpha$};
    \draw[ultra thick, red,-] (0,0.5) to (0,5);  
     \draw[ultra thick, black,-] (4,.5) to (4,2) .. controls ++(0,.8) and ++(0,-0.8) ..   (2,4) to (2,5) ;
    \path [fill=white] (2.8,1.6) rectangle (3.2,3.5);
\draw[ultra thick, black,-] (2,.5) to (2,2) .. controls ++(0,.8) and ++(0,-0.8) ..   (4,4) to (4,5);
    \end{tikzpicture}} \ .
\end{equation}
It is convenient to compute these operators in the basis \eqref{eq:leftbasis} and change basis to the normalized matrix at the end.  Using \eqref{eq:R-move}, we can compute the action of $J_1$ on $\ket{0}$ as 
\begin{equation*}
\begin{split}
    J_1 \ket{0} &= \hackcenter{\begin{tikzpicture}[ scale=1.1]
    \begin{scope}[shift={(.6,.8)}, scale={0.6}]
        \draw [black, ultra thick]     (-1.0,0) to  (-1.0,1);
        \path [fill=white]          (-0.85,.65) rectangle (-1.15, .2);
        \draw [black, ultra thick]   (-1.75,0)  .. controls +(0,.5) and +(0,-.45) ..  (0,1);
    \end{scope}
    \begin{scope}[shift={(.6,0.2)}, scale={.6}]
        \draw [black, ultra thick]   (0,0)  .. controls +(0,.5) and +(0,-.45) ..  (-1.75,1);
        \path [fill=white]          (-0.85,0.4) rectangle (-1.15,0.8);
        \draw [black, ultra thick]     (-1.0,0) to  (-1.0,1);
    \end{scope}
    \begin{scope}
        \draw[black, ultra thick] (0,-.01)-- (0,.2);
        \draw[black, ultra thick] (.6,-.01)-- (.6,.2);
        \draw[black, ultra thick] (1.2,-.01)-- (1.2,.2);
    \end{scope}
    \draw[black, ultra thick] (1.2,0) to (1.2, 1.4);
    \draw[ultra thick, black] (0,0) to (.6,-.6) to (.6,-1.2);
    \draw[ultra thick, black] (0.6,0) to (.3,-.3);
    \draw[ultra thick, black] (1.2,0) to (.6,-.6); 
    \node at (0, 1.6) {$\alpha$};
    \node at (.6, 1.6) {$1$}; 
    \node at (1.2, 1.6) {$1$};  
    \node at (.8,-1) {$\scriptstyle \alpha$};
    \node at (.2,-.6) {$\scriptstyle \alpha + 1$}; 
\end{tikzpicture} } =
\hackcenter{\begin{tikzpicture}[ scale=1.1]
    \begin{scope}[shift={(.3,2)}, scale ={.6}]
        \draw [black, ultra thick]   (.5,-2)  .. controls +(0,.25) and +(0,-.25) ..  (-.5,-1);
        \path [fill=white] (-.25,-1.4) rectangle (.75,-1.6);
        \draw [black, ultra thick]    (-.5,-2) .. controls +(0,.25) and +(0,-.25) ..  (.5,-1) ;
    \end{scope}
    \begin{scope}[shift={(.3,1.4)}, scale ={.6}]
        \draw [black, ultra thick]   (.5,-2)  .. controls +(0,.25) and +(0,-.25) ..  (-.5,-1);
        \path [fill=white] (-.25,-1.4) rectangle (.75,-1.6);
        \draw [black, ultra thick]    (-.5,-2) .. controls +(0,.25) and +(0,-.25) ..  (.5,-1) ;
    \end{scope}
    \begin{scope}
        \draw[black, ultra thick] (0,-.01)-- (0,.2);
        \draw[black, ultra thick] (.6,-.01)-- (.6,.2);
        \draw[black, ultra thick] (1.2,-.01)-- (1.2,.2);
    \end{scope}
    \draw[black, ultra thick] (1.2,0) to (1.2, 1.4);
    \draw[ultra thick, black] (0,0) to (.6,-.6) to (.6,-1.2);
    \draw[ultra thick, black] (0.6,0) to (.3,-.3);
    \draw[ultra thick, black](1.2,0) to (.6,-.6); 
    \node at (0, 1.6) {$\alpha$};
    \node at (.6, 1.6) {$1$}; 
    \node at (1.2, 1.6) {$1$};  
    \node at (.8,-1) {$\scriptstyle \alpha$};
    \node at (.2,-.6) {$\scriptstyle \alpha + 1$}; 
\end{tikzpicture} } 
\\
&= R^{\alpha1}_{\alpha{+}1}R^{1\alpha}_{\alpha{+}1} \hackcenter{\begin{tikzpicture}[ scale=1.1]
    \draw[ultra thick, black] (0,0) to (.6,-.6) to (.6,-1.2);
    \draw[ultra thick, black] (0.6,0) to (.3,-.3);
    \draw[ultra thick, black] (1.2,0) to (.6,-.6); 
    \node at (0,0.2) {$\alpha$};
    \node at (.6,0.2) {$1$}; 
    \node at (1.2,0.2) {$1$};  
    \node at (.8,-1) {$\scriptstyle \alpha$};
    \node at (.2,-.6) {$\scriptstyle \alpha{+}1$}; 
\end{tikzpicture} } = q^{3+\alpha}\ket{0}.
\end{split}
\end{equation*}
Analogously, one finds that $J_1\ket{1}=R^{\alpha1}_{\alpha{-}1}R^{1\alpha}_{\alpha{-}1}\ket{1}=q^{-5-\alpha}\ket{1}$.
Finding the action of $\mathsf{b}_2$ requires a change of basis, as $\ket{0}$ and $\ket{1}$ are not eigenstates. Instead, we change to a basis where $\mathsf{b}_2$  is diagonal as in \eqref{eq:nIsing-braids2} by applying the $F$-move $F^{\alpha11}_\alpha$, computed in \eqref{eq:F_Move_a11}, so that  
\begin{align*}
    \mathsf{b}_2 \ket{0} &= \hackcenter{\begin{tikzpicture}[ scale=1.1]
    \begin{scope}[shift={(.9,1.4)}, scale ={.6}]
        \draw [black, ultra thick]   (.5,-2)  .. controls +(0,.25) and +(0,-.25) ..  (-.5,-1);
        \path [fill=white] (-.25,-1.4) rectangle (.75,-1.6);
        \draw [black, ultra thick]    (-.5,-2) .. controls +(0,.25) and +(0,-.25) ..  (.5,-1) ;
    \end{scope}
    \begin{scope}
        \draw[black, ultra thick] (0,-.01)-- (0,.2);
        \draw[black, ultra thick] (.6,-.01)-- (.6,.2);
        \draw[black, ultra thick] (1.2,-.01)-- (1.2,.2);
    \end{scope}
    \draw[black, ultra thick] (0,0) to (0, .8);
    \draw[ultra thick, black] (0,0) to (.6,-.6) to (.6,-1.2);
    \draw[ultra thick, black] (0.6,0) to (.3,-.3);
    \draw[ultra thick, black](1.2,0) to (.6,-.6); 
    \node at (0, 1) {$\alpha$};
    \node at (.6, 1) {$1$}; 
    \node at (1.2, 1) {$1$};  
    \node at (.8,-1) {$\scriptstyle \alpha$};
    \node at (.2,-.6) {$\scriptstyle \alpha + 1$}; 
\end{tikzpicture} } 
= 
\sum_n   [F^{\alpha 11}_\alpha]_{n(\alpha{+}1)}
\hackcenter{\begin{tikzpicture}[ scale=1.1]
    \begin{scope}[shift={(.9,1.4)}, scale ={.6}]
       \draw [black, ultra thick]   (.5,-2)  .. controls +(0,.25) and +(0,-.25) ..  (-.5,-1);
        \path [fill=white] (-.25,-1.4) rectangle (.75,-1.6);
        \draw [black, ultra thick]    (-.5,-2) .. controls +(0,.25) and +(0,-.25) ..  (.5,-1) ;
    \end{scope}
    \begin{scope}
        \draw[black, ultra thick] (0,-.01)-- (0,.2);
        \draw[black, ultra thick] (.6,-.01)-- (.6,.2);
        \draw[black, ultra thick] (1.2,-.01)-- (1.2,.2);
    \end{scope}
    \draw[black, ultra thick] (0,0) to (0, .8);
    \draw[ultra thick, black] (0,0) to (.6,-.6) to (.6,-1.2);
    \draw[ultra thick, black] (0.6,0) to (.9,-.3);
    \draw[ultra thick, black](1.2,0) to (.6,-.6); 
    \node at (0, 1) {$\alpha$};
    \node at (.6, 1) {$1$}; 
    \node at (1.2, 1) {$1$};  
    \node at (.8,-1) {$\scriptstyle \alpha$};
    \node at (1.0,-.6) {$\scriptstyle n$}; 
\end{tikzpicture} } 
\\
&= 
\sum_n  R^{11}_n [F^{\alpha 11}_\alpha]_{n(\alpha{+}1)} 
\hackcenter{\begin{tikzpicture}[ scale=1.1]
    \begin{scope}
        \draw[black, ultra thick] (0,-.01)-- (0,.2);
        \draw[black, ultra thick] (.6,-.01)-- (.6,.2);
        \draw[black, ultra thick] (1.2,-.01)-- (1.2,.2);
    \end{scope}
    \draw[black, ultra thick] (0,0) to (0, .8);
    \draw[black, ultra thick] (.6,0) to (.6, .8);
    \draw[black, ultra thick] (1.2,0) to (1.2, .8);
    \draw[ultra thick, black] (0,0) to (.6,-.6) to (.6,-1.2);
    \draw[ultra thick, black] (0.6,0) to (.9,-.3);
    \draw[ultra thick, black](1.2,0) to (.6,-.6); 
    \node at (0, 1) {$\alpha$};
    \node at (.6, 1) {$1$}; 
    \node at (1.2, 1) {$1$};  
    \node at (.8,-1) {$\scriptstyle \alpha$};
    \node at (1.0,-.6) {$\scriptstyle n$}; 
\end{tikzpicture} } 
\\
&= 
\sum_{n,m}  [F^{\alpha 11}_\alpha] ^{-1} _{mn}R^{11}_n[F^{\alpha 11}_\alpha]_{n(\alpha{+}1)} \ket{m} 
\end{align*}
where $n\in\{0,2\}$ and $m \in\{\alpha{-}1,\alpha{+}1\}$.

A similar calculation on $\ket{1}$ gives the matrix 
\begin{equation}\label{eq:b2_unnormalized}
    \mathsf{b}_2 =  [F^{\alpha 11}_\alpha]^{-1}\left(\begin{matrix}
    q^{5/2} & 0\\
    0 & q^{1/2}
\end{matrix}\right) [F^{\alpha 11}_\alpha] 
\end{equation}
representing the action of $\b_2$. 
 
While $J_1$ is manifestly unitary, $\mathsf{b}_2$ is not: $[F^{\alpha 11}_\alpha]$ was computed in a non-normalized basis.
%
Let $$M_\H=\text{diag}\left(\sqrt{\braket{0|0}}, \sqrt{\braket{1|1}}\right)=\text{diag}\left(\sqrt{B_0\md_\alpha}, \sqrt{B_1\md_\alpha}\right)$$ using the coefficients from \eqref{eq:B0B1}, 
 so that anyon braiding produces the following gates
\begin{equation}\label{eq:J1b2_normalized}
\begin{split}
J_1 &= \left(\begin{matrix}
        q^{3+\alpha} & 0\\
        0 & q^{-5-\alpha}
    \end{matrix} \right),
\\
        \mathsf{b}_2^{{\rm norm}} &= M_{\H} \mathsf{b}_2 M_{\H}^{-1} =q^{1/2}\left(
        \begin{matrix}
            \frac{1+q^{2}}{1-q^{2\alpha}} & q^{-1}\frac{\sqrt{B_0\md_\alpha}}{\sqrt{B_1\md_\alpha}}\\
            q^{-1}\frac{\sqrt{B_0\md_\alpha}}{\sqrt{B_1\md_\alpha}} & \frac{1+q^{2}}{1-q^{-2\alpha}}
        \end{matrix}
        \right) \ .
        \end{split}
    \end{equation}

With this normalization, the braids $J_1$ and $\mathsf{b}_2^{{\rm norm}}$ are now explicitly unitary, in the sense that they preserve the norm (and orthogonality) of $\ket{0}$ and $\ket{1}$. However, these states do not necessarily have positive norms. If $\braket{0|0}$ and $\braket{1|1}$ are both positive (or both negative --- an overall minus sign is irrelevant), then $\mathsf{b}_2^{{\rm norm}}$ is unitary in the usual sense: $\mathsf{b}_2^\dag \mathsf{b}_2 = \Id$.

\begin{remark}
    The matrices $J_1$ and $\mathsf{b}_2$ have determinants with unit norm, but not equal to one, so we rescale them by the appropriate phases:
    \begin{equation}\label{eq:single_qubit_braids_special_unitary}
\begin{split}
    J_1&= \mathsf{b}_1^2  = -q\left(\mathsf{b}_1^{\alpha 1 1}\right)^2 = \left(\begin{matrix}
        q^{\alpha} & 0\\
        0 & q^{-\alpha} 
    \end{matrix}\right),
\\
              \mathsf{b}_2 &= 
   q^{-\frac{3}{2}}\mathsf{b}_2^{\alpha11}=q^{-1}\left(
 \begin{array}{cc}
  \frac{1+q^{2}}{1-q^{2\alpha}} & q^{-1}\frac{\sqrt{B_{0}}}{\sqrt{B_{1}}}\\
  q^{-1}\frac{\sqrt{B_{0}}}{\sqrt{B_{1}}} & \frac{1+q^{2}}{1-q^{-2\alpha}}
 \end{array}
 \right).
\end{split}
\end{equation}
Note that we dropped $\md_\alpha$ from under the square roots in $\mathsf{b}_2$. This is possible if $\alpha$ is in the range where the metric is positive definite and $B_0$ and $B_1$ have the same sign. Outside of this range, the factors of $\md_\alpha$ can still be cancelled, but a minus sign may arise.  
\end{remark}

The key advantage of augmenting the traditional Ising theory by a neglecton $\alpha$ is the ability to achieve universal single qubit gates.

\begin{theorem}\label{thm:single_qubit_density_result}
Single-qubit universality is achieved by affine braiding of anyons in $\mathcal{H}_1$ via \eqref{eq:J1b2_normalized} for all values of $\alpha$ for which $\mathsf{b}_1^2$ and $\mathsf{b}_2$ are unitary, except for when $\alpha = 2\pm \frac{2}{3}\quad (\text{mod } 4)$ or $\alpha = 2\pm \frac{3}{5} \quad (\text{mod } 4)$.
\end{theorem}

\begin{proof}
We make use of Fact 1 and Theorem 1 of \cite{criteria_universality_SUd}, giving necessary and sufficient conditions for two matrices $A$ and $B \in SU(2)$, to generate $SU(2)$. 
 Let the eigenvalues of $A$ and $B$ be $e^{\pm i\pi\phi_A}$ and $e^{\pm i\pi\phi_B}$ respectively, with $0\leq \phi_a,\phi_b\leq 1$.
Then \cite{criteria_universality_SUd} provides the following three conditions to guarantee single-qubit universality:
\begin{enumerate}
    \item $A$ and $B$ do not commute.
    \item $\phi_A$ and $\phi_B$ are both different from 1/2.
    \item At least one between $\phi_A$ and $\phi_B$ must must not be of the form $\frac{m}{n}$, with $m$ and $n$ are integers such that $m<n\leq 6$.
\end{enumerate}
The first and third conditions are necessary, while the second condition can be relaxed in some cases.

We see that condition 1 is clearly satisfied by $\mathsf{b}_1^2$ and $\mathsf{b}_2$ because they cannot be simultaneously diagonalized. The eigenvalues of $\mathsf{b}_1^2$ and $\mathsf{b}_2$ are  $q^{\pm \alpha} = e^{\pm i\pi \alpha/4}$ and $e^{\pm i\pi 3/4}$ respectively, so condition 2 is always satisfied as long if $\alpha$ is not an integer. The matrix $\mathsf{b}_2$ fails condition 3, but $\mathsf{b}_1^2$ automatically satisfies it for all irrational values of $\alpha$. For rational values, we restrict ourselves to the set $(1,2)\cup(2,3)\cup(5,6)\cup(6,7)$, as $\mathsf{b}_1^2$ and $\mathsf{b}_2$ are periodic in $\alpha$ with period 8.
Condition 3 fails for $\alpha \in \{1+\frac{1}{3}, 1+\frac{2}{5},2+2/3,2+\frac{3}{5}, 5+\frac{1}{3}, 5+\frac{2}{5},6+2/3,6+\frac{3}{5}\}\quad (\text{mod } 8) = \{2\pm \frac{2}{3},2\pm3/5\} \quad (\text{mod }4)$.
\end{proof}

\begin{remark}[Comparison with semisimple Ising]
In the standard semisimple theory of Ising anyons, the single-qubit braiding generators have eigenvalues $e^{\pm i\pi \frac{3}{4}}$, which lie in the Clifford group~\cite{Fan_2010,MR2640343}. Consequently, Ising anyons can implement only Clifford operations and cannot generate arbitrary single-qubit rotations, making them non-universal for single-qubit gates.
\end{remark}

\subsection{Multiqubit operations}\label{sec:entangling-gates}

Using the Jucys-Murphy braids, it is possible to immediately extend the single qubit universality result to any qubit in $\H_n$.  

\begin{proposition}
The matrix representation of the operators $(J_{2i-1},\mathbf{b}_{2i})$ acting on the normalized basis of $i^{th}$ qubit of $\mathcal{H}_n^{{\rm comp}}$ are all given by the same matrices from \eqref{eq:J1b2_normalized}, so that $J_{2i-1} = J_1$ and $\mathbf{b}_{2i} = \mathbf{b}_{2}$, 
for all $1 \leq i \leq n$.  
\end{proposition}

\begin{proof}
To see that $J_{2i-1}$ acts on the $i$th qubit, consider that $J_{2i-1}$ commutes with fusions among the first $2i$ anyons as in Figure~\ref{fig:jucys_murphy}. The result is a local computation identical to the single qubit calculations.
\end{proof}

 \begin{figure}
    \centering
\begin{equation}
\hackcenter{
\begin{tikzpicture}[  scale=0.5, yscale=-1  ]  
\draw [black, very thick]    (-.5,-2) .. controls +(0,.25) and +(0,-.25) ..  (.5,-1) ;
\path [fill=white] (-.25,-1.4) rectangle (.75,-1.6);
\draw [black, very thick]   (.5,-2)  .. controls +(0,.25) and +(0,-.25) ..  (-.5,-1);
\draw[very thick, ] (-1.5,-1) to (-1.5,-2);
\draw[very thick, ] (-2.5,-1) to (-2.5,-3);
\draw [black, very thick]    (-1.5,-3) .. controls +(0,.25) and +(0,-.25) ..  (-.5,-2) ;
\path [fill=white] (-1.25,-2.4) rectangle (-.25,-2.6);
\draw [black, very thick]   (-.5,-3)  .. controls +(0,.25) and +(0,-.25) ..  (-1.5,-2);
 \draw[very thick, ] (.5,-2) to (.5,-6);
 \draw [black, very thick]    (-3.25,-4) .. controls +(0,.45) and +(0,-.45) ..  (-1.5,-3) ;
 \path [fill=white] (-2.35,-3.35) rectangle (-2.65,-3.8);
  \draw [black, very thick]   (-2.5,-7) to  (-2.5,-1);
 \path [fill=white] (-2.35,-4.2) rectangle (-2.65,-4.6);
 \draw [black, very thick]   (-1.5,-5)  .. controls +(0,.5) and +(0,-.45) ..  (-3.25,-4);
 \draw[very thick, postaction={}] (-.5,-5) to (-.5,-3);
 \draw [black, very thick]    (-1.5,-6) .. controls +(0,.25) and +(0,-.25) ..  (-.5,-5) ;
 \path [fill=white] (-1.25,-5.4) rectangle (-.25,-5.6);
 \draw [black, very thick]   (-.5,-6)  .. controls +(0,.25) and +(0,-.25) ..  (-1.5,-5);
 \draw [black, very thick]    (-.5,-7) .. controls +(0,.25) and +(0,-.25) ..  (.5,-6) ;
 \path [fill=white] (-.25,-6.4) rectangle (.75,-6.6);
  \draw [black, very thick]   (.5,-7)  .. controls +(0,.25) and +(0,-.25) ..  (-.5,-6);
   \draw[very thick, ] (-1.5,-6) to (-1.5,-7);
   \draw[very thick,  ] (1.5,-1) to (1.5,-7);
\draw[very thick,] (-2,0) to [out=90, in=220] (-1.25,1);
\draw[very thick, ] (-.5,-1) to [out=90, in=-30] (-1.25,1);
\draw[very thick, ] (-1.2,1) to [out=90, in=220]  (-.65,2);
\draw[very thick,  black] (-2.5,-1)to [out=90, in=210] (-2,0);
\draw[very thick, ] (-1.5,-1) to [out=90, in=-30] (-2,0);
\draw[very thick,] (.5,-1) to [out=90, in=-30] (-.65,2);
\draw[very thick,] (-.65,2) to [out=90, in=220] (-.15,3);
\draw[very thick, ] (1.5,-1) to [out=90, in=-30] (-.15,3) to(-.15,4) ;
\node at (-2.75,-.8) {$\scriptscriptstyle  \alpha$};
\node at (-1.8,-.8) {$\scriptscriptstyle  1$};
\node at (-0.8,-.8) {$\scriptscriptstyle  1$};
\node at (.2, -.8) {$\scriptscriptstyle  1$};
\node at (1.2,-.8) {$\scriptscriptstyle  1$};
\node at (-1.3,1.8) {$\scriptscriptstyle \alpha $};
\node at (-2.4,.6) {$\scriptscriptstyle \alpha \pm 1$};
\node at (-1.2,2.6) {$\scriptscriptstyle \alpha \pm 1$};
\node at (-.4,3.7) {$\scriptscriptstyle  \alpha $};
\end{tikzpicture} }
\quad = \quad
    \hackcenter{
\begin{tikzpicture}[  scale=0.5,   yscale=-1 ]
\draw[very thick  ] (0,0) to [out=90, in=220] (.75,1);
\draw[very thick,  ] (2.2,-6) to [out=90, in=-50] (.75,1);
\draw[very thick, ] (.75,1) to  (.75,2);
\draw[very thick, ] (-.5,-1)to [out=90, in=210] (0,0);
\draw[very thick, ] (.5,-1) to [out=90, in=-30] (0,0);
\draw [black, very thick]    (-.5,-2) .. controls +(0,.25) and +(0,-.25) ..  (.5,-1) ;
\path [fill=white] (-.25,-1.4) rectangle (.75,-1.6);
\draw [black, very thick]   (.5,-2)  .. controls +(0,.25) and +(0,-.25) ..  (-.5,-1);
\draw [black, very thick]    (-.5,-3) .. controls +(0,.25) and +(0,-.25) ..  (.5,-2) ;
 \path [fill=white] (-.25,-2.4) rectangle (.75,-2.6);
 \draw [black, very thick]   (.5,-3)  .. controls +(0,.25) and +(0,-.25) ..  (-.5,-2);
\draw[very thick,] (-1.3,-5) to [out=90, in=220] (-.5,-4);
\draw[very thick, ] (.2,-6) to [out=90, in=-30] (-.5,-4);
\draw[very thick,  ] (-.5,-4) to  (-.5,-3);
\draw[very thick, black ] (-1.8,-6)to [out=90, in=210] (-1.3,-5);
\draw[very thick, ] (-.8,-6) to [out=90, in=-30] (-1.3,-5);
\draw[very thick, ] (1.2,-6) .. controls +(0,1.75) and +(0,-1) ..  (.5,-3);
\node at (-2,-5.8) {$\scriptscriptstyle  \alpha$};
\node at (-1,-5.8) {$\scriptscriptstyle  1$};
\node at (-0,-5.8) {$\scriptscriptstyle  1$};
\node at (1, -5.8) {$\scriptscriptstyle  1$};
\node at (2,-5.8) {$\scriptscriptstyle  1$};
\node at (-.8,-3.4) {$\scriptscriptstyle \alpha $};
\node at (-1.6,-4.2) {$\scriptscriptstyle \alpha \pm 1$};
\node at (-.5,0.6) {$\scriptscriptstyle \alpha \pm 1$};
\node at (.4,1.6) {$\scriptscriptstyle  \alpha $};
\end{tikzpicture} }
\end{equation}
\caption{\justifying The Jucys-Murphy braid $J_3$ acts as the identity on the first qubit, and as $\mathsf{b}_1^2$ on the second qubit.}
    \label{fig:jucys_murphy}
\end{figure}


The eigenbasis of the mutually commuting Jucys-Murphy operators is also advantageous from the perspective of studying positivity of the inner product, essentially reducing to our single qubit computations.  We illustrate with the two qubit space  $\H_2= \Hom(\alpha,\alpha\otimes 1\otimes 1\otimes 1\otimes 1)$. This space is six-dimensional, with a basis given by 
\begin{equation} \label{eq:H2basis}
\begin{array}{cc}
\ket{00} = \hackcenter{\begin{tikzpicture}[scale=0.85]
  \draw[ultra thick, black] (0,0) to (1.2,-1.2);
  \draw[ultra thick, black] (0.6,0) to (.3,-.3);
  \draw[ultra thick, black] (1.2,0) to (.6,-.6);
  \draw[ultra thick, black] (1.8,0) to (.9,-.9);
  \draw[ultra thick, black] (2.4,0) to (1.2,-1.2) to (1.2,-1.8);
  \node at (0,0.2) {$\alpha$};
  \node at (.6,0.2) {$1$};
  \node at (1.2,0.2) {$1$};
  \node at (1.8,0.2) {$1$};
  \node at (2.4,0.2) {$1$};
  \node at (1.2,-2) {$\alpha$};
  \node at (.18,-.6) {$\scriptstyle \alpha{+}1$};
  \node at (.5,-.9) {$\scriptstyle \alpha$};
  \node at (.78,-1.2) {$\scriptstyle \alpha+{1}$};
\end{tikzpicture}} 
&
\ket{10} = \hackcenter{\begin{tikzpicture}[scale=0.85]
  \draw[ultra thick, black] (0,0) to (1.2,-1.2);
  \draw[ultra thick, black] (0.6,0) to (.3,-.3);
  \draw[ultra thick, black] (1.2,0) to (.6,-.6);
  \draw[ultra thick, black] (1.8,0) to (.9,-.9);
  \draw[ultra thick, black] (2.4,0) to (1.2,-1.2) to (1.2,-1.8);
  \node at (0,0.2) {$\alpha$};
  \node at (.6,0.2) {$1$};
  \node at (1.2,0.2) {$1$};
  \node at (1.8,0.2) {$1$};
  \node at (2.4,0.2) {$1$};
  \node at (1.2,-2) {$\alpha$};
  \node at (.18,-.6) {$\scriptstyle \alpha{-}1$};
  \node at (.5,-.9) {$\scriptstyle \alpha$};
  \node at (.78,-1.2) {$\scriptstyle \alpha{+}1$};
\end{tikzpicture}} 
\\[2ex]
\ket{01} = \hackcenter{\begin{tikzpicture}[scale=0.85]
  \draw[ultra thick, black] (0,0) to (1.2,-1.2);
  \draw[ultra thick, black] (0.6,0) to (.3,-.3);
  \draw[ultra thick, black] (1.2,0) to (.6,-.6);
  \draw[ultra thick, black] (1.8,0) to (.9,-.9);
  \draw[ultra thick, black] (2.4,0) to (1.2,-1.2) to (1.2,-1.8);
  \node at (0,0.2) {$\alpha$};
  \node at (.6,0.2) {$1$};
  \node at (1.2,0.2) {$1$};
  \node at (1.8,0.2) {$1$};
  \node at (2.4,0.2) {$1$};
  \node at (1.2,-2) {$\alpha$};
  \node at (.18,-.6) {$\scriptstyle \alpha{+}1$};
  \node at (.5,-.9) {$\scriptstyle \alpha$};
  \node at (.78,-1.2) {$\scriptstyle \alpha{-}1$};
\end{tikzpicture}} 
&
\ket{11} = \hackcenter{\begin{tikzpicture}[scale=0.85]
  \draw[ultra thick, black] (0,0) to (1.2,-1.2);
  \draw[ultra thick, black] (0.6,0) to (.3,-.3);
  \draw[ultra thick, black] (1.2,0) to (.6,-.6);
  \draw[ultra thick, black] (1.8,0) to (.9,-.9);
  \draw[ultra thick, black] (2.4,0) to (1.2,-1.2) to (1.2,-1.8);
  \node at (0,0.2) {$\alpha$};
  \node at (.6,0.2) {$1$};
  \node at (1.2,0.2) {$1$};
  \node at (1.8,0.2) {$1$};
  \node at (2.4,0.2) {$1$};
  \node at (1.2,-2) {$\alpha$};
  \node at (.18,-.6) {$\scriptstyle \alpha{-}1$};
  \node at (.5,-.9) {$\scriptstyle \alpha$};
  \node at (.78,-1.2) {$\scriptstyle \alpha{-}1$};
\end{tikzpicture}}  
\\[2ex]
\ket{NC_1} = \hackcenter{\begin{tikzpicture}[scale=0.85]
  \draw[ultra thick, black] (0,0) to (1.2,-1.2);
  \draw[ultra thick, black] (0.6,0) to (.3,-.3);
  \draw[ultra thick, black] (1.2,0) to (.6,-.6);
  \draw[ultra thick, black] (1.8,0) to (.9,-.9);
  \draw[ultra thick, black] (2.4,0) to (1.2,-1.2) to (1.2,-1.8);
  \node at (0,0.2) {$\alpha$};
  \node at (.6,0.2) {$1$};
  \node at (1.2,0.2) {$1$};
  \node at (1.8,0.2) {$1$};
  \node at (2.4,0.2) {$1$};
  \node at (1.2,-2) {$\alpha$};
  \node at (.18,-.6) {$\scriptstyle \alpha{+}1$};
  \node at (.5,-.9) {$\scriptstyle \alpha{+}2$};
  \node at (.78,-1.2) {$\scriptstyle \alpha{+}1$};
\end{tikzpicture}} 
&
\ket{NC_2} = \hackcenter{\begin{tikzpicture}[scale=0.85]
  \draw[ultra thick, black] (0,0) to (1.2,-1.2);
  \draw[ultra thick, black] (0.6,0) to (.3,-.3);
  \draw[ultra thick, black] (1.2,0) to (.6,-.6);
  \draw[ultra thick, black] (1.8,0) to (.9,-.9);
  \draw[ultra thick, black] (2.4,0) to (1.2,-1.2) to (1.2,-1.8);
  \node at (0,0.2) {$\alpha$};
  \node at (.6,0.2) {$1$};
  \node at (1.2,0.2) {$1$};
  \node at (1.8,0.2) {$1$};
  \node at (2.4,0.2) {$1$};
  \node at (1.2,-2) {$\alpha$};
  \node at (.18,-.6) {$\scriptstyle \alpha{-}1$};
  \node at (.5,-.9) {$\scriptstyle \alpha{-}2$};
  \node at (.78,-1.2) {$\scriptstyle \alpha{-}1$};
\end{tikzpicture}}
\end{array}
\end{equation}
As in the case of quantum computing with Fibonacci anyons~\cite{Cui-leak}, the two-qubit Hilbert space has dimension greater than four, so a four-dimensional \emph{computational space} must be identified. Care must be taken to ensure that braiding operations preserve this subspace, as transitions into the orthogonal complement (the \emph{non-computational space}) constitute \emph{leakage}.

The basis vectors in the computational space are chosen due to their resemblance to the single-qubit basis. This makes the computation of the norm of computational basis states particularly easy: it amounts to applying \eqref{eq:norm_of_ket0} and \eqref{eq:norm_of_ket1} twice:
\begin{equation}\label{eq:two_qubit_norm}
    \begin{split}
        \braket{00|00}=B_0^2 \md_\alpha, \quad
        \braket{01|01}=B_0B_1\md_\alpha,\\
        \braket{10|10}=B_0B_1\md_\alpha,\quad
        \braket{11|11}=B_1^2\md_\alpha.
    \end{split}
\end{equation}
Critically, this means that the signature of the inner product on the computational space is completely determined by the positivity properties of $B_0$, $B_1$, and $\md_{\alpha}$ as in the single qubit setting. 

\begin{proposition} \label{prop:posdef}
For $\alpha$ in the intervals $(1,2) \mod 4$, the inner product on $\H_n^{{\rm comp}}$  will be negative-definite and positive-definite in the range $(2,3) \mod 4$.  Furthermore, for $\alpha$ as in Theorem~\ref{thm:single_qubit_density_result}, braiding of anyons will achieve single qubit universality for each qubit in $\H_n$. 
\end{proposition}

It is important to note that while the computational space $\H_n^{{\rm comp}}$  is guaranteed to be positive definite for all $\alpha$ in the range described in Proposition~\ref{prop:posdef}, the noncomputational space $\H_n^{{\rm NC}}$ will generally contain at least one vector with a signature differing from the computational space. 
It is widely believed that the ability to perform leakage-free entangling of two qubits prohibits any given anyon model from achieving universal quantum computation~\cite{Cui-leak}.  Hence, to entangle qubits in our model will require leakage into the noncomputational space with a mixed signature.   It is therefore critical that we are able to reduce the leakage to an arbitrarily small value to stay within a positive-definite quantum mechanical framework.  
 
We reduce the leakage using a procedure developed by Cui, Tian, Vasquez, Wang, and Wong \cite{Cui-leak} and Reichardt \cite{Reichdist} in the context of Fibonacci anyons.  Here, we extend the result to include mixed signature unitaries. 
\begin{lemma}\label{lemma:Cui}
    Let $U$ be a $2\times2$ unitary or indefinite unitary matrix with off-diagonal entries less than one in absolute value: $|U_{01}|<1$, and let 
\[
D(\theta) := \mathrm{diag}(e^{-i\theta/2}, e^{i\theta/2}),
\]
    be a z-rotation by $\theta$ where
    $-\frac{\pi}{2}<\theta < \frac{\pi}{2}$ and $\theta\neq 0$.
    Then the recursive sequence
    \begin{equation*}
    \begin{split}
        U_0 &= U,\\
        U_{k+1} &= U_kD(\theta) U_k^{-1} D(\theta)U_kD(\theta)^{-2}
    \end{split}
    \end{equation*}
    converges to a diagonal gate.
\end{lemma}

Lemma~\ref{lemma:Cui} can be applied to construct entangling gates with arbitrarily small leakage into the non-computational subspace.  The braid generator $\b_3$, acting on the two-qubit space $\mathcal{H}_2$, has a block structure consisting of two $2\times2$ blocks that couple $\ket{00} \leftrightarrow \ket{NC_1}$ and $\ket{11} \leftrightarrow \ket{NC_2}$, along with two $1\times1$ blocks acting on $\ket{01}$ and $\ket{10}$. The braid $(J_2 J_3^{-1})^2$ acts diagonally on each two-dimensional block via $D(2\pi\alpha)$ and $D(-2\pi\alpha)$ respectively, and as the identity on the remaining subspace. We are now able to approximate single-qubit operations and nearest-neighbor entangling gates, allowing the approximation of any multi-qubit gate.

\begin{theorem}
The image of the affine braid group $\Br_{2n}^{\mathrm{aff}}$ under the representation on $\mathcal{H}_n$ is dense in the unitary group $\mathrm{PSU}(\mathcal{H}_n^{\mathrm{comp}})$  acting on the computational subspace $\mathcal{H}_n^{\mathrm{comp}}$ for all non-integer $\alpha \in (2 - \tfrac{1}{3}, 2 + \tfrac{1}{3}) \mod 4$.   
\end{theorem}

\begin{proof}
Using the recursive construction from Lemma~\ref{lemma:Cui}, with $U = \b_3$ and $D(\theta) = (J_2 J_3^{-1})^2$, one obtains approximately diagonal (and hence nearly leakage-free) gates on $\mathcal{H}_2$, provided that $|\braket{NC_1 | \b_3 | 00}| < 1$ and $|\braket{NC_2 | \b_3 | 11}| < 1$, and for $\pi\alpha \neq 0$ in the interval $(-\frac{\pi}{2}, \frac{\pi}{2}) \mod 2\pi$. The resulting gates are entangling, which can be verified by direct inspection.

This method yields entangling gates with arbitrarily small leakage for all non-integer $\alpha \in (2 - \tfrac{1}{3}, 2 + \tfrac{1}{3})$, ensuring universality in the regime where the Hermitian structure is positive-definite on the computational subspace. The effective range of $\alpha$ may be extended by identifying braids other than $(J_2 J_3^{-1})^2$with reduced leakage amplitudes.
\end{proof}

\subsection{Decoupling from the indefinite space} \label{sec:decoupling_from_indefinite_space}

 Given a general diagonal gate, the algorithm from Lemma~\ref{lemma:Cui}  does not eliminate leakage efficiently; while the off-diagonal terms are suppressed exponentially, the number of operations required to implement the algorithm also grows exponentially with each step, and a number of operations that is polynomial (rather than logarithmic) in the inverse leakage is needed \cite{Reichardt_2005,Cui-leak}. Having an inefficiently-implemented generator is undesirable in quantum compilation, and one of the assumptions of the Solovay-Kitaev theorem is that generators be efficiently computable.
 
While certain values of the rotation angle $\theta$ are known to suppress leakage at a doubly exponential rate~\cite{Reichdist,Cui-leak}, we show that for specific values of $\theta$ it is possible to completely eliminate certain off diagonal entries.    These values depend on the norm of the off-diagonal components of the matrix being corrected. We are able to implement these gates in our model and exactly decouple from the negative-definite sector of the non-computational subspace. 

In the case of low-leakage entangling gates on $\mathcal{H}_2$, the negative-norm direction is represented by the vector $\ket{NC_1}$. We will show that there exists a value of $\alpha$ such that, after a single iteration of the algorithm in Lemma~\ref{lemma:Cui} applied to $\b_3$, the resulting gate has zero amplitude for leakage into the subspace spanned by $\ket{NC_1}$. That is, the computational space is fully decoupled from the negative-definite component of the Hilbert space, ensuring unitary evolution within the physical subspace.

If $U_0 =\left(\begin{matrix}
    a_0  & -\bar{b_0} \\
    b_0  & \bar{a_0}
\end{matrix}\right)$
is an indefinite unitary matrix, then the norm of the off-diagonal term of $U_1$ from Lemma~\ref{lemma:Cui} is given by \cite{Cui-leak}
\begin{equation}\label{eq:cui_reichardt_expansion}
    |b_1|=|b_0||(2-2\cos\theta)(1-|b_0|^2)-1|
\end{equation}
This quantity vanishes if 
$\cos\theta = \frac{2 |b_0|^2-1}{2
   \left(|b_0|^2-1\right)}$.  In the case of Fibonacci anyons, both $\theta$ and $b_0$ are fixed values, so the RHS of \eqref{eq:cui_reichardt_expansion} is never zero.
   However, in the non-semisimple Ising theory, there exist values of $\alpha$ so that this quantity vanishes.  If we pick $D(\theta)$ and $U_0$ to be the restriction of $(J_2J_3^{-1})^2$ and $\b_3$ to the subspace spanned by $\ket{00}$ and $\ket{NC_1}$ respectively (this is where the mixing between positive and negative normed states occur), then $\theta = 2\pi \alpha$ and 
\begin{equation*}
|b_0|^2 = \left|\left(\frac{\sqrt{B_0}}{\sqrt{B_1}} \right)^2\right| = \frac{1}{2}\left| 1-\cot ^2\frac{\pi }{4}( \alpha +1)\right|,
\end{equation*}
and leakage can be eliminated if $\cos \left(2\pi\alpha \right)= \cos ^2\left(\frac{1}{4} \pi  (\alpha +1)\right)$. This equation has a solution\footnote{The equation can be solved exactly by expanding the right hand side using trigonometric identities and solving the resulting cubic equation (in $\sin \frac{\pi \alpha}{2}$) to obtain \\$\alpha_0 = 2-\frac{2 \sin ^{-1}\left(\frac{1}{3}-\frac{2}{3} \cos \left(\frac{1}{3} \cos^{-1}\left(-\frac{5}{32}\right)\right)\right)}{\pi }$.}
\begin{equation*}
\alpha_0\sim 2.145
\end{equation*}
that lies in the range $(2,3)$ where the computational space is positive-definite, and $\ket{NC_1}$ is negative-definite. Hence, a single application of the prescription from Lemma~\ref{lemma:Cui} at $\alpha_0$ completely eliminates any leakage into the negative-definite space. We are thus left with a five-dimensional, positive-definite Hilbert space. 

For higher iterations $U_k$ of the recursive algorithm, the analytic expression for the off-diagonal terms $b_k$ becomes more complex, but given the continuous nature of $\alpha$, it is possible to find values where these terms are identically zero.

While Reichardt's algorithm can completely suppress leakage into the negative-norm state $\ket{NC_1}$ at the specific value $\alpha = \alpha_0$, unitary mixing between $\ket{11}$ and the non-computational state $\ket{NC_2}$ persists. Moreover, leakage suppression via this method is only polynomially efficient for generic $\alpha$, including $\alpha_0$~\cite{Reichdist}. However, because the non-unitary leakage into the negative-definite direction is fully eliminated, we are able to apply more powerful techniques from~\cite{AAEL} to establish more efficient universality.

The key tools are the \emph{bridge lemma} and the \emph{decoupling lemma} (see Section~7 of Aharonov et al.~\cite{AAEL}),
 which we now briefly recall. The bridge lemma states that if a set of operators acts densely on two orthogonal subspaces $A$ and $B$, with $\dim B > \dim A$, and includes at least one unitary that mixes $A$ and $B$, then the action is dense on $A \oplus B$. The decoupling lemma ensures that when the group acts densely on two subspaces of unequal dimension, one can approximate arbitrary unitaries on one subspace while simultaneously approximating the identity on the other.

In our setting, the braids $\b_3$ and $J_1$ generate (not independent) dense actions on the subspaces $\{\ket{00}, \ket{01}\}$ and $\{\ket{10}, \ket{11}\}$, respectively, while $J_2$ acts densely on the one-dimensional subspace spanned by $\ket{NC_2}$ (provided $\pi \alpha_0 \notin \mathbb{Q}\pi$). The braid $\b_3$ serves as a bridge mixing $\ket{11}$ with $\ket{NC_2}$, allowing us to invoke the bridge lemma to conclude that the action is dense on the three-dimensional space $\{\ket{10}, \ket{11}, \ket{NC_2}\}$. Applying the decoupling lemma then ensures that this density is independent of the density over $\{\ket{00}, \ket{01}\}$.

Finally, the braid $\b_4$ acts as a bridge between these two sectors, completing the argument via a second application of the bridge lemma. We thus obtain density in all of $\mathrm{SU}(5)$ acting on the computational and adjacent non-computational states, completing the proof of universality. The Solovay-Kitaev theorem can then be applied to efficiently approximate a non-leaking entangling gate on the computational space.  

\section{Robustness of the generators}\label{sec:robustness}

While we do not propose an experimental realization of this system in this work, it is reasonable to assume the value of $\alpha$ cannot be fine-tuned in a realistic experimental setup. It is then necessary to ensure that the gates are a continuous and ``well-behaved" function of $\alpha$.
We analyze the robustness of the encoding by computing the distance between the ``unperturbed" gate at  $\alpha$ and the ``perturbed" gate at $\alpha + \epsilon$. We denote the perturbed version of a braid $\b$ by $\tilde{\b}$, and we assume that both $\alpha$ and $\alpha+\epsilon$ lie in the interval $(2,3)$.

Diagonal single-qubit gates change in a straightforward way: $\tilde{\mathsf{b}}_1^2 =\left(\begin{matrix}
        q^{\alpha+\epsilon} & 0\\
        0 & q^{-\alpha-\epsilon} 
    \end{matrix}\right)$, and the phase invariant distance between the perturbed and unperturbed gates is 
\begin{equation*}
    D(\b_1^2,\tb_1^2)= \sqrt{2-|\Tr[{\b_1^2}^\dag  \tb_1^2]|}=\sqrt{2-2 \cos \left(\frac{\pi  \epsilon
   }{4}\right)}
\end{equation*}

\begin{figure}
    \centering
    \includegraphics[width= \linewidth]{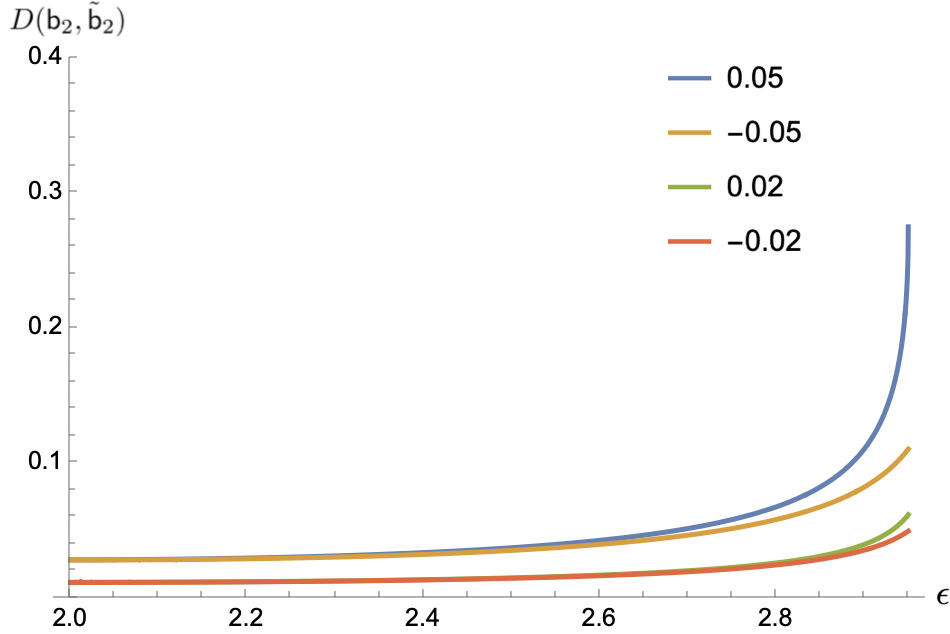}
    \caption{\justifying Distance $D(\b_2, \tilde{\b}_2)$ for select values of $\epsilon$. The rapid growth on the right-hand side of the graph occurs when $\alpha+\epsilon$ approaches $3$. The growth is finite and goes to zero as $\epsilon$ goes to zero.}\label{fig:robustness_single_qubit}
\end{figure}

The phase invariant distance between $\b_2$ and $\tilde{\b}_2$ (restricted to a single qubit subspace) takes the form  
\begin{align}
 &D(\b_2,\tilde{\b}_2) = \scriptstyle \sqrt{2-2\left(\sqrt{\frac{B_0}{B_1}\cdot\frac{\tilde{B}_0}{\tilde{B}_1}}
+\cos\left(\frac{\pi\epsilon}{4}\right)\sqrt{1-\frac{B_0}{B_1}}\sqrt{1-\frac{\tilde{B}_0}{\tilde{B}_1}}\right)}  \nonumber 
\end{align}
where $\tilde{B}_0$ and $\tilde{B}_1$ are the norms of the single-qubit states \eqref{eq:B0B1} at $\alpha+\epsilon$ before normalization. 
Note that the right-hand-side is bounded. 
We plot  $D(\b_2, \tilde{\b}_2)$ for select values of $\epsilon$ in Fig. \ref{fig:robustness_single_qubit} to illustrate the stability of $\b_2$ under perturbations of $\alpha$.

The entangling gate is a greater source of errors. We consider the entangling gate $U_1$ described in Section \ref{sec:decoupling_from_indefinite_space}, obtained by applying the procedure in Lemma \ref{lemma:Cui} to $\b_3$. This gate mixes the computational and non-computational spaces, and is indefinite-unitary (the basis vector $\ket{NC_1}$ has negative norm for $\alpha \in (2,3)$). 

\begin{figure}
    \centering
    \includegraphics[width=\linewidth]{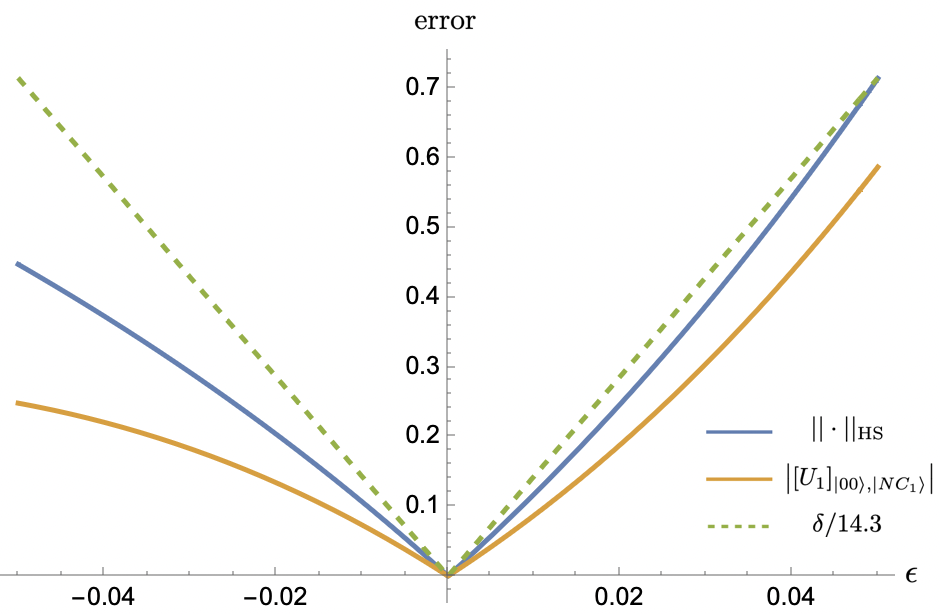}
    \caption{\justifying  The Hilbert-Schmidt norm distance between target matrix and effective matrix at $\alpha = \alpha_0 + \epsilon$ (blue) and norm of the non-unitary mixing term (orange) as a function of $\epsilon$ are bounded by $|\epsilon|< \frac{\delta}{14.3}$ (dashed)} 
    \label{fig:robustness_entangling_a0}
\end{figure}

We consider the robustness of the decoupling of the negative-definite subspace at $\alpha = \alpha_0\sim 2.145$. At a generic $\alpha$, the block structure of $U_1$ is such that $\ket{00}$ and $\ket{NC_1}$ are mixed. At $\alpha=\alpha_0$, the non-unitary mixing is exactly suppressed, and $U_1$ acts as a $5$ by $5$ unitary on the space spanned by the computational space and $\ket{NC_2}$. This is the target gate we wish to realize. We quantify the robustness in two ways: the operator-norm distance between the target braid and the perturbed braid restricted to $SU(5)$, and the norm of the entry that mixes the positive and negative definite subspaces $[U_1]_{\ket{00} ,\ket{NC_1}}$ at  $\tilde{\alpha} = \alpha_0+\epsilon$. Note that $[U_1]_{\ket{00} ,\ket{NC_1}}$ is zero at $\alpha = \alpha_0$, so this measure directly quantifies the non-unitary ``mixing".  
Both measures can be made smaller than a chosen small precision $\delta$ 
by ensuring that $|\epsilon|< \frac{\delta}{14.3}$.
These measures of non-unitary robustness are depicted in Fig. \ref{fig:robustness_entangling_a0}. Note that the error, while growing faster than in the unitary case as a function of $\epsilon$, is approximately linear for small $\epsilon$, ensuring that the value of $\alpha$ does not need to be tuned to exponential precision to achieve accurate braids.

\subsection{Alternative encodings} \label{subsec:encodings}

A nice feature of the traditional Ising anyon theory is the ability to efficiently perform exact Clifford gates~\cite{Fan_2010}.  The encodings we consider above do not easily allow these results to be translated for efficient Clifford gate construction.  In this section, we present an alternative encoding.

Consider the Hilbert space $\H'_n := \Hom(\alpha{+}1, 1 ^{\otimes 2n+1} \otimes \alpha)$.   There is a natural choice of basis on this Hilbert space given by fusion trees where the Ising anyons first fuse via left-hanging fusion trees and the result fuses with the single $\alpha$ anyon, shown below for $n=1$    
\[
\hackcenter{\begin{tikzpicture}[scale=0.85]
  \draw[ultra thick, black] (.6,-.6) to (1.5,-1.5); 
  \draw[ultra thick, black] (1.2,-.6) to (.9,-.9);
  \draw[ultra thick, black] (1.8,-.6) to (1.2,-1.2);
  \draw[ultra thick, black] (2.4,-.6) to (1.5,-1.5) to (1.5,-2.1); 
  \node at (.6,-.35) {$1$};
  \node at (1.2,-.35) {$1$};
  \node at (1.8,-.35) {$1$};
  \node at (2.4,-.35) {$\alpha$}; 
  \node at (1.5,-2.35) {$\alpha{+}1$};  
  \node at (.8,-1.2) {$\scriptstyle v_1$};
  \node at (1.1,-1.5) {$\scriptstyle v_1'$};
\end{tikzpicture}}
\]
In this encoding, even our single-qubit space contains a noncomputational basis vector. The single-qubit space of the alternative encoding is 3-dimensional with the computational space spanned by $(v_1,v_1') = (0,1)$ and $(v_1,v_1') = (2,1)$, labelled by $\ket{0^A}$ and $\ket{1^A}$ respectively. The non-computational space is spanned by $(v_1,v_1') = (2,3)$ and is denoted $\ket{NC_1^A}$.   The computational space is positive-definite for all values of $\alpha$, and the whole space is positive-definite when $1<\alpha<2$. 

The affine braid group $\Br_{2n+1}^{\rm aff}$ acts on $\H'_n$ where the braid generators $\mathsf{b}_i$ for $i>1$ act on the computational space by the standard Ising braid matrices.   Hence, we can execute exact Clifford gates in this encoding following \cite{Fan_2010} up to an overall unimportant global phase
\begin{align}
  H &= \frac{1}{\sqrt{2}}  \left(
  \begin{array}{cc}
    1 & 1 \\
    1 & -1 \\
  \end{array}
\right) \simeq \mathbf{b}_1 \mathbf{b}_2 \mathbf{b}_1,
\\
  S &=    \left(
  \begin{array}{cc}
    1 & 0 \\
    0 & i \\
  \end{array}
\right) \simeq \mathbf{b}_1^{-1},  
\quad 
  X =    \left(
  \begin{array}{cc}
    0 & 1 \\
    1 & 0 \\
  \end{array}
\right) \simeq \mathbf{b}_2^{2} ,
\\ 
  Y &=    \left(
  \begin{array}{cc}
    0 & -i \\
    i &  0 \\
  \end{array}
\right) \simeq \mathbf{b}_1^2\mathbf{b}_2^{-2} ,
\quad
 Z =    \left(
  \begin{array}{cc}
    1& 0 \\
   0 &  -1 \\
  \end{array}
\right) \simeq \mathbf{b}_1^2 .
\end{align}
Here we show that this encoding can be enriched by a non-Clifford phase gate of arbitrarily low leakage, making this encoding single-qubit universal. Furthermore, there are values of $\alpha$ for which the leakage vanishes entirely, the model is capable of performing exact single qubit Clifford gates plus one non-Clifford gate.

The scheme for generating the diagonal gate is the same as in Section~\ref{sec:entangling-gates}: we start with a leaking gate $U_0^A$ and an appropriate phase gate. We take $U_0^A= \b_3$ and $D = J_3$, restricted to $\mathrm{span}\{(2,1),(2,3)\}$. On this subspace, $D$ acts as $\sim\left(\begin{matrix}
    1 & 0\\
    0 & i
\end{matrix}\right)$, so it does not fit the form of Lemma 4.2 of~\cite{Cui-leak}, nor its generalizations~\cite{Reichdist}. Nevertheless, an analysis analogous to that of Section~\ref{sec:decoupling_from_indefinite_space} finds values of alpha for which the leakage is eliminated:
\begin{equation*}
    \begin{split}
        \alpha_1 &= \pm\left(\frac{1}{2}+\frac{2 \tan
   ^{-1}\left(\frac{5}{\sqrt{7}}\right)}{\pi }\right) \mod 8\\
   & \sim \pm 1.1902 \mod 8.
    \end{split}
\end{equation*}
At this value of $\alpha$, the sequence in Lemma~\ref{lemma:Cui} yields a non-leaking gate $U_1^A$ that acts as  $\sim \left(\begin{matrix}
    e^{i\bar{\theta}} & 0\\
    0 & e^{-i\bar{\theta}}\\
\end{matrix}\right)$ with 
\begin{equation*}
\begin{split}
\bar{\theta} & = -\frac{\pi }{4}+\tan ^{-1}\left(\frac{5}{\sqrt{7}}\right)-\frac{1}{2} \tan ^{-1}\left(\frac{1}{57} \left(64+11
   \sqrt{7}\right)\right)\\
   & \sim 0.212016
\end{split}
\end{equation*}
on the computational space. This gate is not Clifford, and we thus achieve single-qubit universality. A numerical analysis of the robustness of the phase gate analogous to the one presented in Section~\ref{sec:robustness} finds that the operator norm distance between the target gate and the braid restricted to the computational space can be made smaller than a threshold $\delta$ by requiring $|\epsilon|<\frac{\delta}{4.8}$.
\begin{figure}
    \centering
    \includegraphics[width=\linewidth]{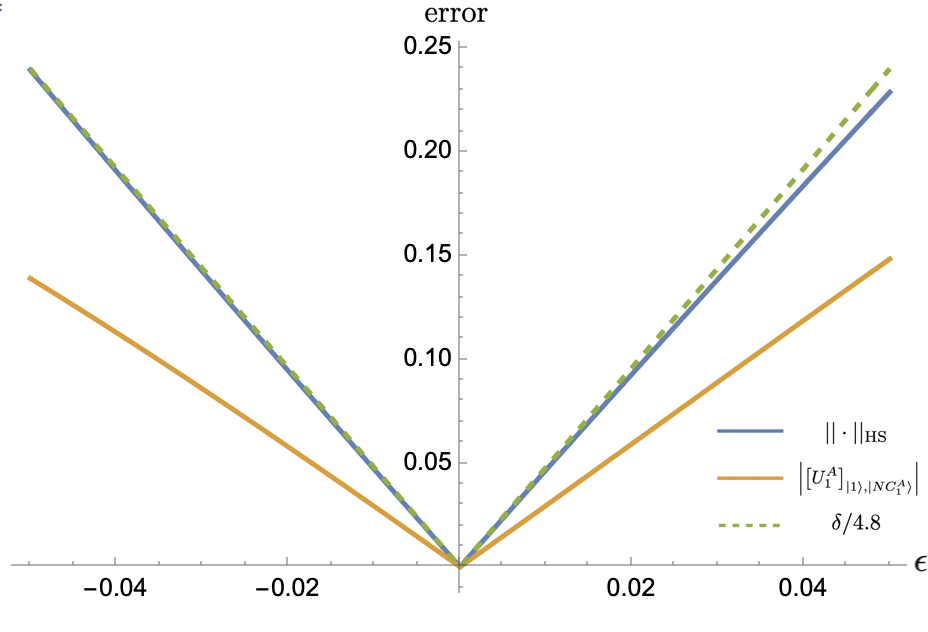}
    \caption{\justifying  Hilbert-Schmidt norm difference between the non-leaking target phase gate and the braid at $\alpha = \alpha_1+\epsilon$ (blue). The distance can be made smaller than a threshold $\delta$ if $\epsilon$ can be chosen so that $|\epsilon| <\frac{\delta}{4.8}$ (dashed line)} 
    \label{fig:robustness_alternative}
\end{figure}

It is unclear whether this model can be extended to multiple qubits. The fusion of four or more $2$-dimensional representations contains indecomposable subrepresentations such as the module $P_2$ discussed in Appendix \ref{sec:representations_of_Uqsl2}, making computations difficult and standard methods inapplicable.

\section*{Outlook and Conclusions}

This work demonstrates that non-semisimple extensions of the Ising anyon model provide a viable path to universal topological quantum computation using braiding alone. By introducing the notion of \emph{neglectons}---quantum trace zero representations that gain physical relevance through renormalization---we recover new anyon types that enrich the computational capabilities of the system. Our construction embeds the computational subspace entirely within the positive-definite sector of a Hilbert space equipped with an indefinite Hermitian form, allowing for low-leakage unitary evolution and, at special values of the deformation parameter $\alpha$, complete decoupling from the negative-norm sector.

From a computational standpoint, we provide explicit constructions of universal single-qubit and two-qubit gates, leveraging a generalized Reichardt-style compilation strategy adapted to the non-semisimple setting.  We have also demonstrated that these gates are robust to small perturbations in $\alpha$. These results show that the non-universality of standard Ising anyons is not a fundamental limitation, but a consequence of the semisimplification procedure. Once quantum trace zero representations are reinstated, universality is restored.

This perspective opens several new directions. On the theoretical side, the interplay between indefinite unitarity and quantum information merits deeper exploration, especially with respect to noise tolerance, error correction, and entanglement dynamics in non-semisimple models. On the experimental side, it raises the question of whether physical systems, possibly in the fractional quantum Hall regime or engineered lattice models, can realize neglectons and the associated fusion and braiding rules. The potential to engineer such systems could broaden the scope of platforms suitable for fault-tolerant quantum computing.

More broadly, our results suggest a re-evaluation of models traditionally deemed non-universal. When treated within the framework of non-semisimple TQFTs, they may yield unexpected computational power. This shift underscores the importance of developing a fuller understanding of non-semisimple modular structures, both mathematically and physically, as part of the evolving landscape of topological quantum computing.

\section*{Acknowledgments}
The PIs are grateful to Shawn Cui for helpful discussions.  A.D.L., F.I., and S.K. are partially supported by NSF
grants DMS-2200419 and the
Simons Foundation collaboration grant on New Structures in Low-dimensional topology. S.K. is supported by the NSF Graduate Research Fellowship DGE-1842487. J.S. is
partially supported by the Simons Foundation Collaboration Grants for Mathematicians, NSF grant DMS-1807161 and PSC CUNY Award
64012-00 52.

 \section*{Data Availability}
The data that support the findings of this article are openly available at \cite{data_availability}.

\appendix

\section{Introduction to quantum $\mathfrak{sl}_2$ }\label{sec:representations_of_Uqsl2}
To fully appreciate the way standard anyon theories arise from quantum group representations, 
it is useful to first study the representation theory of the quantum group at a generic $q$. 

Let $\Uq$ be the algebra generated by $E,F, K, K^{-1}$ satisfying 
\begin{equation}\label{eq:UqRelations}
\begin{aligned}
  KK^{-1}&= K^{-1}K=1, &  [E,F]&=\frac{K-K^{-1}}{q-q^{-1}},\\
  KEK^{-1}&=q^{-2} E, &KFK^{-1}&=q^{-2}F, & \\
\end{aligned}
\end{equation}
with coproducts, counits, and antipodes:
\begin{equation}\label{eq:UqCoproducts}
\begin{aligned}
  \Delta(E)&= 1\otimes E + E\otimes K,
  &\varepsilon(E)&= 0,
  &S(E)&=-EK^{-1},
  \\
  \Delta(F)&=K^{-1} \otimes F + F\otimes 1,
  &\varepsilon(F)&=0,& S(F)&=-KF,
    \\
  \Delta(K)&=K\otimes K,
  &\varepsilon(K)&=1,
  & S(K)&=K^{-1}.\\
\end{aligned}
\end{equation}

These relations match those in \eqref{E:RelDCUqsl} and \eqref{E:HopfAlgDCUqsl}, excluding the ones involving $H$ and the constraint $E^r = F^r =0$,   and with $q$ a generic complex number. 

Setting  $K = e^{-\hbar H/2}$ and $q=e^{-\hbar /2}$ and taking the limit as $\hbar \to 0$, the algebra $\Uq$ becomes the usual enveloping algebra $U(\slt)$ of the Lie algebra $\slt$.  The coproduct, counit, and antipode deform the standard structures on the usual enveloping algebra $U(\slt)$.   This is where the ``quantum" in quantum group comes from~\cite{Tu}.

We are interested in the finite-dimensional representations of $\Uq$ that admit a `weight decomposition' as $q$-eigenspaces for the operator $K$.  That is, 
\begin{equation}
    V = \bigoplus_{\lambda}  V^{\lambda}, \qquad Kv =q^{\lambda} v, \;\; v \in V^{\lambda}.
\end{equation}
The $\lambda$-eigenspaces $V^{\lambda}$ are called \emph{weight spaces} and the vectors therein are called \emph{weight vectors of weight $\lambda$}. We call $V$ a \emph{weight representation}, or \emph{weight module}, if $V$ splits as a direct sum of weight spaces.  Finally, we say that a weight vector $v$ of a $\Uq$-representation is a \emph{highest weight vector} if it is annihilated by $E$, i.e. $E v=0$.    Note that the commutation relations from \eqref{E:RelDCUqsl} imply that $E(V^{\lambda}) \subset V^{\lambda+2}$ and $F(V^{\lambda}) \subset V^{\lambda-2}$. 
 
At generic $q$ (i.e., not a root of unity), the algebra $\Uq$ is semisimple with an infinite number of irreducible representations (also known as simple modules) $S_n$ indexed by non-negative integers $n\in \Z_{\geq 0}$.  These  $(n+1)$-dimensional representations $S_n = \{s^n_0, s^n_1, \dots, s^n_n\}$ are generated by a highest weight vector $s_0^n$ of highest weight $n$ with action of $\Uq$ defined by
\begin{equation} \label{eq:Sn}
\Jz s_i = q^{(n-2i)} s_i, \quad \Jp s_i = [i][n-i] s_{i-1}, \quad \Jm  s_i = s_{i+1}, 
\end{equation}
where for $n\in \N$, we define
 $$\qn{x}=q^x-q^{-x},\quad \qn{n}!=\qn{n}\qn{n-1}\cdots\qn{1},$$ $$\qN{x}=\frac{\qn x}{\qn1}, \quad \qN{n}!=\qN{n}\qN{n-1}\cdots\qN{1}.$$  

The representations $S_n$ are the $q$-deformations of the traditional spin-$\lambda/2$ representations\footnote{Here we use integer eigenspaces, rather than half-integer, following the mathematical convention.} of $\mathfrak{sl}_2$. In particular, at generic $q$, these representations satisfy the usual Clebsch-Gordan fusion rule 
\begin{equation}\label{eq:CSfusion}
    S_m \otimes S_n = S_{m+n} \oplus S_{m+n-2} \oplus \dots \oplus S_{|m-n|}. 
\end{equation}
However, when $q$ is specialized to a root of unity, the category of $\Uq$ representations is no longer semisimple and this tensor product decomposition into simples is no longer valid.

Simple objects of modular tensor categories describe the allowed anyon types in models of topological quantum computation.  Decompositions of the form \eqref{eq:CSfusion} give the allowed fusion rules for how anyons indexed by given simples can be fused together into other allowed anyon types.   For simple objects $a$ and $b$ decomposing as $a\otimes b = \bigoplus_c N_{ab}^{c} c$, the fusion coefficients $N_{ab}^c$ specify the allowed types that can result from the fusion of $a$ and $b$.  Dually, a particle of type $c$ can split into a pair $a \otimes b$, whenever the coefficient $N_{ab}^c$ is nonzero.   

We can represent these fusion and splitting processes using a graphical calculus describing the worldlines of these particles during their fusion or splitting.  The decomposition of a tensor product into a direct sum $a\otimes b = \bigoplus_c N_{ab}^c c$ implies the existence of a map of representations from $c$ into the product of $a\otimes b$ 
\begin{equation}\label{eq:embedding}
    Y_c^{ab}: c \rightarrow a \otimes b  \quad\hackcenter{\begin{tikzpicture}[scale=0.8]
        \draw[very thick, black ] (0, 0.2) to (0, 0.7);
        \draw[very thick, black, <-] (-.5,1.3) to (0, 0.7);
        \draw[very thick, black, <-] (.5,1.3) to (0, 0.7);
        \end{tikzpicture}} 
        \quad
        \xy
            (0,7)*+{\scs  a \otimes b}="1";
            (0,-7)*+{\scs c}="2";
            {\ar_{Y^{ab}_c} "2";"1"};
        \endxy 
\end{equation}
describing the splitting of the anyon $c$.  If $c$ has multiplicity one inside $a\otimes b$, such a map is unique up to a scalar.

\begin{remark}
    The module $S_0$ is the monoidal identity: for any module $V$ we have that $S_0 \otimes V \cong V \cong V \otimes S_0$,  and can be interpreted as the vacuum particle. We are free to add and remove vacuum world lines from space-time diagrams.
\end{remark}

\begin{notation}\label{not:upper_and_lowercase}
We use capital letters (e.g.\ $V, W$) for representations of the quantum group, not necessarily simple, and lowercase letters (e.g.\ $a, b, c$) as placeholders for simple representations to emphasize their correspondence with anyon types.
\end{notation}

For any simple object $a$, Schur's lemma implies that $\End(a)=\C\Id_a$ so that for any $f \in \End(a)$ we have $f = \brk{f}_a \Id_a$ for some scalar $\brk f_a=\lambda\in\C$ (see\footnote{More generally, for a  module $V$ and a map of representations $f\in\End(V)$, we write
$\brk f_V=\lambda\in\C$ if $f-\lambda\Id_V$ is nilpotent.  This more general notion is relevant for extending the modified trace from Section~\ref{subsec:modtrace} to not necessarily simple modules like the projective $P_2$ introduced in Figure~\ref{fig:P2_weight_diagram}.}).

\subsection{Quantum dimensions}\label{subsec:quantum_dimensions}
 
To define the quantum dimension, we need to introduce duals.  Duals give antiparticles to a given anyon type.
Let $\{v_i\}$ be a basis of $V$ and $\{v_i^*\}$ be
a dual basis of $V^*=\Hom(V,\C)$, so that $\Uq$ acts on $V^{\ast}$ by $(x\cdot  v_i^{\ast})(w) := v_i^{\ast}(S(x)\cdot w)$ for $x \in \Uq$ and $w\in V$.   Then
\begin{align*}
  \coev_V \colon & \C \rightarrow V\otimes V^{*}   
 \text{ : } 1 \mapsto \sum
v_i\otimes v_i^*,  
\quad &  \hackcenter{
\begin{tikzpicture}[scale=0.8]
\draw[very thick, black, <-] (-.5,1.3) .. controls +(0,-.75) and +(0,-.75) .. (.5,1.3);
\end{tikzpicture}} 
\xy
  (0,7)*+{V \otimes V^{\ast}}="1";
 (0,-7)*+{\C}="2";
 {\ar_{e_{V}} "2";"1"};
 \endxy 
\\
  \ev_V \colon & V^*\otimes V\rightarrow \C  
  \text{ : }
  f\otimes w \mapsto f(w), \quad & 
 \hackcenter{
\begin{tikzpicture}[scale=0.8]
\draw[very thick, black, <-] (-.5,.2) .. controls +(0,.75) and +(0,.75) .. (.5,.2);
\end{tikzpicture}}  
\xy
  (0,7)*+{\C}="1";
 (0,-7)*+{V^{\ast} \otimes V}="2";
 {\ar_{e_{V}} "2";"1"};
 \endxy 
\\
  \coev_V' \colon & \C \rightarrow V^*\otimes V   
 \text{ : } 1 \mapsto
  \sum v_i^* \otimes K^{r-1}v_i,  
 &  \hackcenter{
\begin{tikzpicture}[scale=0.8]
\draw[very thick, black, ->] (-.5,1.3) .. controls +(0,-.75) and +(0,-.75) .. (.5,1.3);
\end{tikzpicture}} 
\xy
  (0,7)*+{V^{\ast} \otimes V}="1";
 (0,-7)*+{\C}="2";
 {\ar_{e_{V}} "2";"1"};
 \endxy 
\\
  \ev_V' \colon & V\otimes V^*\rightarrow \C  
  \text{ : }
 v\otimes f \mapsto f(K^{1-r}v),  & 
 \hackcenter{
\begin{tikzpicture}[scale=0.8]
\draw[very thick, black, ->] (-.5,.2) .. controls +(0,.75) and +(0,.75) .. (.5,.2);
\end{tikzpicture}}  
\xy
  (0,7)*+{\C}="1";
 (0,-7)*+{V \otimes V^{\ast}}="2";
 {\ar_{e_{V}} "2";"1"};
 \endxy 
\end{align*}
are duality morphisms, and we have included their standard string diagram depictions. These maps can be interpreted as $i_V$ describing the production of a particle-antiparticle pair of type $V$, while $e_V$ describes the annihilation of an antiparticle with its dual particle type. 

The \emph{quantum dimension} $\qdim(V)$ of an object $V$ in $\cat$ is then defined by  

\[
\qdim(V) := \;
  \hackcenter{
\begin{tikzpicture}[scale=0.8]
\draw[very thick, black, <-] (-.5,0) .. controls +(0,-.75) and +(0,-.75) .. (.5,0);
\draw[very thick, black, ] (-.5,0) .. controls +(0,.75) and +(0,.75) .. (.5,0);
\node at (-.85,0) {\small $V$};
\end{tikzpicture}} 
  \;\; = \;\; \brk{\tev_V\circ \coev_V}_\C=\sum  v_i^*(K^{1-r}v_i) \ .
\]

\begin{figure}
    \centering
    \begin{tikzpicture}[scale=0.9]
    \draw[stealth-] (.6,0.3) --  (0.6+.5 ,0.3+.5);
    \draw[-stealth] (.6,-0.3) -- (.6+.5, -0.3-.5);
    \draw[-stealth] (5.9, 0.7) -- (5.9 +.5, 0.7-.5);
    \draw[stealth-] (5.9, -0.7) -- (5.9+.5, -0.7+.5);
    \draw[stealth-stealth] (2, 1) -- (3, 1);
    \draw[stealth-stealth] (4, 1) -- (5, 1);
    \draw[-stealth] (1.75,0.75) -- (3.25, -0.75);
    \draw[-stealth] (3.75,0.75) -- (5.25, -0.75);
    \draw[stealth-stealth] (2, -1) -- (3, -1);
    \draw[stealth-stealth] (4, -1) -- (5, -1);
    \node at (0.2, 0) {\large $w^L_{-4}$};
    \node at (6.8, 0) {\large $w^R_4$};
    \node at (1.5, 1) {\large$w^H_{-2}$};
    \node at (3.5, 1) {\large$w^H_{0}$};
    \node at (5.5, 1) {\large$w^H_{2}$};
    \node at (1.5, -1) {\large$w^S_{-2}$};
    \node at (3.5, -1) {\large$w^S_{0}$};
    \node at (5.5, -1) {\large$w^S_{2}$};
    \draw[stealth-] (7,-0.75) -- (8, -0.75);
    \draw[-stealth] (7,0.75) -- (8, 0.75);
    \node at (7.5,1) {$E$};
    \node at (7.5,-0.5) {$F$};
    \end{tikzpicture}
    \caption{\justifying The weight structure of the module $P_2$ when $r=4$. The subscripts under each vector denote the weights. The module $P_2$ is the projective cover of $S_2$, having it as a submodule.  The quotient $P_2/S_2$ has $\C^H_{4}$ and $\C^H_{-4}$ as submodules.  Quotienting again by these leaves an additional $S_2$.}  \label{fig:P2_weight_diagram}
\end{figure}

\subsection{Semisimplification }
There are many different forms of the quantum group for $\mf{sl}_2$ depending on the choices made for the quantum parameter $q$ and the amount of the center that is killed.   These various forms have vastly different representation theories,  which may or may not be semisimple, have a finite or infinite number of simple objects, and may or may not be braided.  See \cite{CGP2} and the references therein for more discussion on the various forms of the quantum group.  

One form of interest is the semi-restricted quantum group $\UqMed$ obtained from $\Uq$ at an (even) root of unity $q=e^{\frac{2\pi i}{2 r}}$ modulo the relations
\begin{equation}\label{eq:Ubar}
    E^r =F^r =0.
\end{equation}
Its category of representations $\UqMed\text{-mod}$ is not semisimple, nor braided, and has an infinite number of non-isomorphic simple representations.  
Nevertheless, one can obtain a semisimple modular category (corresponding to a $SU(2)_k$ Chern-Simons TQFT at level $k=r-2$ ) from $\UqMed-\text{mod}$ through a process called \emph{semsimplification}. 
We will furthermore see in the next section  that $\UqMed$ admits a modest modification $\Ubar$ that results in the braided category of highest weight that governs the non-semisimple Ising anyons that are the main subject of this work.

The representation theory of $\UqMed$ removes certain cyclic modules that are not highest weight representations that exist in the representation theory of $\Uq$,  as the new relations \eqref{eq:Ubar} force representations to be highest weight \cite{CGP2}.   However, $\UqMed$  admits an infinite class of irreducible $r$-dimensional representations. Let $\alpha \in (\C \backslash \Z) \cup r\Z$ and define the $r$-dimensional module $V_\alpha = \{v_0^\alpha, v_1^\alpha, ... ,v^\alpha_{r-1}  \}$, generated by a vector $v_0^{\alpha}$ of highest weight $\alpha+r-1$ with action 
 \begin{equation} \label{eq:Va}
 \Jz v^\alpha_i = q^{(\alpha +r -1 -2i)} v^\alpha_i, \; 
 \Jp v^\alpha_i = [i][i-\alpha] v^\alpha_{i-1}, \; 
 \Jm v^\alpha_{i} = v^\alpha_{i+1}.
\end{equation}
Since $q^{2r}=1$, many of these $V_{\alpha}$ are isomorphic $V_{\alpha} \cong V_{\alpha{+}2r}$ by the map sending $v_0^{\alpha} \to v_{0}^{\alpha{+}2r}$.  
For all $\alpha\in \C$, the quantum dimension of $V_\alpha$ is zero:
\begin{align*}
\qdim(V_\alpha)&= \sum_{i=0}^{r-1} v_i^*(K^{1-r}v_i)=
 \sum_{i=0}^{r-1} q^{(r-1)(\alpha + r-1-2i)} 
 \\&=
 q^{(r-1)(\alpha + r-1)}\frac{1-q^{2r}}{1-q^{2}}=0.
 \end{align*}
 
Representations with quantum dimension zero and morphisms that factor through them are sometimes called \emph{negligible}, as the vanishing of the quantum dimension can be thought of as them having vanishing amplitude and are therefore nonphysical.  To arrive at a finite semisimple tensor category from the representations of $\UqMed$, one must quotient the category $\UqMed$-representations by negligible morphisms. This has the effect of killing all the representations $V_\alpha$ as well as the representations $S_n$ for $n\geq r-1$, thus truncating the fusion rule from~\eqref{eq:CSfusion}. 
\begin{example}[Ising]
     When $r=4$ the only surviving modules are $S_0, S_1$, and $S_2$. The fusion rules then become $S_1 \otimes S_1 = S_0\oplus S_2$, $S_1 \otimes S_2 = S_1$, and $S_2 \otimes S_2 = S_0$, with $S_0$ as the monoidal identity. We see that these fusion rules are indeed the same as the ones for Ising anyons presented in Table \ref{tab:Ising-fusion}.
\end{example}

\begin{table}
\begin{tabular}{|c|c|c|c|}
  \hline
    & $\;\; \Uq \;\; $ & $ \;\; \UqMed \;\; $ & \;\; Ising Category \;\; \\ \hline 
  $S_1\otimes S_1$    & $S_0 \oplus  S_2$ & $S_0 \oplus  S_2$ & $S_0 \oplus S_2$\\ \hline 
   $S_1\otimes S_2 $  & $S_1 \oplus  S_3$ & $S_1 \oplus S_3$ & $S_1$\\ \hline 
   $ S_2\otimes S_2$    & $S_0\oplus  S_2\oplus S_4$ & $S_0 \oplus P_2$ & $S_0 $\\ \hline 
\end{tabular}
    \caption{\justifying A table illustrating the various fusion processes in various forms of the quantum group at $r=4$. The first column is the familiar Clebsch-Gordon fusion rules in the context of quantum $\slt$. The second column illustrates the appearance of non-semisimple representations, where $S_3$ now has quantum dimension zero, and the fusion of $S_2 \otimes S_2$ produces the non-semisimple module $P_2$. The third column represents the Ising fusion rules after semisimplification. }
    \label{tab:Ising-fusion}
\end{table}

\section{The unrolled quantum group} \label{sec:unrolled_Uqsl2}

While the semisimple modular tensor categories resulting from the semisimplification process are braided ribbon categories, the category of $\UqMed$ representations is not.   To leverage the non-semisimple representation theory, we remedy this by introducing a modest modification $\Ubar$ of $\UqMed$.  This algebra has an additional generator $H$ that can be thought of as the logarithm of $K$, that can be used to define braiding and twist operations.  This new generator 
breaks the isomorphism $V_{\alpha} \cong V_{\alpha +2r}$ and introduces an infinite cyclic group of invertible one dimensional representations that play a key role in applications to low-dimensional topology.   

Let $\Ubar$  be the algebra $\UqMed$ with an additional generator $H$ satisfying
\begin{align*}
  HK&=KH,
& [H,E]&=2E, & [H,F]&=-2F.
\end{align*}
The algebra $\Ubar$ is a Hopf algebra where the coproduct, counit and
antipode are defined in
\eqref{E:HopfAlgDCUqsl} and by
\begin{align*}
  \Delta(H)&=H\otimes 1 + 1 \otimes H,
  & \varepsilon(H)&=0,
  &S(H)&=-H.
\end{align*} 
This version of the quantum group is called the \textit{unrolled quantum group}.

Let $V$ be a finite-dimensional $\Ubar$ representation. Then the weight decomposition into $q$-eigenspaces for $K$ corresponds to standard eigenspaces for the operator $H:V\to V$, and we can define weight representations as before.    We call $V$ a \emph{weight representation}, or \textit{weight module},  if $V$ splits as a direct sum
of weight spaces $V = \bigoplus_{\lambda}V^{\lambda}$ and $q^H=K$ as operators on $V$; in this case, $Kv=q^{\lambda} v$ for any vector $v$ of weight $\lambda$.  Let $\cat$ be the category
of finite-dimensional weight $\Ubar$-representations. 

If $V$ and $W$ are two (not necessarily simple) $\Ubar$-representations, we define the \emph{$\Hom$-space} $\Hom(V,W)$ as the set of equivariant maps from $V$ to $W$. Formally, if $\rho_V$ and $\rho_W$ are the representations of $\Ubar$ on $V$ and $W$ respectively, we define: 
\begin{equation*}
    \Hom(V,W) = \{ f: V\rightarrow W ~|~   f(\rho_V(u)v) = \rho_W(u)f(v)  \}.
\end{equation*}
where $f$ is a linear map and for all   $u\in \Ubar$, $v\in V$.  Since $\Ubar$ is a Hopf algebra, $\cat$ is a tensor category where
the unit $\unit$ is the 1-dimensional trivial representation $\C$.  Moreover,
$\cat$ is $\C$-linear: $\Hom$-sets are $\C$-vector spaces, the composition and
tensor product of morphisms are $\C$-bilinear, and
$\End(\unit)=\C\Id_\unit$.  When it is clear, we denote the unit
$\unit$ by $\C$.

 \subsection{Representations} \label{subsec:reps}
 The irreducible representation of $\Ubar$ where $K$ and $H$ act diagonally and the eigenvalues of $K$ and $H$ are related by $K=q^H$ has been completely classified.  
 At $q$ an even root of unity,  we see immediately that the constraint $\Jp^r=\Jm^r=0$ forces irreducible representations to have dimension less than or equal to $r$.

The irreducible representations fall into the following classes:
\begin{itemize}
    \item For $n \in \{0, 1, \dots, r-1\}$, the $(n+1)$-dimensional module $S_n = \{s^n_0, s^n_1, \dots, s^n_n\}$ is defined by the action in~\eqref{eq:Sn}, with $H$ acting as $H s^n_i = (n - 2i) s^n_i$. The module $S_0$ serves as the monoidal identity.

    \item The infinite family of $r$-dimensional simple projective modules $V_\alpha = \{v^\alpha_0, v^\alpha_1, \dots, v^\alpha_{r-1}\}$, defined for $\alpha \in (\mathbb{C} \setminus \mathbb{Z}) \cup r\mathbb{Z}$, are all simple and projective. Their structure is given by the action in~\eqref{eq:Va}, with $H$ acting as
   $
    H v^\alpha_i = (\alpha + r - 1 - 2i)\, v^\alpha_i$.

    \item There is also a family of one-dimensional simple modules, denoted $\mathbb{C}^H_{kr} = \{v^{kr}_0\}$ for $k \in \mathbb{Z}$. The action of $\Ubar$ on $\mathbb{C}^H_{kr}$ is given by
   $
    H v^{kr}_0 = kr\, v^{kr}_0$ and $\Jp v^{kr}_0 = 0= \Jm v^{kr}_0$.
    Tensoring $S_n$ with $\mathbb{C}^H_{kr}$ yields the simple module $S_n \otimes \mathbb{C}^H_{kr}$, which has the same structure as $S_n$ but with all weights shifted by $kr$. This generates an infinite cyclic family of shifted simple modules.
\end{itemize}

\begin{proposition}
\label{prop:classification}
    All irreducible $\Ubar$ representations are of one of the following two forms \cite{CGP2}. 
    \begin{itemize}
        \item $S_n\otimes\C^H_{k r}$ for $n\in \{0,1,...,r-2 \}$, $k \in \Z,$
        \item $V_\alpha$ for $\alpha \in (\C \backslash \Z) \cup r\Z$. 
    \end{itemize}
\end{proposition}
 The module $S_{r-1}$ fits into the above classification as it is isomorphic to $V_0$. 

\begin{notation}
    We will often denote the modules $S_n$ simply by the integer $n$ and the modules $V_\alpha$ by its index $\alpha$.
\end{notation}

The duality, braiding, and ribbon structures introduced above extend to these $\Ubar$-representations equipping the category $\cat$ with the structure of a braided ribbon category.

\begin{example}[$r=4$ Ising theory] \label{exampe:Ianyons}
Suppose $q$ is a primitive $8^{th}$ root of unity corresponding to $r=4$.  Simple objects $S_0=\unit$, $S_1$, and $S_2$ exist as usual with $\qdim(S_n)=[n+1]$. Although the module $S_3$ remains simple, it now has quantum dimension zero since $[4]=0$ at this root of unity.

 The $(n+1)$-dimensional modules $S_n$ for $n>3$ no longer exist because the quantum group relations $E^4 = F^4=0$ prevent the satisfaction of the $S_n$ relations provided in~\eqref{eq:Sn}. Instead of generating a 5-dimensional module like $S_4$, acting on a highest weight vector $v$ of highest weight 4 produces a 4-dimensional module spanned by $\{ v, Fv, F^2v, F^3v \}$  since $F^4=0$.  A weight diagram of this $4$-dimensional module is shown below.
\[
\begin{tikzpicture}[scale=.9, every node/.style={circle,draw,fill=white,inner sep=1pt}]
  \foreach \x in {-2,0,2,4}
    \node[label=above:{\x}] at (\x,0) {};

  \draw[-latex] (-3.5,0) -- (5.5,0);

  \foreach \x in {-3,-2,...,4}
    \draw[shift={(\x,0)}, thick] (0pt,3pt) -- (0pt,-3pt);
\end{tikzpicture}
\]
This module is isomorphic to the module $V_1$ defined above with highest weight $1+(r-1)=4$.  Since $1 \notin  (\C \backslash \Z) \cup r\Z$ from Proposition~\ref{prop:classification}, $V_1$ is not simple.  In fact, it is not even semisimple.  It does contain a submodule isomorphic to $S_2$.  This can be seen by observing that the vector $Fv$ is a highest weight vector since $E(Fv) = FEv + [4]v = 0$, where the last equality follows since $v$ is a highest weight vector and $[4]=0$ at an $8^{th}$ root of unity.  Taking a quotient $V_1/S_2$ leaves a one-dimensional simple $\C_{1r}^H$ of highest weight 4.

There are non-semisimple modules that can be endowed with a Hermitian form: these are the projective covers of the simple modules $S_i$ which are denoted by $P_i $ for $i\in \{0,1,...,r-2\}$. In the context of this work, where we take $r=4$, these modules appear when fusing four or more $S_1$ modules. In practice, we are able to do all the relevant computations in a basis where the $P_i$ representations are not explicitly involved, so we shall not cover them in detail here. See~\cite{CGP2,GLPMS} for a detailed description of these modules and their Hermitian structures and Fig.~\ref{fig:P2_weight_diagram} for a schematic of the weight structure of $P_2$ when $r=4$.
The following list of modules forms a Hermitian subcategory (closed under direct sums and tensor products): 
$S_i, V_\alpha , \C_{kr},$ and $P_i$~\cite{GLPMS}.

\end{example}

\subsection{Braiding and twists}\label{sec:hermitian_structures}
Anyons can be physically braided around each other. In the mathematical description of this process, we make use of the $R$-matrix defined by Ohtsuki~\cite{Oh}. The $R$-matrix is an operator $R=R_{VW}$ on $V\otimes W$ defined by
\begin{equation}
  \label{eq:R}
  R=q^{H\otimes H/2} \sum_{n=0}^{r-1} \frac{\{1\}^{2n}}{\{n\}!}q^{n(n-1)/2}
  E^n\otimes F^n 
\end{equation}
where $q^{H\otimes H/2}$ is the operator given by
$$q^{H\otimes H/2}(v\otimes v') =q^{\lambda \lambda'/2}v\otimes v'$$
for weight vectors $v$ and $v'$ of weight $\lambda$ and
$\lambda'$ respectively. The $R$-matrix gives rise to a braiding 
\begin{equation} \label{eq:braiding}
\hackcenter{
\begin{tikzpicture} [ scale=.55, rotate=-90]
\draw [very thick, directed=1, black] (2,0) .. controls ++(-.5,0) and ++(.5,0) .. (0,1);
\draw [very thick, black] (2,1) [out=180, in=30] to (1.2, .6);
\draw [very thick, directed=1, black] (.8,.35) [out=210,in=0] to (0,0);
\node at (2.3,1) {$\scs W$};
\node at (2.3,0) {$\scs V$};
\node at (-.3,1) {$\scs V$};
\node at (-.3,0) {$\scs W$};
\end{tikzpicture}} \qquad
\xy
  (0,7)*+{\scs  W \otimes V}="1";
 (0,-7)*+{\scs V \otimes W}="2";
 {\ar_{c_{VW}} "2";"1"};
 \endxy 
\end{equation}
defined by $v\otimes w \mapsto
c_{VW}(v\otimes w) :=\tau(R(v\otimes w))$ where $\tau=\tau_{VW}$ is the permutation $x\otimes
y\mapsto y\otimes x$.
Notice that $R_{VW}\rightarrow \text{Id}_{V\otimes W}$ as $q\rightarrow 1$, recovering the usual symmetry operator $\tau$ that acts as an intertwiner of representations for traditional $\slt$. Figure~\ref{fig:tau} illustrates that importance of the $R$-matrix in detecting topology.  In particular, the maps $c_{VW}$ collectively satisfy the Yang-Baxter equation.
\[
\hackcenter{
\begin{tikzpicture}[scale=.5, xscale=-1, yscale=-1]
\draw[very thick, black,<-] (.75,0) .. controls ++(0,.8) and ++(0,-.8) ..   (-.75,2) to (-.75,3);
\draw[very thick, black,<-] (2.25,0) to (2.25,1) .. controls ++(0,.8) and ++(0,-.8) ..  (.75,3);
\path [fill=white] (-.2,.7) rectangle (.2,1.25);
\path [fill=white] (1.3,1.74) rectangle (1.7,2.3);
\draw[very thick, black,<-] (-.75,0) .. controls ++(0,1) and ++(0,-1) ..   (2.25,3) to (2.25,4.5);
\draw[very thick, black] (.75,3) .. controls ++(0,.8) and ++(0,-0.8) ..   (-.75,4.5);
\path [fill=white] (-.2,3.6) rectangle (.2,4);
\draw[very thick, black] (-.75,3) .. controls ++(0,.8) and ++(0,-0.8) ..   (0.75,4.5);
\node at (-.75,4.8) {$V_3$};
\node at (.75,4.8) {$V_2$};
\node at (2.25 ,4.8) {$V_1$};
\node at (-.75,-.3) {$V_1$};
\node at (.75,-.3) {$V_2$};
\node at (2.25 ,-.3) {$V_3$};
\end{tikzpicture} }
 \quad
\vcenter{\xy
(0,-9)*+{\scs  V_1\otimes V_2 \otimes V_3}="3";
(0,0)*+{\scs  V_1\otimes V_3 \otimes V_2}="2";
(0,10)*+{\scs  V_3\otimes V_2\otimes V_1}="1";
 (0,20)*+{\scs V_3\otimes V_2\otimes V_1}="0";
 {\ar@{->}^{ \1 \otimes c_{23}  } "3";"2"};
{\ar@{->}^{ c_{13}\otimes \1    } "2";"1"};
{\ar@{->}^{ \1 \otimes  c_{21}  } "1";"0"};
 \endxy}
  =  
 \vcenter{\xy
(0,-9)*+{\scs  V_1\otimes V_2 \otimes V_3}="3";
 (0,0)*+{\scs  V_2\otimes V_1 \otimes V_3}="2";
(0,10)*+{\scs  V_2\otimes V_3 \otimes V_1}="1";
(0,20)*+{\scs V_3\otimes V_2 \otimes V_1}="0";
{\ar@{->}_{  c_{12} \otimes \1  } "3";"2"};
{\ar@{->}_{ \1\otimes c_{13}   } "2";"1"};
{\ar@{->}_{  c_{23}\otimes \1  } "1";"0"};
 \endxy}
  \;\;
\hackcenter{
\begin{tikzpicture}[scale=.5]
\draw[very thick, black] (.75,0) .. controls ++(0,.8) and ++(0,-.8) ..   (-.75,2) to (-.75,3);
\draw[very thick, black] (2.25,0) to (2.25,1) .. controls ++(0,.8) and ++(0,-.8) ..  (.75,3);
\path [fill=white] (-.2,.7) rectangle (.2,1.25);
\path [fill=white] (1.3,1.74) rectangle (1.7,2.3);
\draw[very thick, black,->] (-.75,0) .. controls ++(0,1) and ++(0,-1) ..   (2.25,3) to (2.25,4.5);
\draw[very thick, black,->] (.75,3) .. controls ++(0,.8) and ++(0,-0.8) ..   (-.75,4.5);
\path [fill=white] (-.2,3.6) rectangle (.2,4);
\draw[very thick, black,->] (-.75,3) .. controls ++(0,.8) and ++(0,-0.8) ..   (0.75,4.5);
\node at (-.75,-.3) {$V_1$};
\node at (.75,-.3) {$V_2$};
\node at (2.25 ,-.3) {$V_3$};
\node at (-.75,4.8) {$V_3$};
\node at (.75,4.8) {$V_2$};
\node at (2.25 ,4.8) {$V_1$};
\end{tikzpicture} }
\]

\begin{figure}
$
\hackcenter{
\begin{tikzpicture} [ scale=.55, rotate=-90]
\draw [very thick, directed=1, black] (2,0) .. controls ++(-.5,0) and ++(.5,0) .. (0,1);
\draw [very thick, black] (2,1) [out=180, in=30] to (1.2, .6);
\draw [very thick, directed=1, black] (.8,.35) [out=210,in=0] to (0,0);
\draw [very thick, directed=1, black] (4,0) .. controls ++(-.5,0) and ++(.5,0) .. (2,1);
\draw [very thick, black] (4,1) [out=180, in=30] to (3.2, .6);
\draw [very thick, directed=1, black] (2.8,.35) [out=210,in=0] to (2,0);
\node at (4.3,1) {$\scs W$};
\node at (4.3,0) {$\scs V$};
\node at (-.3,1) {$\scs W$};
\node at (-.3,0) {$\scs V$};
\end{tikzpicture}} \;  := \; 
\vcenter{\xy
      (0,20)*+{\scs V \otimes W}="0";
  (0,7)*+{\scs W \otimes V}="1";
 (0,-6)*+{\scs V \otimes W}="2";
{\ar_{\tau_{VW}} "2";"1"};
 {\ar_{\tau_{WV}} "1";"0"};
 \endxy}
 \quad
 \vcenter{\xy
  (0,20)*+{\scs  v \otimes w}="0";
  (0,7)*+{\scs  w \otimes v}="1";
 (0,-6)*+{\scs v \otimes w}="2";
{\ar@{|->} "2";"1"};
{\ar@{|->} "1";"0"};
 \endxy}
 \;  = \; 
\vcenter{\xy
  (0,20)*+{\scs V \otimes W}="0";
 (0,-6)*+{\scs V \otimes W}="2";
 {\ar_{Id_{V \otimes W}} "2";"0"};
 \endxy}
  = \; 
\hackcenter{
\begin{tikzpicture} [ scale=.55, rotate=-90]
\draw [very thick, directed=1, black] (4,0) -- (0,0);
\draw [very thick, directed=1, black] (4,1) -- (0,1);
\node at (4.3,1) {$\scs W$};
\node at (4.3,0) {$\scs V$};
\node at (-.3,1) {$\scs W$};
\node at (-.3,0) {$\scs V$};
\end{tikzpicture}}
$
\caption{\justifying In $U(\slt)$the flip map $\tau_{VW} \maps V \otimes W \to W \otimes V$ given by $v\otimes w \mapsto w \otimes v$ preserves the action of $U(\slt)$.  This makes $U(\slt)$ less useful for studying knots, as the identity depicted above shows that $U(\slt)$ cannot detect knottedness.  When we pass to the quantum group $\Uq$, the twist map is no longer a map of representations.   Deforming the twist map by including the $R$-matrix from \eqref{eq:R} makes $c_{VW} \maps V\otimes W \to W\otimes V$ a map of representations that will not, in general, satisfy the above identity, i.e. $c_{WV} \neq c_{VW}^{-1}$.} 
    \label{fig:tau}
\end{figure}
 
Finally, let $\theta$ be the operator given by
\begin{equation}  \label{eq:halftwist}
\theta=K^{r-1}\sum_{n=0}^{\ro-1}
\frac{\{1\}^{2n}}{\{n\}!}q^{n(n-1)/2} S(F^n)\qr^{-H^2/2}E^n
\end{equation}
where $q^{-H^2/2}$ is an operator defined on a weight vector $v_\lambda$ by
$q^{-H^2/2}\cdot v_\lambda = q^{-\lambda^2/2}v_\lambda.$
Using the $\theta$ operator, one can define \textit{twist maps} $\theta_V:V\rightarrow V$ by $v\mapsto \theta^{-1}v$ (see \cite{jM,Oh}). 
\[
\xy
  (0,7)*+{V}="1";
 (0,-7)*+{V}="2";
 {\ar_{\theta_{V}} "2";"1"};
 \endxy 
\hackcenter{
\begin{tikzpicture}[scale=.45]
\draw[very thick,->, black] (.75,0) .. controls ++(-.2,.8) and ++(0,-.8) ..   (-.75,2) to (-.75,3);
\path [fill=white] (-.2,.7) rectangle (.2,1.25);
\draw[very thick, black](-.75,-1) to  (-.75,0) 
    .. controls ++(0,.8) and ++(-.2,-.8) ..   (.75,2) 
    .. controls +(.3,.8) and +(0,.8) .. (2.25,2) to (2.25,0)
    .. controls +(0,-.8) and +(.2,-.8) .. (.75,0);
\end{tikzpicture} } 
\]

With these structures, we can define give alternative and equivalent definitions to the right duality morphisms
\begin{equation}\label{E:d'b'}
\begin{split}
   \hackcenter{
\begin{tikzpicture}[scale=0.8]
\draw[very thick, black, ->] (-.5,.2) .. controls +(0,.75) and +(0,.75) .. (.5,.2);
\end{tikzpicture}} 
\xy
  (0,7)*+{\C}="1";
 (0,-7)*+{V \otimes V^{\ast}}="2";
 {\ar_{\tev_V} "2";"1"};
 \endxy 
&=\ev_{V}c_{VV^*}(\theta_V\otimes\Id_{V^*})
\\
  \hackcenter{
\begin{tikzpicture}[scale=0.8]
\draw[very thick, black, ->] (-.5,1.3) .. controls +(0,-.75) and +(0,-.75) .. (.5,1.3);
\end{tikzpicture}} 
\xy
  (0,7)*+{V^{\ast} \otimes V}="1";
 (0,-7)*+{\C}="2";
 {\ar_{\tcoev_V} "2";"1"};
 \endxy 
&=(\Id_{V^*}\otimes\theta_V)c_{VV^*}\coev_V
\end{split}
\end{equation}
which are compatible with the left duality morphisms $\{\coev_V\}_V$ and
$\{\ev_V\}_V$.

The definition of quantum dimension $\qdim(V)$ of an object $V$ in $\cat$ is unchanged:  
\[
\qdim(V) := \;
  \hackcenter{
\begin{tikzpicture}[scale=0.8]
\draw[very thick, black, <-] (-.5,0) .. controls +(0,-.75) and +(0,-.75) .. (.5,0);
\draw[very thick, black, ] (-.5,0) .. controls +(0,.75) and +(0,.75) .. (.5,0);
\node at (-.85,0) {\small $V$};
\end{tikzpicture}} 
  \;\; = \;\; \brk{\tev_V\circ \coev_V}_\C=\sum  v_i^*(K^{1-r}v_i) \ .
\]
One can compute that  
$
\qdim(S_n) = [n+1]$. 
 Again, in the limit as $q\to 1$, this quantum dimension reduces to $(n+1)$, the dimension of the spin-$n/2$ representation.  However, notice that for $q=e^\frac{2\pi\sqrt{-1}}{2 r}$, we have
\[
\qdim(S_{r-1}) = [r] = \frac{q^{r} - q^{-r}}{q-q^{-1}}  = q^{-r}\left(\frac{ q^{2r}-1}{q-q^{-1}}  \right) = 0, 
\]
since $q$ is a $2r^{th}$ root of unity.  Furthermore, at this root of unity, the representations $S_n$ for $n>r-1$ are no longer simple modules.    

\subsection{Hermitian structures}\label{sec:hermitian_structures}

In topological quantum computing,  a collection of anyons with a given topological charge is described by $\Hom$-spaces in a modular tensor category $\cat$.  While the linear structure of the $\Hom$-spaces of the category of non-semisimple Ising anyons is clear, it is not immediately obvious how to define a positive definite inner product on such spaces. A naive guess would be to simply pick a set of basis vectors, such as a set of fusion trees, and to require them to be orthonormal. Such an approach does not work, as braiding and fusion in such bases will not be unitary.    The reason for this failure is that this inner product is incompatible with the tensor product structure.  

The unitarity of traditional semisimple anyon theories, such as $SU(2)_k$ theories, is already quite subtle from the representation-theoretic perspective and leverages intricate results for the representation theory of the quantum groups~\cite{Kauffman_2006,Fan_2010}, see also~\cite[Section XII A.9]{Tu} and \cite{Kir96,Wen98} for general quantum groups.    
The category of representations must have an adjoint map $\dagger$ associating to any map $f\maps V \to W$ an adjoint map $f^{\dagger} \maps W \to V$ satisfying the following conditions
 \begin{equation}
(f^{\dag})^{\dag}=f, \quad (f \otimes g)^{\dag} = f^{\dag} \otimes g^{\dag}, \quad  (f\circ g)^{\dag} = g^{\dag} \circ f^{\dag}. 
 \end{equation}
These relations imply $ \Id_V^{\dagger} = \Id_V$.  In other words, $\dagger$ is an object preserving contravariant involution on $\cat$.
    For unitarity of braiding and particle-antiparticle pair creation, we also require this dagger map to satisfy $c_{V\otimes W}^{\dagger} = (c_{VW})^{-1}$,  $\theta_V^{\dagger}= (\theta_V)^{-1}$, $\coev_V^{\dagger} = \ev_V c_{VV^{\ast}}(\theta_V \otimes \Id_{V^{\ast}})$, and $\ev_V^{\dagger} =
 (\Id_{V^{\ast}} \otimes \theta_V) c_{VV^{\ast}} \coev_V$.     Such relations make $\cat$ into a \emph{Hermitian ribbon category} in the sense of~\cite[Section 5.1]{Tu}.
  A key result is that a Hermitian ribbon category $\cat$ equipped with a categorical trace as in Section~\ref{subsec:modtrace} will have a natural Hermitian pairing on $\Hom(V,W)$ by
\begin{equation} \label{eq:trace-pairing}
    \langle f, g\rangle := \mt({f^\dag} g).
\end{equation}
This means that all $\Hom$-spaces in $\cat$ will be equipped with a Hermitian inner product (with respect to a possibly indefinite inner product) and that braiding anyons will be unitary with respect to this inner product.  
In the case of semisimple quantum group representations, such a dagger map was defined by Kirillov \cite{Kir96} and Wenzl \cite{Wen98}, and they proved that the pairing in \eqref{eq:trace-pairing} defines a non-degenerate Hermitian pairing making the associated categories Hermitian.  The unitarity of these theories implies that the resulting forms will be positive definite.

One way to construct the $\dagger$ map on a category of representations is to observe that the Hopf algebra $\Ubar$ itself admits a natural dagger map given by
\begin{equation}\label{eq:hermiticity_relations}
    E^{\dagger} = F, \quad F^{\dagger}=E, \quad K^{\dagger} = K^{-1}, \quad H^{\dagger} = H    
\end{equation}
satisfying\footnote{These relations imply that $\dagger$ is an algebra antiautomorphism that is also a coalgebra antihomomorphism. } 
\begin{equation*}
\begin{split}
    (ax)^{\dagger} =\bar{a}x^{\dagger}, 
  \;\; (xy)^{\dagger}=y^{\dagger}x^{\dagger}, 
  \\
      (x^{\dagger})^{\dagger}  =x \quad \Delta(x^{\dagger}) =(^{\dagger}\otimes^{\dagger})(\tau(\Delta x)),
\end{split}
\end{equation*}
 $a \in \C \text{ and } x,y \in \Ubar$.    It then makes sense to study  representations $V$ of $\Ubar$ that are equipped with Hermitian inner products $\langle \cdot,\cdot \rangle_V$ compatible with this operation, meaning that
\begin{equation} \label{eq:comptible}
    \langle x^{\dagger} v, v' \rangle_V = \langle v, x v'\rangle_V  
\end{equation}
for any $x \in \Ubar$ and $v,v' \in V$. 
When two representations $V$ and $W$ of  a Hopf algebra like $\Ubar$ are each equipped with compatible Hermitian forms, then the adjoint $f^{\dagger}$ of the map $f \maps V \to W$   is shown in \cite{Wen98,GLPMS,GLPMS3} to be given  by the formula
\begin{equation}
\langle w, fv\rangle_W = \langle f^{\dagger}w, v \rangle_V,
\end{equation}
for $v\in V$ and $w\in W$.  Thus, the problem of defining an inner product on anyon spaces reduces to defining compatible Hermitian inner products for $\Ubar$-representations. This is possible if we restrict the allowed values of $\alpha$ to be real.  

To define a compatible Hermitian form on a simple $\Ubar$-module $V$, \eqref{eq:comptible} implies that the form $\langle \cdot, \cdot\rangle_V$ must in particular satisfy 
 \begin{equation}\label{eq:hermitian_form}
     \langle E v, w\rangle_V^{'} = \langle v, Fw\rangle_V^{'}
 \end{equation}
for all $v,w \in a$. A generic Hermitian form $\langle \cdot , \cdot \rangle_V^{'}$, will not  satisfy \eqref{eq:hermitian_form}, but a new form can be defined by $\langle\cdot, \cdot\rangle_V:=\langle\cdot, \eta_V\cdot\rangle_V^{'}$ where $\eta_V$ is a Hermitian operator. Then $\langle\cdot, \cdot\rangle_V$ satisfies \eqref{eq:hermitian_form} if and only if 
\begin{equation}\label{eq:hermitian_matrix_representation}
    {\rho_V(E)}^\dag \eta_V = \eta_V \rho_V(F) .
\end{equation}
Where $\rho_V(E) $ and $\rho_V(F) $ are the representations of $E$ and $F$ on $V$, respectively, and $\dag$ is the Hermitian conjugation defined by $\langle \cdot, \cdot\rangle_V^{'}$.

To see this, fix a basis $\{v_0, ...,v_k\}$ of $V$, where $v_i := F^i v_0$.  Then, it is clear that the form $\langle \cdot, 
 \cdot \rangle_V^{'}$ defined by $\langle v_i, 
 v_j\rangle_V^{'}=\delta_{ij}$ will not satisfy \eqref{eq:hermitian_form}, because of the asymmetry in the definition of $F v_i = v_{i+1}$ and $E v_i= \beta_iv_{i-1}$, where we denote by $\beta_i$ the coefficient that results from the action of $E$ on $v_i$. Then, an explicit calculation shows that $\eta_V=\text{diag}(1, \beta_1,\beta_1\beta_2, ... \prod_{i=1}^{k} \beta_i)$.
\begin{example}
    Let $r=4$ and $V=V_\alpha$ with basis as in \eqref{eq:Va}. Then 
\begin{equation*}
\begin{split}
    \Jp_{V_\alpha} &= \left(
\begin{array}{cccc}
    0 &[1] [1-\alpha] & 0 & 0 \\
    0 & 0 & [2] [2-\alpha ] & 0 \\
    0 & 0 & 0 & [3] [3-\alpha] \\
    0 & 0 & 0 & 0 \\
\end{array}\right),
\\
\Jm_{V_\alpha} &= \left(\begin{array}{cccc}
    0 & 0 & 0 & 0 \\
    1 & 0 & 0 & 0 \\
    0 & 1 & 0 & 0 \\
     0 & 0 & 1 & 0 \\\end{array}\right)
     \end{split}
\end{equation*}
and $\beta_1 = [1] [1-\alpha], \beta_2=[2] [2-\alpha], \beta_3 =[3] [3-\alpha]$. 
Thus, 
\begin{equation*} \scriptstyle 
    \eta_{V_\alpha} = \text{diag}\left(1,[1] [1-\alpha],[1] [1-\alpha][2] [2-\alpha],[1] [1-\alpha][2] [2-\alpha][3] [3-\alpha]\right).
\end{equation*}
\end{example}

\begin{remark}
One might imagine that it is possible to choose a basis that is more symmetric for the action of $E$ and $F$.  Note, however, that the entries of $\eta_V$ are not necessarily positive, nor do they necessarily have the same sign (as it happens in the case of   $V=V_\alpha$ described above), so the Hermitian form is not always positive-definite.
\end{remark}

The definition of a Hermitian form on tensor products is more subtle, and we refer the reader to \cite[Section 4]{GLPMS} for a detailed treatment. Here, we summarize the main results of \cite{GLPMS}, which builds on earlier work by Wenzl~\cite{Wen98}. Given two (not necessarily simple) modules $V$ and $W$ equipped with inner products $\langle \cdot, \cdot \rangle_V$ and $\langle \cdot, \cdot \rangle_W$, respectively, a natural but naive choice of inner product on the tensor product $V\otimes W$ is defined by
\begin{equation*}
    \langle v_1 \otimes w_1, v_2\otimes w_2\rangle_{V\otimes W}^{n}:=\langle v_1, v_2\rangle_V \langle w_1, w_2\rangle_W .
\end{equation*}
Such an inner product is, however, not compatible with the action of $\Ubar$ on $V\otimes W$, and the braiding is not unitary.

\begin{table}[]
\begin{tabular}{|c|c|c|}
\hline
\multicolumn{1}{|l|}{} & $S_i$ & $V_\alpha$ \\ \hline
$\theta$ & $q^{\frac{i(i+2-2r)}{2}}$ & $q^{\frac{\alpha^2-(r-1)^2}{2}}$ \\ \hline
$\sqrt{\theta}$ & $q^{\frac{i(i+2-2r)}{4}}$ & $q^{\frac{\alpha^2-(r-1)^2}{4}}$ \\ \hline
\end{tabular}
\caption{\justifying The twist and half-twist maps on $S_i$ for $i =0, ...,r-1$ and $V_\alpha$ for $\alpha \in (\C/ \Z)\cup r\Z$.}
\label{tab:twist}
\end{table}
In \cite{Wen98,GLPMS,GLPMS3}, it is shown that the crucial ingredient for extending Hermitian forms to tensor products is a square root of the half-twist defined in \eqref{eq:halftwist}.  In \cite{GLPMS,GLPMS3} such a square root can be defined and is given on simple $\Ubar$-modules in Table~\ref{tab:twist}.  The square root 
$\sqrt{\theta}_{W\otimes V}$ is computed by taking the decomposition into simples of $W{\otimes}V= \bigoplus\limits_i T_i$ and applying $\sqrt{\theta}$ to each simple in the decomposition: 
\begin{equation*}
\sqrt{\theta}_{W{\otimes}V}=U\left(\bigoplus\limits_i \sqrt{\theta}_{T_i}\right)U^{-1}
\end{equation*}
where $U^{-1}$ is the matrix that block-diagonalizes $W\otimes V$ into a direct sum (see Example~\ref{ex:X_Map}). 
Let $\tau_{WV}:W{\otimes} V\rightarrow V{\otimes} W$ be the flip map where $w\otimes v \mapsto v\otimes w$.
We can then define the maps $X_{VW}$ and $\eta_{VW}$ as 
\begin{equation}\label{eq:X_Map}
\begin{split}
X_{VW} &=\left(\sqrt{\theta}_{W\otimes V}\right)^{-1}c_{VW}\left(\sqrt{\theta}_V{\otimes}\sqrt{\theta}_W\right)\\
\eta_{VW} &=\tau_{WV}X_{VW} .
\end{split}
\end{equation}

This construction makes $\langle \cdot , \eta_{VW}\,\cdot \rangle_{V \otimes W}^n$ a Hermitian form on $V \otimes W$ satisfying \eqref{eq:hermitian_matrix_representation}. 

\begin{example}[$S_1\otimes S_1$]\label{ex:X_Map}
    As a first example, we  compute $X_{VW}$ in the case when $V=W=S_1$ and $r=4$ that arises in the traditional semisimple theory showing how the inner product simplifies.
    From Table~\ref{tab:twist}, we have that $\sqrt{\theta}_{1}= q^{-5/4} \Id_{1}$ while $c_{11} = \tau_{11}R_{11}$ can be calculated using \eqref{eq:R} to find that
    \begin{equation*}
        c_{11}=\left(
\begin{array}{cccc}
 q^{1/2} & 0 & 0 & 0 \\
 0 & 0 & q^{-1/2} & 0 \\
 0 & q^{-1/2} & (q-q^{-1})q^{-1/2} & 0 \\
 0 & 0 & 0 & q^{-1/2} \\
\end{array}
\right) \ .
    \end{equation*}

It now remains to compute $\sqrt{\theta}_{1{\otimes}1}$. In order to apply $\sqrt{\theta}$ to each simple in the decomposition $1{\otimes} 1 = 0 \oplus 2$ we must find a basis that makes the decomposition manifest. This process is completely analogous to finding the singlet and triplet subrepresentations of a composite system of two spin-$1/2$ particles using Clebsch-Gordan coefficients in the theory of angular momentum: one finds the highest weight vectors within $1 {\otimes} 1$ and acts on them with $\Delta \Jm$. For example, to find the subrepresentation of $1{\otimes}1$ isomorphic to $0$, one looks for a vector $w \in 1{\otimes} 1$ satisfying $\Delta\Jp_{1{\otimes}1}w =0, \, \Delta H_{1{\otimes}1} w=0$. There is a unique solution (up to a scalar): $w= s_0^1{\otimes}s_1^1 - q^{-1}s_1^1{\otimes}s_0^1$. Example \ref{ex:two_into_11} presents an analogous calculation for the subrepresentation of $1{\otimes}1$ isomorphic to $2$. Using these two results, we construct the transformation from the Clebsch-Gordan basis to the tensor product basis:
\begin{equation*}
    U=\left(
\begin{array}{cccc}
 0 & 1 & 0 & 0 \\
 1 & 0 & q^{-1} & 0 \\
 -q^{-1} & 0 & 1 & 0 \\
 0 & 0 & 0 & q+q^{-1} \\
\end{array}
\right).
\end{equation*}
Then, the half-twist on $1{\otimes}1$ is defined as 
\begin{equation*}
    \sqrt{\theta}_{1\otimes1}=U\left(\sqrt{\theta}_0{\oplus}\sqrt{\theta}_2\right)U^{-1}
\end{equation*}
where $\sqrt{\theta}_0 = \Id_0=1$ and $\sqrt{\theta}_2 = q^{-2}\Id_2$. 
Putting everything together, one finds that $X_{11} = \Id_{1\otimes 1}$, so that the Hermitian form on $1{\otimes 1}$ is just $\tau_{11} X_{11}= \tau_{11}$. Note that, in general, $X_{VW}$ is not the identity.
\end{example}

In Example~\ref{example:Ydag-alpha}, we compute an example in the non-semisimple setting where the $X$ map is nontrivial.

\section{Modified trace} \label{app:trace}
 
To make the notion of modified trace more precise,  let $\Proj$ be the full subcategory of $\cat$ consisting of projective
$\Ubar$-modules.  The subcategory $\Proj$ is an ideal (see also
\cite{GKP1}).  
In particular, if $X$ is in $\cat$ and $Y$ is in $\Proj$,
then $X \otimes Y$ is in $\Proj$.

For any objects $V,W$ of $\cat$ and any endomorphism
$
f \;\; = \;\;  
\hackcenter{
\begin{tikzpicture}[scale=0.7]
\draw[very thick, black, ->] (0,0) to (0,2);
\node at (0,2.3) {$V$};
\node at (0,-0.3) {$V$}; 
\draw[very thick, black, ->] (1,0) to (1,2);
\node at (1,2.3) {$W$};
\node at (1,-0.3) {$W$};
\node[draw, thick, fill=black!20,rounded corners=4pt,inner sep=4pt] (X) at (.5,1) {$\quad f \quad $};
\end{tikzpicture} }
\; $ of  $V\otimes
W$,
set
\begin{align}
    \label{E:trL}
\ptr_{L}(f) & \;\; = \;\;
\hackcenter{
\begin{tikzpicture}[scale=0.8]
\draw[very thick, black] (0,2) .. controls ++(0,.5) and ++(0,.5) ..(-1,2) to (-1,0) .. controls ++(0,-.5) and ++(0,-.5) .. (0,0);
\draw[very thick, black, ->] (0,0) to (0,2);
\node at (.1,-0.5) {$V$}; 
\draw[very thick, black, ->] (1,0) to (1,2);
\node at (1,2.3) {$W$};
\node at (1,-0.3) {$W$};
\node[draw, thick, fill=black!20,rounded corners=4pt,inner sep=4pt] (X) at (.5,1) {$\quad f \quad $};
\end{tikzpicture} }\\
& = (\ev_{V}\otimes \Id_{W})(\Id_{V^{*}}\otimes
f)(\tcoev_{V}\otimes \Id_{W}) \in \End(W),
\\
\ptr_{R}(f) &=
\hackcenter{
\begin{tikzpicture}[scale=0.8]
\draw[very thick, black] (1,2) .. controls ++(0,.5) and ++(0,.5) ..(2,2) to (2,0) 
    .. controls ++(0,-.5) and ++(0,-.5) .. (1,0);
\draw[very thick, black, ->] (0,0) to (0,2);
\node at (0,2.3) {$V$};
\node at (0,-0.3) {$V$}; 
\draw[very thick, black, ->] (1,0) to (1,2);
\node at (.85,-0.5) {$W$};
\node[draw, thick, fill=black!20,rounded corners=4pt,inner sep=4pt] (X) at (.5,1) {$\quad f \quad $};
\end{tikzpicture} }\\
& \;\; = \;\;
(\Id_{V}\otimes \tev_{W})  (f \otimes \Id_{W^{*}})
(\Id_{V}\otimes \coev_{W}) \in \End(V).
\end{align}
These operations can be thought of as a left and right partial trace of such an endomorphism $f$, respectively. The compatibility between a modified trace and these partial traces is given in the following definition. 

\begin{definition}\label{D:trace}  A \emph{trace on $\Proj$} is a family of linear functions
$$\{\mt_V:\End(V)\rightarrow K\}$$
where $V$ runs over all objects of $\Proj$, such that the following two
conditions hold:
\begin{enumerate}
\item  If $U\in \Proj$, and $W\in \ob$, then for any $f\in \End(U\otimes W)$, we have
\begin{equation}\label{E:VW}
\mt_{U\otimes W}\left(f \right)=\mt_U \left( \ptr_R(f)\right).
\end{equation}
\item  If $U,V\in \Proj$, then for any morphisms $f:V\rightarrow U $, and $g:U\rightarrow V$  in $\cat$, we have
\begin{equation}\label{E:fggf}
\mt_V(g\circ f)=\mt_U(f \circ g).
\end{equation}
\end{enumerate}
\end{definition}
There exists up to a scalar a unique trace on $\Proj$. 

\section{Non-semisimple $F$ symbols} \label{app:F}
For the traditional Ising category, the $F$-symbols describing the change of basis for $((1,1)_a,1)_1$ with $a=0,2$ into the space $(1,(1,1)_b)_1$ gives the nontrivial $F$-symbols
\[
((1,1)_a,1)_1 = \sum_b \left(F^{111}_{1} \right)_{ba} (1,(1,1)_b)_1 ,
\]
\[
\left(F^{111}_{1} \right) = \frac{1}{\sqrt{2}} \left(
  \begin{array}{cc}
1 & 1 \\
    1& -1 \\
  \end{array}
\right).
\]
For the new non-semisimple anyons, fusion rules can be computed by comparing the images of a highest weight vector of $d$ under the maps on both sides of \eqref{eq:F_move}. 

\begin{example} \label{ex:alpha11-F}
We compute non-semisimple $F$-symbols $F^{\alpha 11}_\alpha$.
\begin{equation}
F^{\alpha11}_{\alpha}
~=~
\begin{blockarray}{ccc}
\alpha{+}1 & \alpha{-}1 \\
\begin{block}{[cc]c}
F_{0(\alpha{+}1)} & F_{0(\alpha{-}1)} & 0\\
F_{2(\alpha{+}1)} & F_{2(\alpha{-}1)} & 2\\
\end{block}
\end{blockarray}
\end{equation}
The first column amounts to solving the following linear equation.

\begin{equation}\label{eq:F_example}
\begin{split}
\hackcenter{\begin{tikzpicture}[ scale=1.1]
  \draw[ultra thick, black] (0,0) to (.6,-.6) to (.6,-1.2);
  \draw[ultra thick, black] (0.6,0) to (.3,-.3);
  \draw[ultra thick, black] (1.2,0) to (.6,-.6);
   \node at (0,0.2) {$\alpha$};
   \node at (.6,0.2) {$1$}; 
   \node at (1.2,0.2) {$1$};
   \node at (.8,-1) {$ \alpha$};
   \node at (.2,-.6) {$\scriptstyle \alpha{+}1$};
\end{tikzpicture} }
&~=~[F_{\alpha}^{\alpha 11}]_{0(\alpha{+}1)}
\hackcenter{\begin{tikzpicture}[ scale=1.1]
  \draw[ultra thick, black] (0,0) to (.6,-.6) to (.6,-1.2);
  \draw[ultra thick, black] (0.6,0) to (.9,-.3);
  \draw[ultra thick, black] (1.2,0) to (.6,-.6);
   \node at (0,0.2) {$\alpha$};
   \node at (.6,0.2) {$1$}; 
   \node at (1.2,0.2) {$1$};
   \node at (.8,-1) {$ \alpha$};
   \node at (.9,-.6) {$\scriptstyle 0$};
\end{tikzpicture} } \\
&~+~[F_{\alpha}^{\alpha 11}]_{2(\alpha{+}1)}
\hackcenter{\begin{tikzpicture}[ scale=1.1]
  \draw[ultra thick, black] (0,0) to (.6,-.6) to (.6,-1.2);
  \draw[ultra thick, black] (0.6,0) to (.9,-.3);
  \draw[ultra thick, black] (1.2,0) to (.6,-.6);
   \node at (0,0.2) {$\alpha$};
   \node at (.6,0.2) {$1$}; 
   \node at (1.2,0.2) {$1$};
   \node at (.8,-1) {$ \alpha$};
   \node at (.9,-.6) {$\scriptstyle 2$};
\end{tikzpicture} }\ .
\end{split}
\end{equation}
We compute these maps on the image of 
the highest-weight vector $v_0^{\alpha}$ of $V_\alpha$. The map on the left-hand-side yields
\begin{equation*}
\begin{split}
    &\left(Y^{\alpha 1}_{\alpha{+}1} \otimes \text{Id}_1 \right)\circ Y^{(\alpha{+}1)1}_\alpha (v_0^\alpha)\\
    &= \left(Y^{\alpha 1}_{\alpha{+}1} \otimes \text{Id}_1 \right)\left(v_0^{\alpha{+}1}\otimes s^1_1 + \frac{q^{-1}}{[\alpha]}v^{\alpha{+}1}_1 \otimes s^1_0 \right)\\
    &= v_0^\alpha\otimes s_0^1\otimes s_1^1 + \frac{-q^{-\alpha}}{[\alpha]} v_0^\alpha\otimes s_1^1 \otimes s_0^1 + \frac{q^{-1}}{[\alpha]} v_1^\alpha\otimes s_0^1\otimes s_0^1 .
\end{split}
\end{equation*}
Similarly, the RHS computes to:
\begin{equation}
\begin{split}
    &[F_{\alpha}^{\alpha 11}]_{0(\alpha{+}1)}(v_0^\alpha\otimes s_0^1\otimes s_1^1 + q^3 v_0^\alpha\otimes s_1^1\otimes s_1^0) \\
    &+ [F_{\alpha}^{\alpha 11}]_{2(\alpha{+}1)} ( -q^3v_0^\alpha\otimes s_0^1\otimes s_1^1 + v_0^\alpha\otimes s_1^1\otimes s_1^0 \\
    &+\sqrt{2}\frac{q^{-2}}{[\alpha-1]} v_1^\alpha\otimes s_0^1\otimes s_0^1 ).
\end{split}
\end{equation}
By equating these two equations, we can verify that \eqref{eq:F_example} indeed has a unique solution up to a scalar, given by 
\begin{equation*}
    [F_{\alpha}^{\alpha 11}]_{0(\alpha{+}1)} = \frac{1}{2}
    \left(1+\text{cot}\left(\frac{\pi \alpha}{4}\right)\right)
\end{equation*}
and
\begin{equation*}[F_{\alpha}^{\alpha 11}]_{2(\alpha{+}1)} = -\frac{1}{2}q
    \left(\text{cot}\left(\frac{\pi \alpha}{4}\right) - 1\right).
\end{equation*}
The other entries of $F^{\alpha 11}_\alpha$ can be calculated in an analogous manner to yield:
\begin{equation}\label{eq:F_Move_a11}
    [F^{\alpha11}_1]=
\begin{bmatrix}
 \frac{1}{2} \left(\cot \left(\frac{\pi  \alpha }{4}\right)+1\right) & -\frac{1}{\sqrt{2}} \\
 -\frac{1}{2} q \left(\cot \left(\frac{\pi  \alpha }{4}\right)-1\right) & \frac{1}{2}+\frac{i}{2} \\
\end{bmatrix}.
\end{equation}

Note that $[F^{\alpha11}_1]$ is not unitary in this basis. In general, this procedure will not yield a unitary matrix with respect to the inner product on $\Hom(\alpha,\alpha \otimes 1 \otimes 1)$ defined in Section~\ref{sec:hermitian_structures}, as we have made no assumption on the norm of the basis vectors involved. We will return to this matter in Section~\ref{subsec:normal}.
\end{example}

Here, we provide certain (unnormalized) $F$-symbols that arise within the non-semisimple Ising framework.

\begin{align*}
F^{\alpha 11}_\alpha &= \left(\begin{array}{cc}
F_{0(\alpha{+}1)} & F_{0(\alpha{-}1)}\\
F_{2(\alpha{+}1)} & F_{2(\alpha{-}1)}
\end{array}\right) \\
&= \frac{1}{\sqrt{2} (q^{2\alpha} - 1)}\left(\begin{array}{cc}
q(q^{2\alpha} + q^2) & -(q^{2\alpha} - 1)\\
q^{2\alpha} - q^2 & q(q^{2\alpha} - 1)
\end{array}\right),\\
F^{\alpha21}_{\alpha{+}1} &= \left(\begin{array}{cc}
F_{1\alpha} & F_{1(\alpha{+}2)} \\
F_{3\alpha} & F_{3(\alpha{+}2)}
\end{array}\right) \\
&= \frac{1}{q^{2\alpha }+q^2}\left(\begin{array}{cc}
 (q^2-1)(q^{2\alpha }+q^2) & \left(q^2+1\right) \left(q^{2 \alpha }+1\right) \\
 (q^2+1)(q^{2\alpha }+q^2) &q^{2\alpha }-q^2
\end{array}\right),\\
F^{\alpha21}_{\alpha{-}1} &= \left(\begin{array}{cc}
F_{1\alpha} & F_{1(\alpha{-}2)} \\
F_{3\alpha} & F_{3(\alpha{-}2)}
\end{array}\right) \\
&= \frac{1}{q^{2 \alpha }-q^2}
\left(\begin{array}{cc}
 \left(q^2+1\right) \left(q^{2 \alpha }+q^2\right) & -2(q^{2 \alpha }-q^2) \\
 q^{2 \alpha }+1 & q^2(q^{2 \alpha }-q^2)
\end{array}\right),\\
F^{\alpha12}_{\alpha{+}1} &= \left(\begin{array}{cc}
F_{1(\alpha{+}1)} & F_{1(\alpha{-}1)} \\
F_{3(\alpha{+}1)} & F_{3(\alpha{-}1)}
\end{array}\right) \\
&= \frac{1}{q^{2 \alpha }-1}
\left(\begin{array}{cc}
 q^2 \left(q^{2 \alpha }+1\right) & -q(q^{2 \alpha }-1) \\
 \sqrt{2}\left(q^{2 \alpha }-q^2\right) & q^2(q^{2 \alpha }-1)
\end{array}\right),\\
F^{\alpha12}_{\alpha{-}1} &= \left(\begin{array}{cc}
F_{1(\alpha{+}1)} & F_{1(\alpha{-}1)} \\
F_{3(\alpha{+}1)} & F_{3(\alpha{-}1)}
\end{array}\right) \\
&= \frac{1}{q^{2 \alpha }-1}
\left(\begin{array}{cc}
 q \left(q^2+1\right) \left(q^{2 \alpha }+q^2\right) & -(q^{2 \alpha }-1) \\
 q^{2 \alpha }+1 & q(q^{2 \alpha }-1)
\end{array}\right).
\end{align*}

\section{Non-semisimple $R$-symbols} 
As in the previous section, these $R$-symbols can be directly computed by comparing the image of a highest weight vector under the action of each map.  
The $R$-symbols for the traditional Ising category are given by
\[
R^{ab}_c = (-1)^{(a+b-c)/2} \left(ie^{i\pi/8} \right)^{-(a(a+2)+b(b+2)-c(c+2))/2}
\] 
for $a,b,c \in \{ 0, 1,2\}$.  The new non-semisimple 
$R$-symbols are given as follows

\begin{equation}
\begin{array}{rcl@{\quad}rcl}
R^{\alpha2}_{\alpha{+}2}      &=& q^{3+\alpha}, 
& R^{2\alpha}_{\alpha{+}2}    &=& q^{3+\alpha}, \\
R^{\alpha1}_{\alpha{+}1}  &=& q^{(3+\alpha)/2}, 
& R^{1\alpha}_{\alpha{+}1}  &=& q^{(3+\alpha)/2}, \\
R^{\alpha2}_{\alpha}        &=& s_\alpha q^{1-\alpha}, 
& R^{2\alpha}_{\alpha}      &=& s_\alpha q^{3+\alpha}, \\
R^{\alpha1}_{\alpha{-}1}  &=& s_\alpha q^{-(1+3\alpha)/2}, 
& R^{1\alpha}_{\alpha{-}1}  &=& s_\alpha q^{(7+\alpha)/2}, \\
R^{\alpha2}_{\alpha{-}2}      &=& t_\alpha q^{1-3\alpha}, 
& R^{2\alpha}_{\alpha{-}2}    &=& t_\alpha q^{5+\alpha}, \\
R^{21}_{3}     &=& q, 
& R^{12}_{3}     &=& q, \\
R^{21}_{1}        &=& q, 
& R^{12}_{1}      &=& q^3, \\
R^{11}_{2}      &=& q^{1/2}, 
& R^{11}_{0}    &=& q^{5/2},
\end{array}
\end{equation}
where
\begin{equation}
\begin{split}
 s_\alpha &:=\begin{cases} 
      +1, & \alpha \in (0,1) \cup (5,8) \setminus \mathbb{Z} \mod 8\\
      -1, & \alpha \in (1,5)\setminus \mathbb{Z} \mod 8
   \end{cases}
\\
t_\alpha &:=\begin{cases} 
      +1, & \alpha \in (0,1) \cup (2,4) \setminus \mathbb{Z} \mod 4\\
      -1, & \alpha \in (1,2) \setminus \mathbb{Z} \mod 4
   \end{cases}
\end{split}
\end{equation}
 
\section{Bubble pops} \label{app:bubble}

Here, we provide the relevant bubble pop data.
\begin{equation}\label{eq:B-data}
\begin{aligned}
& B^{\alpha0}_{\alpha} 
= B^{\alpha1}_{\alpha{+}1} 
= B^{\alpha2}_{\alpha{+}2} 
= B^{11}_{2} 
= B^{21}_{3} 
= B^{12}_{3} 
= 1,  \\
%
&B^{11}_{1} 
= \left(B^{21}_{1}\right)^{-1} 
= B^{12}_{1} 
= -\sqrt{2}, \quad 
B_{\alpha{-}1}^{\alpha1} 
= \frac{\sqrt{2}}{-1 + \cot\left(\frac{\pi\alpha}{4}\right)}, \\[0.5ex]
%
& B_{\alpha}^{(\alpha{+}2)2} 
= 2 \cot\left(\frac{\pi \alpha}{4} \right), \quad
B_{\alpha}^{\alpha2} 
= \frac{\sqrt{2} \cos\left(\frac{\pi\alpha}{2}\right)}{1 - \sin\left(\frac{\pi\alpha}{2}\right)},\\ 
%
&B_{\alpha{+}1}^{\alpha 3} 
= \frac{\sqrt{2}}{1 - \tan\left(\frac{\pi\alpha}{4}\right)},
B_{\alpha{-}1}^{\alpha 3} 
= \frac{2 + 2 \tan\left(\frac{\pi\alpha}{4}\right)}{-1 + \cot\left(\frac{\pi\alpha}{4}\right)}.
\end{aligned}
\end{equation}

\begin{remark}
   In general, if the conjugates of trivalent vertices are represented by an upside-down trivalent vertex as in \eqref{eq:trivalent_conjugate}, the conjugate of a more complex fusion tree will equal the horizontal reflection of the graph.
\end{remark}

\begin{example} 
    Let us compute ${Y^{11}_0}^\dag$. Recall that $Y^{11}_0 = 
    \begin{bmatrix}
        0 & 1 & -q^{-1} & 0
    \end{bmatrix}^T$ and we showed in Example~\ref{ex:X_Map} that $\tau_{11} X_{11} = \tau_{11}$ in the tensor product basis. Hence, \eqref{eq:trivalent_conjugate} gives:
    \begin{equation}
    \begin{split}
        \left[\Yd\right]_{01, 0} &= \sum_{mn}\left[Y^{11}_0\right]_{0,mn} \langle v^1_m\otimes v^1_n, v^1_0\otimes v_1^1\rangle_{1\otimes 1}\\
        &= \sum_{mn}\left[Y^{11}_0\right]_{0,mn} \langle v^1_m\otimes v^1_n, \tau_{11}\left(v^1_0\otimes v_1^1\right)\rangle_{1\otimes 1}^n\\
        &= \sum_{mn}\left[Y^{11}_0\right]_{0,mn} \langle v^1_m, v^1_1\rangle_1 \langle v^1_n, v^1_0\rangle_1\\
        &= \left[Y^{11}_0\right]_{0,10} \langle v^1_1, v^1_1\rangle_1 \langle v^1_0, v^1_0\rangle_1\\
        &= \left[Y^{11}_0\right]_{0,10}[1]_q[1]_q \;\; =\;\;  1
    \end{split}
    \end{equation}
\end{example}

\begin{example} \label{example:Ydag-alpha}
The computation of ${Y^{\alpha1}_{\alpha{-}1}}^\dag$is similar to the previous example, except that  $X_{\alpha1}$ is not the identity map. Analogous to Example~\ref{ex:X_Map}, the matrix $ \sqrt{\theta}_{\alpha} \otimes \sqrt{\theta}_{1}$ can be computed from the values in Table~\ref{tab:twist}, the braiding $c_{\alpha1}$ is calculated using \eqref{eq:R}, and $\sqrt{\theta}_{1 \otimes \alpha}$ can be written down explicitly via its diagonalized form 
\begin{equation*}
\begin{tikzcd}
1 \otimes \alpha \arrow[rr, "\sqrt{\theta}_{1\otimes \alpha}"] \arrow[d, "U^{-1}", swap] & & 1 \otimes \alpha \\
(\alpha{+}1)\oplus(\alpha{-}1) \arrow[rr, "\sqrt{\theta}_{\alpha{+}1} \oplus \sqrt{\theta}_{\alpha{-}1}", swap] & & (\alpha{+}1)\oplus(\alpha{-}1) \arrow[u, "U", swap]
\end{tikzcd}
\end{equation*}
where $U=\begin{bmatrix}
    Y^{1\alpha}_{\alpha{+}1} & Y^{1\alpha}_{\alpha{-}1}
\end{bmatrix}$. It is often helpful to verify the correctness of the $X_{ab}$ map by checking $X_{ab}X_{ba}=\Id=X_{ba}X_{ab}$.

Following the commutative diagram gives the following matrix:
\vspace{3mm}
\begin{widetext}
\begin{equation}
 {Y^{\alpha1}_{\alpha{-}1}}^\dag=\vcenter{\hbox{%
    \hackcenter{
\begin{tikzpicture}[ scale=1.1, yscale=-1.0]
  \draw[ultra thick, black] (0,0) to (.6,-.6) to (.6,-1.2);
  \draw[ultra thick, black] (1.2,0) to (.6,-.6);
   \node at (0,0.2) {$\alpha$};
   \node at (1.2,0.2) {$1$};
   \node at (.6,-1.4) {$\alpha{-}1$};
\end{tikzpicture} }}}
~=~ \
\begin{blockarray}{ccccccccc}
\scs{v_0^\alpha \otimes s_0^1} & \scs{v_0^\alpha \otimes s_1^1} & \scs{v_1^\alpha \otimes s_0^1} & \scs{v_1^\alpha \otimes s_1^1} & \scs{v_2^\alpha \otimes s_0^1} & \scs{v_2^\alpha \otimes s_1^1} & \scs{v_3^\alpha \otimes s_0^1} & \scs{v_3^\alpha \otimes s_1^1} &  \\
\begin{block}{[cccccccc]c}
 0 & -q & -q^{2-\alpha } & 0 & 0 & 0 & 0 & 0 & \scs{v_0^{\alpha{-}1}}\\
 0 & 0 & 0 & -q & q^{-\alpha}\left(1-q^2\right) & 0 & 0 & 0 & \scs{v_1^{\alpha{-}1}}\\
 0 & 0 & 0 & 0 & 0 & -q & q^{-\alpha } & 0 & \scs{v_2^{\alpha{-}1}}\\
 0 & 0 & 0 & 0 & 0 & 0 & 0 & -q & \scs{v_3^{\alpha{-}1}}\\
\end{block}
\end{blockarray}.
\end{equation}

Following~\eqref{eq:bubble_pop}, the bubble pop can be computed as follows:
\begin{equation}
\begin{split}
{Y^{\alpha1}_{\alpha{-}1}}^\dag {Y^{\alpha1}_{\alpha{-}1}}
&~=~
\hackcenter{\begin{tikzpicture}[ scale=1.1]
  \draw[ultra thick, black] (.6,-.6) to (.6,-1.2);
   \node at (.1,-.2) {$\alpha$};
   \node at (1.1,-.2) {$1$};
   \node at (.6,-1.5) {$\alpha{-}1$};
   \node at (.6,1) {$\alpha{-}1$};
   \draw[ultra thick, black] (.9,-.2) to [out=270, in=45] (.6,-.6);
   \draw[ultra thick, black] (.9,-.2) to [out=90, in=315] (.6,.2);
   \draw[ultra thick, black] (.3,-.2) to [out=270, in=135] (.6,-.6);
   \draw[ultra thick, black] (.3,-.2) to [out=90, in=225] (.6,.2);
   \draw[ultra thick, black] (.6,.2) to (.6,.8);
\end{tikzpicture} }~=~
B_{\alpha{-}1}^{\alpha1}~
\hackcenter{\begin{tikzpicture}[scale=1.1]
  \draw[ultra thick, black] (.6,-1.2) to (.6,.8);
   \node at (1.1,-.2) {$\alpha{-}1$};
\end{tikzpicture} } 
~=~
\frac{\sqrt{2}}{-1+\cot \left(\frac{\pi  \alpha }{4}\right)} \Id_{\alpha{-}1}.
\end{split}
\end{equation}
\end{widetext}
\end{example}

%

%

 \end{document}